\documentclass[10pt,onecolumn]{IEEEtran}

\usepackage{amssymb}
\usepackage{amsmath}
\usepackage{multirow}
\usepackage{epsfig}
\usepackage{rotating}
\usepackage{subfigure}
\usepackage{algorithmic}
\usepackage{algorithm}
\usepackage{psfrag}
\usepackage{graphics}

\newif\ifTEXTDOUBLESPACING\TEXTDOUBLESPACINGtrue

\newtheorem{theorem}{Theorem}[section]

\newtheorem{lemma}[theorem]{Lemma}
\newtheorem{proposition}[theorem]{Proposition}

\newtheorem{definition}[theorem]{Definition}
\newtheorem{example}{Example}[section]

\newcommand{\EP}{{\mathrm{EP}}}
\newcommand{\BER}{{\mathrm{BER}}}
\newcommand{\FER}{{\mathrm{FER}}}
\newcommand{\Aut}{{\mathrm{Aut}}}

\newcommand{\transposed}{{\mathrm{T}}}

\newcommand{\AGDA}{{\mathrm{AGD}}_{A}}
\newcommand{\AGDB}{{\mathrm{AGD}}_{B}}
\newcommand{\W}{{\mathrm{W}}} 
\newcommand{\HS}{{\mathrm{HS}}} 

\newcommand{\ML}{{\mathrm{ML}}}

\def\ve#1{{\mathchoice{\mbox{\boldmath$\displaystyle #1$}}%
              {\mbox{\boldmath$\textstyle #1$}}%
              {\mbox{\boldmath$\scriptstyle #1$}}%
              {\mbox{\boldmath$\scriptscriptstyle #1$}}}}

\newcommand{\Hamminggeneralized}{ {\mathrm{h,\,g}}}
\newcommand{\Hammingstandard}{ {\mathrm{h,\,s}} }
\newcommand{\BCH}{ {\mathrm{BCH}} }
\newcommand{\modstar}{\,\mathrm{mod}^*\,}
\newcommand{\mymod}{\,\mathrm{mod}\,}
\newcommand{\cog}{{\mathrm{cog}}}
\newcommand{\A}{{\mathrm{A}}}
\newcommand{\B}{{\mathrm{B}}}
\newcommand{\C}{{\mathrm{C}}}
\newcommand{\D}{{\mathrm{D}}}

\begin{document}
\title{Permutation Decoding and the Stopping Redundancy Hierarchy
of Cyclic and Extended Cyclic Codes
\thanks{Part of this work was presented at the 44th Annual Allerton Conference on Communication,
Control, and Computing, Monticello, Illinois, Sept. 2006, and the
IEEE International Symposium on Information Theory, Nice, France,
June 2007. This work was supported in part by NSF Grant
CCF-0514921 awarded to Olgica Milenkovic, by a research fellowship
from the Institute for Information Transmission, University of
Erlangen-Nuremberg, Erlangen, Germany, awarded to Stefan Laendner,
and by a German Academic Exchange Service (DAAD) fellowship
awarded to Thorsten Hehn.} }

\author{Thorsten Hehn\ddag, Olgica Milenkovic\dag, Stefan
Laendner\ddag, and
Johannes B. Huber\ddag,\\
\ddag University of Erlangen-Nuremberg, Erlangen, Germany\\
\dag University of Illinois at Urbana-Champaign, USA
}

\date{}
\maketitle
\thispagestyle{empty}

\begin{abstract}
We introduce the notion of the stopping redundancy hierarchy of a
linear block code as a measure of the trade-off between
performance and complexity of iterative decoding for the binary
erasure channel. We derive lower and upper bounds for the stopping
redundancy hierarchy via Lov{\'a}sz's Local Lemma and
Bonferroni-type inequalities, and specialize them for codes with
cyclic parity-check matrices. Based on the observed properties of
parity-check matrices with good stopping redundancy
characteristics, we develop a novel decoding technique, termed
automorphism group decoding, that combines iterative message
passing and permutation decoding. We also present bounds on the
smallest number of permutations of an automorphism group decoder
needed to correct any set of erasures up to a prescribed size.
Simulation results demonstrate that for a large number of
algebraic codes, the performance of the new decoding method is
close to that of maximum likelihood decoding.
\end{abstract}


\ifTEXTDOUBLESPACING
\baselineskip=1.8\baselineskip
\fi
\section{Introduction}\label{sec:introduction}

The error-correcting performance of linear block codes under
iterative decoding depends jointly on a number of combinatorial
properties of their parity-check matrices and corresponding Tanner
graphs. These parameters include the minimum distance, weight
distribution, girth, and diameter of the code graph. One exception
is the binary erasure channel (BEC) transmission scenario, where
only one class of combinatorial objects, termed \emph{stopping
sets}, completely characterizes the failure events during
\emph{edge removal} iterative decoding~\cite{dietal02}.

A stopping set is a collection of variable nodes in the Tanner
graph of a code such that all check nodes in the subgraph induced
by the variable nodes and their neighbors have degree at least
two. The size of the smallest stopping set, as well as the
distribution of stopping set sizes, depends on the particular form
of the parity-check matrix used for decoding. Since including a
large number of rows in the parity-check matrix of a code ensures
increased flexibility in meeting predefined constraints on the
structure of stopping sets, several authors recently proposed the
use of redundant parity-check matrices for improving the
performance of iterative decoders
\cite{schwartzetal06,ghaffaretal07,hanetal07,hehnetal06, hanetal07a}. The
effects of augmenting the sets of parity-checks in matrices from
random ensembles were also studied in \cite{milenkovicetal06, wadayama07}.

Redundant rows in a parity-check matrix improve the performance of
the decoder, but they also increase the overall time and hardware
complexity of decoding. This motivates the study of achievable
trade-offs between the number of redundant rows and the size of
the smallest stopping set(s) in the chosen parity-check matrix. In
this context, Schwartz and Vardy~\cite{schwartzetal06} introduced
the notion of the \emph{stopping redundancy} of a linear block
code. The stopping redundancy represents the smallest number of
codewords that span the dual code and constitute the rows of a
matrix with no stopping sets of size smaller than the minimum
distance of the code. The same authors also derived lower and
upper bounds on the stopping redundancy, the latter growing
exponentially with the co-dimension of the code for most examples
considered. This finding raised the question if there exist codes
for which one can significantly decrease the number of redundant
rows in a parity-check matrix by slightly decreasing the size of
its smallest stopping set.

Some results in this direction were independently derived by Weber
and Abdel-Ghaffar~\cite{weberetal05}, and Hollmann and Tolhuizen
\cite{hollmannetal07}, who posed and partly solved the more
complicated question of determining the smallest redundancy needed
for decoding all \emph{correctable erasure patterns}.

We extend the scope of the work in~\cite{hollmannetal07} by
introducing the notion of the \emph{stopping redundancy hierarchy}
of a code. The stopping redundancy hierarchy characterizes the
achievable trade-off between the stopping distance of a
parity-check matrix and its number of rows. We focus our attention
on codes with parity-check matrices of cyclic form. In particular,
we analyze cyclic and extended cyclic codes, which have a rich
mathematical structure and large minimum distance. We also improve
a set of bounds presented in~\cite{hollmannetal07}, pertaining to
the family of BCH codes.


For cyclic codes, we derive the best known constructive upper
bounds on the stopping redundancy hierarchy, using parity-check
matrices of cyclic form. Our construction methods imply that,
rather than adding redundant rows to the parity-check matrix, one
can change the belief propagation algorithm instead, by combining
it with \emph{permutation decoding}. Permutation decoding was
first proposed for decoding of cyclic codes over the binary
symmetric channel in~\cite{macwilliamsetal77}. In the new setting,
permutation decoders can be seen as creating ``virtual'' redundant
rows in the parity check-matrix of a code. The permutation
algorithm itself operates on properly designed
\emph{non-redundant} parity-check matrices, and it has the
property that it sequentially moves collections of erasures
confined to stopping sets into positions that do not correspond to
stopping sets.

The contributions of our work are three-fold. The first
contribution consists in introducing the notion of the stopping
redundancy hierarchy of a code and providing general upper and
lower bounds on the elements of this ordered list. The second
contribution lies in deriving upper and lower bounds on the stopping
redundancy hierarchy of parity-check matrices of cyclic form. For
the class of cyclic codes, we characterize the relationship
between the stopping redundancy hierarchy and the minimum-weight
codewords of their dual codes. As a third contribution, we
demonstrate how the stopping redundancy hierarchy can be studied
in the context of \emph{permutation decoding}. More specifically,
we describe a new decoding strategy, termed \emph{automorphism
group decoding}, which represents a combination of iterative
message passing and permutation decoding. In connection with
iterative permutation decoders, we study a new class of invariants
of the automorphism group of a code, termed $s$-\textbf{S}topping
set \textbf{A}utomorphism group \textbf{D}ecoder (\textbf{s-SAD})
sets.

The algorithms described above are tested on the $[23,12,7]$ Golay
code and the extended $[24,12,8]$ Golay code, a set of primitive
BCH codes, as well as on a representative subclass of
quadratic-residue codes~\cite{macwilliamsetal77}. Our findings
indicate that automorphism group decoders exhibit near-maximum
likelihood (ML) performance for short to moderate length cyclic
codes.

The paper is organized as follows. Section~\ref{sec:definition}
contains a summary of the terminology used throughout the paper
and introduces the notion of the stopping
redundancy hierarchy of a linear block code. A collection
 of general upper and lower bounds on the hierarchy is presented in Section~\ref{sec:bounds}.
Section~\ref{sec:cyclic} outlines several methods for studying the
stopping distance of cyclic codes. Section~\ref{sec:analysis}
introduces the automorphism redundancy, as well as the
notion of \textbf{s-PD} and \textbf{s-SAD} sets. Automorphism
group decoders are described in Section~\ref{sec:algos}.
Section~\ref{sec:results} provides a sampling of results regarding
the performance of automorphism group decoders. Concluding remarks
are given in Section~\ref{sec:conclusions}.

\section{Definitions and Terminology}
\label{sec:definition}

We state next the definitions and terminology used throughout the
paper. We restrict our attention to binary, linear block-codes
$\mathcal{C}$ with parameters $[n,k,d]$, used for signalling over
the {\emph{binary erasure channel}} (BEC). The erasure probability
$\EP$ of a BEC is assumed to satisfy $0 <\EP <1$. The iterative
edge removal decoder operates on a suitably chosen parity-check
matrix $\ve{H}$ of the code $\mathcal{C}$. For such a
communication scenario, decoding errors are confined to subsets of
codeword-coordinates termed \emph{stopping sets} \cite{dietal02}.

Stopping sets can be formally defined by referring either to the
parity-check matrix of the code, or its underlying Tanner graph.
For the former approach, we first introduce the notion of a
\emph{restriction} of a matrix.

\begin{definition} Let the columns
of $\ve{H}$ be indexed by $J=\left\{0,\dots, n-1\right\}$. For a
set $I\subseteq J$, $|I| \geq 1$, the {\emph{restriction}} of
$\ve{H}$ to $I$ is the matrix that consists of the set of columns
of $\ve{H}$ indexed by the elements of $I$.
\end{definition}

\begin{definition} For a given parity-check matrix $\ve{H}$ of $\mathcal{C}$,
a stopping set of size $\sigma$ is a subset $I$ ($|I|=\sigma$) of
columns of $\ve{H}$, such that the restriction of $\ve{H}$ to $I$
avoids rows of Hamming weight one.
\end{definition}

Alternatively, let $\ve{H}$ be a fixed parity-check matrix of
$\mathcal{C}$, and let the bipartite graph $G=(L \cup R,E)$ be
such that the columns of $\ve{H}$ are indexed by the
\emph{variable nodes} in $L$, and the rows of $\ve{H}$ are indexed
by \emph{check nodes} in $R$. For two vertices $i \in L$ and $j
\in R$, $(i,j) \in E$ iff $H_{j,i}=1$. The graph $G$ constructed
in this manner is called the Tanner graph of the code
$\mathcal{C}$ with parity-check matrix $\ve{H}$. For $S \subset
L$, we use $\Gamma(S)$ to denote the set of neighbors of $S$ in
$R$.

Stopping sets can be defined using the notion of Tanner graphs as
follows.

\begin{definition}
Given a bipartite graph $G=(L \cup R,E)$, we say that $S\subset L$
is a \emph{stopping set} if the degree of each vertex in
$\Gamma(S)$, restricted to the subgraph induced by $S\cup
\Gamma(S)$, is greater than one.
\end{definition}

The \emph{stopping distance} of a parity-check matrix $\ve{H}$,
and its underlying Tanner graph, is the cardinality of its
smallest stopping set. Note that in the classical sense, the
stopping distance does not actually represent a distance measure.
Nevertheless, we adopt this terminology since it is by now
standardly used in the coding theory literature.

The sizes and the number of stopping sets in a Tanner graph
depends on the particular choice of the parity-check matrix. It is
straightforward to see that adding rows to a fixed parity-check
matrix $\ve{H}$ may only increase its stopping distance. To
maintain the integrity of the code, the added rows must represent
linear combinations of the base vectors in $\ve{H}$, and we refer
to such rows as \emph{redundant rows (redundant parity-checks)}. A
parity-check matrix $\ve{H}$ containing redundant parity-check
equations is henceforth termed a \emph{redundant parity-check
matrix}. The phrase ``parity-check
 matrix'' is reserved for a matrix with the smallest possible number of
rows, i.e.\ for a matrix of full row-rank; similarly, the symbol
$m$ is reserved for the row-dimension of the (possibly) redundant
parity-check matrix $\ve{H}$. Finally, we say that a matrix
$\ve{H}$ (or, a subset of rows in $\ve{H}$) \emph{resolves} a set
of coordinates $I$ if the restriction of $\ve{H}$ (or the subset
of rows of $\ve{H}$) to $I$ does not correspond to a stopping set.
Observe that resolvability does not imply that \emph{all}
coordinates in $I$ can be retrieved if $I$ is not a stopping set.
In the latter case, i.e.\ when $I$ is not a stopping set itself
but ``contains'' one or more stopping sets, only a subset of its
coordinates can be retrieved.


We are now ready to define the stopping redundancy hierarchy of a
code.

\begin{definition} Let $\mathcal{C}$ be a linear
code with minimum distance $d$. For $\ell \leq d$, the $\ell$-th
stopping redundancy of $\mathcal{C}$ is the smallest non-negative
integer $\rho_{\ell}(\mathcal{C})$ such that there exists a
(possibly redundant) parity-check matrix $\ve{H}$ of $\mathcal{C}$
with $\rho_{\ell}(\mathcal{C})$ rows and stopping distance at
least $\ell$. The ordered set of integers
\[
\left(\rho_1(\mathcal{C}),\rho_2(\mathcal{C}),\rho_3(\mathcal{C}),\ldots,\rho_d(\mathcal{C})\right)
\]
is called the \emph{stopping redundancy hierarchy} of
$\mathcal{C}$. The \emph{order} of an element in the stopping
redundancy hierarchy is its position in the list. The integer
$\rho_d(\mathcal{C})$ is denoted the stopping redundancy of $\mathcal{C}$,
and was first introduced in~\cite{schwartzetal06}.
\end{definition}


For codes with minimum distance $d \geq 3$, 
no two columns of the parity-check matrix are identical nor is any
of the columns equal to the all-zero vector. Therefore,
$\rho_1(\mathcal{C})=\rho_2(\mathcal{C})=\rho_3(\mathcal{C})=n-k$.
Consequently, only stopping redundancies of order larger than
three are considered. Finding the stopping redundancy of a given
order is very likely a complicated problem, since it was recently
shown that computing the stopping distance of a matrix $\ve{H}$ is
NP-hard and NP-hard to
approximate~\cite{krishnanetal06,mcgregoretal07}. This is why we
henceforth focus only on deriving upper and lower bounds on the
elements of the hierarchy.

In the exposition to follow, we also frequently refer to the
notion of {\emph{(un)correctable erasure patterns}}
\cite{hollmannetal07}, defined below. Only general properties of
the code, but not the particular choices of their parity-check
matrices, determine whether a pattern is correctable or not.

\begin{definition}
Let the entries of a codeword be indexed by $J=\left\{0,\dots,
n-1\right\}$, and define the support of the codeword as the set of
its non-zero coordinates. A correctable erasure pattern is a set
$I\subseteq J$ that does not properly contain the support of any
codeword $c\in{\mathcal{C}}$. An uncorrectable erasure pattern is
a set of positions that does not correspond to a correctable
erasure pattern.
\end{definition}

It is clear from the definition of uncorrectable erasure patterns
that they represent erasure configurations that cannot be
successfully reconstructed via ML decoding. An uncorrectable
erasure pattern contains a stopping set, but the converse claim is
not true.

\section{Bounds on the Stopping Redundancy Hierarchy}
\label{sec:bounds}

\subsection{Lower Bounds on the Stopping Redundancy Hierarchy}
\label{sec:lower_bounds_on_the_stopping_redundancy_hierarchy}

We present next lower bounds on the $\ell$-th stopping redundancy,
$\ell=4,\ldots,d,$ of arbitrary binary linear codes. The first two
bounds represent generalizations and extensions of lower bounds on
the stopping redundancy derived in~\cite{schwartzetal06}. For
completeness, we briefly state this result below.

\begin{theorem}(\cite{schwartzetal06})\label{mt1}
Let $\mathcal{C}$ be a binary linear code with parameters
$[n,k,d]$, and let
\begin{equation} \label{eq:ch-w}
\omega_{\sigma}=\max \{ \lceil (n+1)/\sigma \rceil-1,d^{\perp} \},\;
\sigma=1,\ldots,d-1,
\end{equation}
where $d^{\perp}$ denotes the minimum distance of
$\mathcal{C}^{\perp}$. Then
\begin{equation} \label{uper}
\rho_{d}(\mathcal{C}) \geq
\max\left(n-k,\,\max_{\sigma}\left(\frac{\binom{n}{\sigma}}{\omega_{\sigma}\;
\binom{n-\omega_{\sigma}}{\sigma-1}}\right)\right),\,\sigma\in\{1,\dots,d-1\}.\notag
\end{equation}


\end{theorem}
It is instructive to briefly repeat the arguments leading to the
results in Theorem~\ref{mt1}.


The bound is established by first noting that there exist
$\binom{n}{\sigma}$ subsets of columns of $\ve{H}$ of cardinality
$\sigma$. A row of Hamming weight $w$ can resolve exactly
$w\,\binom{n-w}{\sigma-1}$ stopping sets of size $\sigma$. A
simple application of the union bound shows that
\begin{equation}
\binom{n}{\sigma} \leq \sum_{i=1}^{\rho_{d}(\mathcal{C})}\, w(i)
\binom{n-w(i)}{\sigma-1} \leq \rho_{d}(\mathcal{C})
\,\max_i\left(w(i)\,\binom{n-w(i)}{\sigma-1}\right), \notag
\end{equation}
where $w(i)$ denotes the weight of the $i$-th row in $\ve{H}$. The
result follows by noting that this expression has to hold for all
$\sigma \leq d-1$ and that the upper bound is maximized by
choosing the row weights $w(i)$ according to
Equation~\eqref{eq:ch-w}.

\begin{theorem}\label{ja1} Let $\mathcal{C}$ be a binary $[n,k,d]$ linear
code, and let
\begin{equation}
\omega_\sigma=\max \{ \lceil (n+1)/\sigma \rceil-1,d^{\perp} \},\;
\sigma=1,\ldots,d-1, \notag
\end{equation}
where $d^{\perp}$ denotes the minimum distance of
$\mathcal{C}^{\perp}$. Then
\begin{equation} \label{eq:thm2.2}
\rho_{\ell}(\mathcal{C}) \geq \max\left(n-k,\,\max_{\sigma}\left(
\frac{\binom{n}{\sigma}}{\omega_\sigma\;\binom{n-\omega_{\sigma}}{\sigma-1}}\right)\right),\,\sigma\in\{1,\dots,\ell-1\}. \notag
\end{equation}

\end{theorem}
Note that Theorem~\ref{ja1} is a straightforward generalization of
Theorem~\ref{mt1}. Hence, the proof of this result follows the
proof of Theorem~\ref{mt1} very closely and is therefore omitted.
Both Theorem~\ref{mt1} and \ref{ja1} can be shown to be very loose
through numerous examples.

Tighter lower bounds on the stopping redundancy hierarchy can be
derived by means of a more precise count of the number of subsets
of columns of $\ve{H}$ that do not correspond to stopping sets.
More specifically, the number of such subsets equals the
cardinality of the union of all subsets for which the restriction
of one individual row of $\ve{H}$ does not have weight one. The
result of Theorem \ref{ja1} represents a weak bound on the
stopping redundancy hierarchy due to the use of the inherently
loose union bound. More accurate bounds can be calculated by
invoking the principle of inclusion-exclusion, stated below.

\begin{theorem}[\cite{dohmen03}]
\label{PIE} Let $|A|$ denote the cardinality of the set $A$, and
let $V$ be the set $\{1,\dots,|V|\}$. For a family of sets $\{ A_v
\}_{v \in V}$, the principle of inclusion-exclusion (PIE) asserts
that
\begin{equation}\label{eq:PIE}
\left|\bigcup_{v \in V} A_v\right| = \sum \limits_{ \scriptsize
\begin{array}{c}I \subseteq V \\ I \neq \emptyset \end{array}}
(-1)^{|I|-1} \left|\bigcap_{v \in I} A_v\right| .\notag
\end{equation}
\end{theorem}


Let $\Sigma_{\sigma, i}$ denote the set of stopping sets of size
$\sigma$ resolved by the $i$-th row of $\ve{H}$. Also, let
$R=\{1,2,\ldots,m\}$ denote the set of row indices of $\ve{H}$,
and let $D_j$ denote the set of all $j$-subsets of $R$.

The number of stopping sets of size $\sigma$ resolved by $\ve{H}$ can be found from
the PIE equation as
\begin{equation} \label{sum}
|\bigcup_{i=1}^m \Sigma_{\sigma, i}| = \sum\limits_{
\begin{array}{c}I \subseteq R \\ I \neq \emptyset \end{array}}
(-1)^{|I|-1} |\cap_{i \in I} \Sigma_{\sigma, i} | =
\sum\limits_{j=1}^{m}  (-1)^{j-1} S_{\sigma,j}, \;\;\;\;\;\;\;\;\;
\textrm{where} \;\;\;\;\;\;\; S_{\sigma,j}=\sum\limits_{T \in
D_{j}}s_{\sigma, T}. \notag
\end{equation}
Here, $s_{\sigma, T}$ denotes the number of stopping sets of size
$\sigma$ resolved by the rows indexed by elements of the set $T$.

There exists a simple relationship between the number of stopping
sets resolved by a given parity-check matrix and the $\ell$-th
stopping redundancy hierarchy of the underlying code. The latter
equals the smallest possible dimension $m$ of a matrix that
resolves all stopping sets of size $\sigma$ smaller than $\ell$,
i.e.\ an $m$ such that
\begin{equation}
|\bigcup_{i=1}^m \Sigma_{\sigma, i}| = \binom{n}{\sigma}, \;\; 1
\leq \sigma < \ell. \notag
\end{equation}

The exact calculation of the number of stopping sets resolved by
an arbitrary collection of $j$ rows is very complex, unless the
code has certain regularity properties. One such property is that
all collections of $j$ codewords from ${\mathcal{C}}^{\bot}$ resolve the same number of
stopping sets. In this case, $S_{\sigma,j}=\binom{m}{j}\cdot
s_{\sigma}$, where $s_{\sigma}$ denotes the number of stopping
sets of size $\sigma$ resolved by the rows indexed with
$1,\ldots,j$.

The PIE is frequently applied in the form of an upper or lower
bound: this is accomplished by neglecting terms involving
intersections of sets including more than $b$ terms, i.e.\ by
retaining only those terms for which $|I|\leq b$~\cite{dohmen03}.
Whenever the sign of the first omitted term is negative, the
resulting expression represents an upper bound on the left hand
side of the PIE formula. Otherwise, the obtained formula
represents a lower bound. Inequalities obtained in this manner are
known as \emph{Bonferroni} inequalities.

There exists a rich body of work on Bonferroni
inequalities~\cite{dohmen03,kwerel75}, and one of its forms most amenable
for stopping set analysis is given below.
\begin{equation}\label{eq:bon}
\left|\bigcup_{v \in V} A_v\right| \leq \sum \limits_{ \scriptsize
\begin{array}{c}I \subseteq V \\ |I|=1 \end{array}}
(-1)^{|I|-1} \left|\bigcap_{v \in I} A_v\right| + \frac{2}{|V|}\; \cdot
\sum \limits_{ \scriptsize
\begin{array}{c}I \subseteq V \\ |I|=2 \end{array}}
(-1)^{|I|-1} \left|\bigcap_{v \in I} A_v\right|.
\end{equation}

This  allows us to state the following result.

\begin{proposition}
The number of distinct stopping sets of size $\sigma$ resolved by
a parity-check matrix $\ve{H}$ with $m$ rows satisfies
\begin{equation}
|\bigcup_{i=1}^m \Sigma_{\sigma, i}| \leq \sum\limits_{i=1}^{m}
|\Sigma_{\sigma, i} | - \frac{2}{m}\cdot
\sum\limits_{\begin{array}{c} i,j=1\\ i < j \end{array}}^{m} |
\Sigma_{\sigma, i} \cap \Sigma_{\sigma, j} |. \label{eq:bound2}
\end{equation}
The $\ell$-th order stopping redundancy $\rho_{\ell}(\mathcal{C})$
is lower bounded by the smallest integer $m$ for which the right
hand side of Equation~\eqref{eq:bound2} equals or exceeds
$\binom{n}{\sigma}$, for all $\sigma < \ell$.
\end{proposition}
\begin{proof}
Equation~\eqref{eq:bound2} is a straightforward consequence of
Equation~\eqref{eq:bon}; the second observation follows from the
fact that, in order for a matrix with $m$ rows to have stopping
distance at least $\sigma$, the upper bound on the number of
stopping sets it resolves must exceed the total number of stopping
sets of size $\sigma$ or less.
\end{proof}




The above results will be specialized for codes with cyclic
parity-check matrices in Section \ref{sec:cyclic}. General
expressions for the set intersection cardinalities
of~\eqref{eq:bound2} are presented below. They reveal that the
weights of the rows, as well as the \emph{gaps} (spacings) between
non-zero elements in the rows of a (redundant) parity-check matrix
$\ve{H}$ bear a strong influence on its stopping distance
hierarchy.

\begin{definition}\label{def:intersection_numbers}
Let $O_{l}$ and $Z_{l}$, $l \in \{{1,\dots,m\}},$ denote the sets
of positions of ones and zeros in the $l$-th row of the
parity-check matrix $\ve{H}$ with $m$ rows, respectively. The
\emph{intersection number} $|X_{i} \cap Y_{j}|$ of $\ve{H}$, where
$X$ and $Y$ serve as placeholders for $O$ and $Z$, and where
$i,j\in\{1,\dots,m\}$, is defined as the cardinality of the set
$P\subseteq J$ of column positions $p$ such that $H_{i,p}=X$ and
$H_{j,p}=Y$ for all $p\in P$.
\end{definition}

\begin{lemma}
The number of stopping sets of size $\sigma$, resolved by the
$i$-th row of $\ve{H}$ of weight $\omega(i)$, equals
\begin{equation}
|\Sigma_{\sigma, i}| = \omega(i) \cdot {n-\omega(i) \choose
\sigma-1}.\notag
\end{equation}
The number of stopping sets of size $\sigma$, resolved jointly by
two rows of $\ve{H}$ indexed by $i$ and $j$, equals
\begin{equation}
|\Sigma_{\sigma, i} \cap \Sigma_{\sigma, j}| = |O_i\cap O_j| \cdot
{|Z_i \cap Z_j|
  \choose \sigma-1} + |O_i\cap Z_j| \cdot |Z_i\cap O_j| \cdot {|Z_i \cap Z_j| \choose
  \sigma-2}\notag.
\end{equation}
\label{simulremovedssgeneral}
\end{lemma}

\begin{proof}
The first part follows from the fact that a stopping set of size
$\sigma$ can be resolved by the restriction of a row of weight
one, with the one-entry chosen among $\omega(i)$ options and the
remaining zeros from $\binom{n-\omega(i)}{\sigma-1}$ choices. The
second claim is a consequence of the following observation: a pair
of rows resolves the same stopping set if either both rows share a
``1'' and all the ``0''s in the remaining positions of the
stopping set; or, if within the support of the stopping set the
two rows share all except for two positions for the ``0'' symbols,
and have two non-overlapping positions with the symbol ``1''.
\end{proof}

The expressions in Lemma~\eqref{simulremovedssgeneral} can be
further simplified by assuming that all rows in $\ve{H}$ have
weight $\omega$ (provided that $\mathcal{C}^{\perp}$ has
sufficiently many codewords of weight $\omega$), and by writing
\begin{equation}
|\Sigma_P|=\max_{\{i,j\}} | \Sigma_{\sigma, i} \cap \Sigma_{\sigma, j} |. \notag
\end{equation}
Under these assumptions,
\begin{eqnarray} \label{eq:up}
|\bigcup_{i=1}^m \Sigma_{\sigma, i}| \leq m \cdot \omega \cdot
\binom{n-\omega}{\sigma-1}  - \frac{2}{m} \cdot \binom{m}{2} \cdot
|\Sigma_P|,
\end{eqnarray}
where $|\Sigma_P|$ can be found based on the formula in
Lemma~\ref{simulremovedssgeneral}.

A lower bound on the $\ell$-th stopping redundancy can be obtained
from Lemma~\ref{simulremovedssgeneral} by observing that the upper
bound in~\eqref{eq:up} has to exceed $\binom{n}{\sigma}$ for all
$\sigma \leq \ell$ . As a consequence,
\begin{equation}
\left.\rho_{\ell}(\mathcal{C})\right|_{\omega(i)=\omega} \geq \max_{\sigma < \ell} \; \left \lceil
\frac{\binom{n}{\sigma} - |\Sigma_P|}{\omega \,
\binom{n-\omega}{\sigma-1}-|\Sigma_P|} \right \rceil,
\label{eq:number_of_rows_min}
\end{equation}
where the qualifier $\omega(i)=\omega$ denotes that a constant codeword weight $\omega$ was assumed for this bound.
The previous analysis shows that, in order to resolve a large
number of stopping sets by a (redundant) parity-check matrix
$\ve{H}$ with a small number of rows, one has to minimize the
number of stopping sets resolved jointly by subsets of rows.
Therefore, small sizes for the intersections of the supports of
pairs, triples, etc., of rows are desirable. This is why we
henceforth restrict our attention to matrices $\ve{H}$ that
contain minimum weight codewords of $\mathcal{C}^\perp$ as their
rows.

It is also worth pointing out that for many codes the codewords
of minimum weight or other fixed weight represent a design
\cite{macwilliamsetal77}. In those cases, the intersection numbers
of the minimum weight codewords can be obtained from the
parameters of the design. Examples supporting this observation,
pertaining to Quadratic Residue (QR) codes and the Golay codes
\cite{macwilliamsetal77}, are given in subsequent sections.

\begin{example}
Consider a parity-check matrix of the form shown below:
$$
\ve{H}=\left(\begin{array}{ccccc}
1 & 0 & 0 & 0 & 1  \\
1 & 0 & 0 & 1 & 0
\end{array}\right)\,.
$$
The columns indexed by $\{{1,2,3\}}$ have a unique
row-restriction, namely ``100''. The columns indexed by
$\{{2,4,5\}}$ and $\{{3,4,5\}}$ have row restrictions ``001'' and
``010'', respectively. Consequently, both rows simultaneously
resolve stopping sets indexed by the described set of columns.
More precisely, both rows resolve the three restrictions indexed
by $\{1,2,3\}$, $\{2,4,5\}$, and $\{3,4,5\}$, while all other
restrictions are resolved by at most one of the rows.
\end{example}

\begin{example}
Consider the family of $[2^s-1,s,2^{s-1}]$ simplex codes. Since
simplex codes have constant codeword weight $2^{s-1}$, for all
distinct indices $i$ and $j$, the intersection numbers are given as $|O_{i}\cap O_{j}| = 2^{s-2}$,
$|O_{i}\cap Z_{j}|=|Z_{i}\cap O_{j}|=2^{s-2}$, and $|Z_{i}\cap
Z_{j}|=2^{s}-1-3\cdot 2^{s-2}=2^{s-2}-1$. The bound
in~\eqref{eq:number_of_rows_min} shows that for $i \neq j$, the
dual code of a simplex code, a $[2^s-1,2^s-s-1,3]$ Hamming code $\mathcal{C}(s)$,
satisfies
$$
|\Sigma_{\sigma,i} \cap \Sigma_{\sigma,j}|=|\Sigma_{\sigma,1} \cap
\Sigma_{\sigma,2}| = 2^{s-2}\,\left[\binom{2^{s-2}-1}{\sigma-1} +
2^{s-2}\binom{2^{s-2}-1}{\sigma-2}\right], 
$$
and that, consequently,
\begin{equation}\label{eq:rho_simplex}
\left.\rho_{3}(\mathcal{C}(s))\right|_{\omega_i=2^{s-1}} \geq  \max \limits_{\sigma<3} \left\lceil
\frac{\binom{2^{s}-1}{\sigma} - 2^{s-2}\cdot
\left[\binom{2^{s-2}-1}{\sigma-1} + 2^{s-2}\cdot
\binom{2^{s-2}-1}{\sigma-2} \right]} {2^{s-1}\cdot
\binom{2^{s-1}-1}{\sigma-1} - 2^{s-2}\cdot
\left[\binom{2^{s-2}-1}{\sigma-1} + 2^{s-2}\cdot
\binom{2^{s-2}-1}{\sigma-2} \right]} \right\rceil \;. \notag
\end{equation}
Simple evaluation of the bound reveals that
\begin{equation}
\left.\rho_{3}(\mathcal{C}(s))\right|_{\omega_i=2^{s-1}} \geq \max
\left( \left\lceil 3-2^{2-s} \right\rceil, \left\lceil \, \frac{3-5\cdot
2^{1-s}}{1-2^{1-s}} \, \right\rceil \right), \notag
\end{equation}
which implies the trivial result $\rho_{3}(\mathcal{C}(s)) \geq
3,\, s>1$. This illustrates the fact that PIE-based lower bounds
can be arbitrarily weak, since it is known that
$\rho_{3}(\mathcal{C}(s))=s$.
\end{example}

\subsection{Upper Bounds on the Stopping Redundancy Hierarchy}

Upper bounds on the stopping redundancy hierarchy can be derived
by invoking probabilistic methods~\cite{alonetal00}. Our
derivations follow the framework we developed
in~\cite{milenkovicetal06}, based on Lov{\'a}sz Local
Lemma~\cite{dengetal04,alonetal00}. Related probabilistic
techniques were also used for deriving upper bounds on the
stopping redundancy in~\cite{hanetal07}.

For subsequent derivations, we need the following results, known
as Lov{\'a}sz Local Lemma (LLL), as well as the high-probability
variation of LLL.

\begin{lemma} Let $E_1,E_2,\ldots,E_N$ be a set of events in an
arbitrary probability space. Suppose that each event $E_i$ is
independent of all other events $E_j$, except for at most $\tau$
of them, and that
\begin{equation}
\label{eq:lll}
P\{{E_i\}}\leq p, \;\;\forall\, 1 \leq i \leq N.
\end{equation}

If
\begin{equation} \label{eq:lll_tau_ineq}
{\mathrm{e}}\;p\;(\tau+1) \leq 1,
\end{equation}
where ${\mathrm{e}}$ is the base of the natural logarithm, then
$P\{{\bigcap_{i=1}^N\, \overline{E}_i\}}>0$.

Furthermore, let $0<\epsilon<1$. If
\begin{equation}\label{eq:lll_hpv}
P\{{E_i\}}\leq
\frac{\epsilon}{N}\left(1-\frac{\epsilon}{N}\right)^{\tau},
1 \leq i \leq N,
\end{equation}
then $P\{{\bigcap_{i=1}^N\, \overline{E}_i\}}>1-\epsilon$.
\label{lovasz_local_lemma}
\end{lemma}

Based on Lov{\'a}sz Local Lemma~\eqref{eq:lll} and its
high-probability variation~\eqref{eq:lll_hpv}, one can obtain the
following bounds on the $\ell$-th stopping redundancy of a code,
for $\ell \leq \lfloor \frac{d+1}{2} \rfloor$.
The derivations are straightforward, and based on associating
stopping distance properties of restrictions of a parity-check
matrix with the events $E_i$ described in the statement of LLL.

\begin{theorem} \label{improve} Let $\mathcal{C}$ be an $[n,k,d]$ code. If
$\ell \leq \lfloor \frac{d+1}{2} \rfloor$
and
\begin{equation}
m \geq
\frac{1+\log\sum\limits_{j=1}^{\ell-1} \left(\binom{n}{j} - \binom{n-j}{j}\right)}
{-\log\left(1-\frac{\ell-1}{2^{\ell-1}}\right)}
 + n - k - \ell + 1,%
\label{eq:improve}
\end{equation}
then $\rho_{\ell}(\mathcal{C}) \leq m$. Alternatively, if the
conditions of the theorem are fulfilled, then there exists at
least one parity-check matrix with $m$ rows and stopping distance
at least $\ell$.
\end{theorem}
The proof of this non-constructive bound is given in
Appendix~\ref{app:proof_improve}. An asymptotic estimate of the
above bound is presented in Appendix~\ref{app:asymptotic} for
codes with $d=\mathrm{const.}\cdot n$.

One way to make the finding of Theorem~\ref{improve} more useful
for practical purposes is to generalize it by invoking the high
probability version of LLL, as stated below.

\begin{theorem} \label{highimp} Let $\mathcal{C}$ be an $[n,k,d]$
code, let $\ell \leq \lfloor \frac{d+1}{2} \rfloor$, and assume that $m$ fulfills the condition 
\begin{equation}
m \geq
\frac{\log\frac{\epsilon}{\sum_{j=1}^{\ell-1}\binom{n}{j}} +
\left({\sum\limits_{j=1}^{\ell-1} \left[\binom{n}{j} -
\binom{n-j}{j} -1\right]}\right) \cdot \log
\left(1-\frac{\epsilon}{\sum_{j=1}^{\ell-1}\binom{n}{j}}\right)}
{\log{\left(1-\frac{\ell-1}{2^{\ell-1}}\right)}}.\notag
\end{equation}
Then the probability that a parity-check matrix consisting of $m$
randomly chosen codewords of the dual code (with possible
repetitions of a codeword) has stopping distance $\ell$
is at least $1-\epsilon$. With at most $n-k-\ell+1$
additional rows, the matrix also has rank $n-k$ and represents a
valid parity-check matrix of the code $\mathcal{C}$.
\end{theorem}

The proof of the theorem is given in
Appendix~\ref{app:proof_highimp}. Numerical results for a selected
set of values of $\ell$ is given in Section~\ref{sec:cyclic}.

As the results of Theorems~\ref{improve} and~\ref{highimp} are
non-constructive and conditioned on $\ell \leq \lfloor
\frac{d+1}{2}\rfloor$, we present another method for finding upper
bounds on the $\ell$-th stopping redundancy. This second class of
upper bounds on the stopping redundancy hierarchy of a code is
constructive in nature, and based on the following result.
\begin{theorem} (\cite{schwartzetal06}) \label{mt*}
Let $\mathcal{C}$ be a binary linear code with parameters
$[n,k,d]$, with $d>3$. Then
\begin{equation}
\rho_d(\mathcal{C}) \leq \binom{n-k}{1}+\ldots+ \binom{n-k}{d-2}.
\label{eq:vardy_bound}
\end{equation}
\end{theorem}

The proof of the bound in Theorem~\ref{mt*} is constructive: one
starts with an arbitrary parity-check matrix $\ve{H}$ of the code
$\mathcal{C}$, and then successively adds all sums of not more
than $d-2$ distinct rows of $\ve{H}$.

It is straightforward to invoke Theorem~\ref{mt*} for
upper-bounding the stopping redundancy hierarchy
$\rho_{\ell}(\mathcal{C})$ of a code - more precisely, in terms of
adding all sums of at most $\ell-2$ rows to a given parity-check
matrix. This result is stated below.

\begin{theorem} \label{mt*_ell}
Let $\mathcal{C}$ be a binary linear code with parameters
$[n,k,d]$, with $d>3$. Then
\begin{equation}
\rho_\ell(\mathcal{C}) \leq \binom{n-k}{1}+\ldots+ \binom{n-k}{\ell-2}.
\label{eq:vardy_bound_extended}
\end{equation}
\end{theorem}

Assume that one can identify a sub-code $\mathcal{B}$ of
$\mathcal{C}^{\perp}$ with dual distance
$d_{\mathcal{B}}^{\perp}$. Since the generators of a sub-code form
a subset of the generators of $\mathcal{C}^{\perp}$, one needs to
apply the procedure of adding redundant rows leading to
Theorem~\ref{mt1} or Theorems~\ref{improve}/\ref{highimp} only to
the basis vectors in $\mathcal{B}$ in order to ensure that the
redundant matrix has stopping distance at least
$d_{\mathcal{B}}^{\perp}$. This argument leads to the following
result.
\begin{theorem} \label{subcode}
Let $\Theta$ be the set of all sub-codes of the dual code
$\mathcal{C}^{\perp}$ of a linear $[n,k,d]$ code $\mathcal{C}$
that have support weight\footnote{The support weight of a sub-code
of a code is defined as the number of positions for which at least
one of the codewords of the sub-code is non-zero.} $n$ and dual
distance $\ell$. Furthermore, let the dimensions of the sub-codes
in $\Theta$ be $K_i,i=1,\ldots,|\Theta|$, and define
$K=\min_i\,K_i$. Then
\begin{equation}
\rho_{\ell}(\mathcal{C}) \leq \binom{K}{1}+\ldots+\binom{K}{\ell-2}.
\label{eq:vardy_bound_subcodes}
\end{equation}
\end{theorem}
\begin{proof}
Let $\mathcal{C}_{1}$ be an $[n,n-K_{1},\ell]$ code, and let its
dual code $\mathcal{C}_{1}^{\perp}$ be a sub-code of
$\mathcal{C}^{\perp}$ of support $n$ and dimension $K_1$. From
Equation~\eqref{eq:vardy_bound_extended}, $\rho_{\ell}(\mathcal{C}_{1})\leq
\binom{K_{1}}{1} + \ldots + \binom{K_{1}}{\ell-2}$. As
$\mathcal{C}_{1}^{\perp}$ is a sub-code of $\mathcal{C}^{\perp}$,
the rows of the parity-check matrix of $\mathcal{C}_{1}$ are a
subset of the rows of the parity-check matrix of $\mathcal{C}$. As
a result, the upper bound also holds for
$\rho_{\ell}(\mathcal{C})$.
\end{proof}

\begin{example}
Let $\mathcal{G}$ be the [$24,12,8$] extended Golay code. The code
$\mathcal{G}$ is self-dual and contains a [$24,10,8$] sub-code.
This code is unique \cite{jaffe}, and has minimum dual distance $6$. Therefore
\begin{equation}
\rho_6(\mathcal{G}) \leq 
\binom{10}{1}+\binom{10}{2}+\binom{10}{3}+\binom{10}{4}=385. 
\notag
\end{equation}
This bound is not tight as it by far exceeds the best known upper
bound (found through extensive computer
search~\cite{schwartzetal06}) on $\rho_8(\mathcal{G}) \leq 34$.
\end{example}
\begin{example}
Let $\mathcal{C}_{\text{BCH},1}$ be the $[31,16,7]$ BCH code.
Since BCH codes are nested, this code is a sub-code of a
$[31,21,5]$ BCH code $\mathcal{C}_{\mathrm{BCH},2}$. Consequently,
$\mathcal{C}_{\mathrm{BCH},2}^{\perp} \subseteq
\mathcal{C}_{\mathrm{BCH},1}^{\perp}$, so that
\begin{equation}
\rho_{5}(\mathcal{C}_{\mathrm{BCH},1})  \leq \binom{10}{1} +
\binom{10}{2} + \binom{10}{3} = 175. \notag
\end{equation}
\end{example}

\section{Case Study: Parity-Check Matrices of Cyclic Form}
\label{sec:cyclic}

The lower and upper bounds on the stopping redundancy hierarchy,
presented in the previous section, hold for all linear block
codes. Unfortunately, as illustrated with several examples, these
bounds tend to be very loose. As will be shown in this section,
much tighter upper bounds on the stopping redundancy hierarchy of
certain classes of codes can be obtained constructively, by
focusing on special forms of parity-check matrices. In particular,
we consider parity-check matrices of \emph{cyclic form}. Upper
bounds on the stopping distance of matrices in cyclic form also
represent upper bounds on the stopping redundancy hierarchy of the
codes. This claim is not true for lower bounds on the stopping
distance hierarchy, although the derived bounds still offer
valuable insight into the stopping set properties of matrices in
cyclic form.

We start by specializing the bound of Section \ref{sec:bounds} to
cyclic parity-check matrices, and then proceed with a case study
of BCH codes and codes based on cyclic difference sets.

\subsection{Stopping Sets in Parity-Check Matrices of Cyclic Form}
\label{sec:red_hierarchy_cyclic}

We start this section by introducing cyclic codes and cyclic
parity-check matrices.

\begin{definition}
Let $\mathcal{C}$ be an $[n,k,d]$ binary linear code. A code is
called {\emph{cyclic}} if a cyclic shift of a codeword
$\ve{c}\in\mathcal{C}$ is also a codeword. A (redundant)
parity-check matrix of a cyclic code is said to be of \emph{cyclic
form} if it consists of $m$ cyclic shifts of one given codeword of
the dual code, and provided that it has row-rank $n-k$.
\end{definition}
Observe that parity-check matrices of cyclic form necessarily
satisfy $n-k \leq m \leq n$. Also, note that a code can have a
parity-check matrix of cyclic form without being cyclic -
nevertheless, we focus our attention exclusively on cyclic codes.
A standard form for $\ve{H}$ with $m=n-k$ cyclic row-shifts is
shown below.

\begin{equation} \ve{H} = \left(
\begin{array}{cccccccc}
1         &  \ldots  &  1       &  0       &  0        &  0       &  0        &  0 \\
0         &  1        & \ldots  &  1       &  0        &  0       &  0        &  0 \\
0         &  0        & 1        & \ldots  &  1        &  0       &  0        &  0 \\
0         &  0        & 0        & 1         & \ldots  &  1       &  0        &  0 \\
\vdots  &            &           & \ddots & \ddots &          & \ddots &  0 \\
0         &  0        &  0       &  0        & 0         & 1       & \ldots   & 1\\
\end{array}
\right) 
\notag
\end{equation}

A binary cyclic code can be completely characterized by a
normalized parity-check polynomial $h(x)=h_0 + h_1x +
h_2x^2+\cdots+ h_{k-1}x^{k-1}+ x^k$, of degree $k$, for which $h_i
\in \mathbb{F}_2$, $\forall \,i \in [0,k-1]$. In this case, a
cyclic parity-check matrix $\ve{H}$ of the code can be constructed
by setting the entry in the $i$-th row and $j$-th column to
\[
H[i,j]=\left.\frac{h^{(k-j+i\mymod n)}(x)}{(k-j+i\mymod
n)!}\right|_{x=0},
\]
where $h^{(\zeta)}(x)$ denotes the $\zeta$-th derivative of $h$,
$1\leq i \leq m$, $1\leq j \leq n$. This leads to a parity-check
matrix of the form
\[
\ve{H}=\left(\begin{array}{ccccccc}1&
h_{k-1}&\cdots&h_1&h_0&&\ve{0}\\&\ddots&\ddots&&\ddots\\\ve{0}&&1&h_{k-1}
&\cdots&h_1&h_0\end{array}\right).
\]

\begin{definition}
Let ${\mathcal{C}}$ be a cyclic $[n,k,d]$ linear code. Partition
the set of codewords of the dual code ${\mathcal{C}}^{\perp}$
into cyclic orbits of its codewords. More precisely, let
${\mathcal{C}}^{\perp}$ be the disjoint union $\bigcup_{\ve{c}}
G_{\ve{c}}$ in which $\ve{c}$ $\in {\mathcal{C}}^{\perp}$,
$G_{\ve{c}}=\{{\pi^i\;\ve{c},\; i=1,\ldots,n\}}$, and
where $\pi$ denotes the right-cyclic shift permutation of the
symmetric group $S_n$. A designated element of $G_{\ve{c}}$ is
henceforth referred to as the {\emph{cyclic orbit generator}}
(cog) of $G_{\ve{c}}$. Whenever apparent from the context, the
reference to $G_{\ve{c}}$ will be omitted.
\end{definition}

Note that, in general, the first row of a redundant parity-check
matrix does not have to be defined by a parity-check polynomial.
The first row can be any cog which has the property that a
sufficient number of its cyclic shifts generates the dual code. In
what follows, with a slight abuse of notation, we use $\ve{h}$ to
denote such a cog (i.e.\ the first row of a cyclic parity-check
matrix $\ve{H}$), and let $\ve{h}[i]$ denote its $i$-th entry.

We start by providing an intuitive explanation why (redundant)
parity-check matrices of cyclic form have good stopping distance
properties.

\begin{definition} Without loss of generality,
assume that the first row in a cyclic parity-check matrix $\ve{H}$
has a non-zero symbol in its first position. The \emph{span} of
the first row, denoted by $u$, is the largest value of the index
$j$ for which $H[1,j]=1$.  The \emph{zero-span} $z$ of $\ve{h}$
represents the number of zeros trailing the last non-zero entry.
Clearly, $u + z = n$.
\end{definition}

Consider an arbitrary (redundant) cyclic parity-check matrix
$\ve{H}$ of a code ${\mathcal{C}}$. The matrix consists of a
non-zero cog and $m-1$ consecutive cyclic shifts thereof, where
$n-k \leq m \leq n$. The matrix $\ve{H}$ has $i-1$ leading zeros
and $z-i+1$ tailing zeros in row $i$, $1 \leq i \leq m$.
For any stopping set with largest column index $r \leq z+1$, the
$r$-th row of the matrix has an entry ``1'' in column $r$, and
zeros in all the remaining positions confined to
$\{1,2,\ldots,r-1\}$. Therefore, the rows of $\ve{H}$ resolve all
stopping sets of size $\sigma$ with support contained in the set
$\{1,2,\ldots,z+1\}$. Similarly, for all stopping sets with
coordinates confined to the set $\{ n-z,n-z+1,\ldots,n\}$, let the
smallest non-zero coordinate in the stopping set be indexed by
$l$. In this case, the row indexed by $l-(n-z)$ has a ``1'' in the
leftmost position $l$, and zeros in all remaining positions of the
stopping set. Therefore, the rows of $\ve{H}$ also resolve all
stopping sets of size $\sigma$ with support contained in the set
$\{n-z,n-z+1,\ldots,n\}$.

It can also be easily seen that a redundant cyclic parity-check
matrix with $m=n$ rows resolves all stopping sets with support
confined to $z+1$ consecutive column indices. This straightforward
analysis indicates that parity-check matrices of cyclic form may
have good stopping distance properties. We explore these questions
in more detail in the next section.

\subsection{Lower Bounds on the Stopping Distance of Cyclic Parity-Check Matrices}

We describe next how to bound the stopping distance of cyclic
parity-check matrices by specializing the results of Section \ref{sec:bounds}.

\begin{proposition}
The number of stopping sets of size $\sigma$ resolved by a cyclic
(redundant) parity-check matrix with dimensions $m\times n$, $n-k
\leq m\leq n$, is bounded from above by
\begin{equation}
m \; |\Sigma_{\sigma,1}| - \frac{2}{m}\cdot
\sum\limits_{\kappa=1}^{m-1} (m-\kappa) | \Sigma_{\sigma,1} \cap
\Sigma_{\sigma, ({(1+\kappa)}\modstar m)}
|.\label{eq:bound2_cyclic_1}
\end{equation}
The notation $\modstar$ is reserved for the modulo function, for
which $\,m \,\modstar \, m$ equals $m$, rather than
zero.
\label{bound2_cyclic_1}
\end{proposition}

In the proposition, it is tacitly assumed that the rows of the
parity-check matrix are arranged in such a way that the row
indexed by $({(\kappa+1)}\modstar m)$ is the $\kappa$-th cyclic
shift of the first row.

The result of Proposition~\ref{bound2_cyclic_1} is a
specialization of Equation~\eqref{eq:bound2}. In a cyclic
parity-check matrix, each row resolves the same number of stopping
sets as the positions of the resolved stopping sets are cyclic
shifts of each other. Furthermore, the number of stopping sets
simultaneously resolved by two rows indexed by $i$ and $j$, $i
\neq j$, only depends on the their mutual cyclic shift distance
$\kappa$.

Let the set $XY_{\kappa}(\ve{h})$ denote all pairs of positions
$(a,a+{\kappa})\modstar n$ in $\ve{h}$, $1\leq a \leq n$, with
entry $X$ at position $a\modstar n$ and entry $Y$ at position $a+{\kappa}\modstar n$,
where $X$ and $Y$ again serve as placeholders for $Z$ and $O$. For parity-check matrices of cyclic form, the
cardinalities of these sets equal the intersection numbers given
in Definition~\ref{def:intersection_numbers}, i.e.\
$|XY_{\kappa}(\ve{h})| = |X_{a\modstar n}\cap Y_{a+\kappa\modstar n}|$, $\forall a \in
\{ 1,\ldots,n\}$. If clear from the context, we omit the reference
to $\ve{h}$.

\begin{lemma}\label{lemma:num_stopsets_cyclic_pair_of_rows}
Let $\delta[k]$ be the indicator function, such that $\delta[k]=1$
iff $k=0$. The number of stopping sets resolved jointly by two
rows of a cyclic parity-check matrix $\kappa$ cyclic shifts apart
equals
\begin{equation}
| \Sigma_{\sigma, 1} \cap \Sigma_{\sigma, ({(\kappa+1)}\modstar
m)} | = |OO_{\kappa}|\cdot { |ZZ_{\kappa}| \choose \sigma-1 } +
|OZ_{\kappa}|\cdot |ZO_{\kappa}|\cdot{ |ZZ_{\kappa}| \choose
\sigma -2}\,, \notag
\end{equation}
where
\begin{equation}
|OO_{\kappa}|=\sum\limits_{i=1}^{n}\left\lfloor\frac{1}{2}\sum_{j=1}^{n}
h[j]\left(\delta[j-i]+\delta[j-i-{\kappa}]\right)\right\rfloor,
\notag
\end{equation}

\begin{equation}
|ZZ_{\kappa}|=\sum\limits_{i=1}^{n}\left\lfloor\frac{1}{2}
\sum_{j=1}^{n}(1-h[j])\left(\delta[j-i]
+\delta[j-i-{\kappa}]\right)\right\rfloor, \notag
\end{equation}

\begin{equation}
|ZO_{\kappa}|=|OZ_{\kappa}|=\sum\limits_{i=1}^{n}
\left[\frac{1}{2}\left(\sum_{j=1}^{n}h[j]
\left(\delta[j-i]+\delta[j-i-{\kappa}]\right)\right) \cdot
\left|\sum_{j=1}^{n}(h[j]\left(\delta[j-i]+
\delta[j-i-{\kappa}]\right))-2\right|\right], \notag
\end{equation}
and

\begin{equation}
|\Sigma_{\sigma,1}|={n-\sum_{j=1}^n\,h[j] \choose \sigma-1}
\cdot{\sum_{j=1}^n\,h[j]}.\notag
\end{equation}
Note that in all equations above it is assumed that ${i\choose
j}=0$ for $i<j$. \label{simulremovedsscyclic}
\end{lemma}

The expressions in Lemma~\ref{simulremovedsscyclic} are derived by
simple counting arguments, details of which are omitted. Although
the expressions above cannot be used to directly characterize the
stopping redundancy hierarchy, they represent a useful tool for
evaluating the stopping distance properties of cyclic parity-check
matrices, as illustrated by the examples that follow.


\begin{example} \label{example:qr_codes} Consider the class of
quadratic residue (QR) codes with prime length $n=3\,(\mymod 4)$.
Let the redundant parity-check matrix be of cyclic form, and let
the first row of the parity-check matrix be defined by the
idempotent of the code $\mathcal{Q}^{\bot}$ (defined in
\cite{macwilliamsetal77}), $I_{q}(x) = 1+\sum\limits_{\nu \in
\mathcal{N}} x^{\nu}$. Here, $\mathcal{N}$ denotes the set of
quadratic non-residues in $\mathbb{F}_n$\footnote{For more details
regarding QR codes, the interested reader is referred to Section
16 of~\cite{macwilliamsetal77}.}. The cyclic parity-check matrix
generated by $m \leq n$ cyclic shifts of the idempotent has
$S_{\sigma,2}$ stopping sets resolved by all pairs of rows, where
$$
S_{\sigma,2}=\binom{m}{2}\cdot \frac{n+1}{4} \cdot \left( \binom{\frac{n-3}{4}}{\sigma-1}
+ \frac{n+1}{4}\binom{\frac{n-3}{4}}{\sigma-2}\right).
$$
Be reminded that $S_{\sigma,2}$ denotes the number of stopping sets resolved by all pairs of two rows (cf.\ Section~\ref{sec:lower_bounds_on_the_stopping_redundancy_hierarchy}). As a consequence, for such a cyclic parity-check matrix to have
stopping distance at least $\ell$, 
its number of rows
$\mu_{\ell}$ must satisfy the following inequality
$$
\mu_{\ell}(\mathcal{C}) \geq \max\limits_{\sigma<\ell} \left\lceil
\frac{\binom{n}{\sigma} - M}{\frac{n+1}{2}
\binom{\frac{n-1}{2}}{\sigma-1} - M} \right\rceil,
$$
where
$$
M=\frac{n+1}{4}\cdot \left[\binom{\frac{n-3}{4}}{\sigma-1} +
\frac{n+1}{4}\cdot \binom{\frac{n-3}{4}}{\sigma-2} \right].
$$
A detailed derivation of this result can be found in
Appendix~\ref{sec:app_qr_codes}. Numerical values for the numbers
$\mu_{\ell}$, for three chosen QR codes, are given in
Table~\ref{tab:qr_codes_bounds_m_ell}.
\begin{table}[h!]
\begin{center}
\caption{Lower bounds on $\mu_{\ell}$ for parity-check matrices of
cyclic form generated from the idempotent of QR codes.}\label{tab:qr_codes_bounds_m_ell}
\begin{tabular}{|c|cccccccc|}\hline
 & $\mu_{4}$ & $\mu_{5}$ & $\mu_{6}$ & $\mu_{7}$ & $\mu_{8}$
 & $\mu_{9}$ & $\mu_{10}$ & $\mu_{11}$ \\ \hline
$[23,12,7]$ Golay & 4 & 6 & 10 & 19 & & & & \\
$[31,16,7]$ BCH   & 4 & 6 & 9 & 17 & & &  &\\
$[47,24,11]$ QR   & 4 & 6 & 9 & 15 & 27 & 55 & 117 & 265 \\
\hline
\end{tabular}
\end{center}
\end{table}
The bounds in the table imply that for the $[47,24,11]$ code, the
stopping distance of any redundant parity-check matrix of cyclic
form, defined by the idempotent, cannot exceed seven.
\end{example}

\begin{example}
As a second example, consider the class of cyclic difference set
(CDS)
codes~\cite{macwilliamsetal77},~\cite{linetal04},~\cite{tomlinsonetal04},
formally defined below.

\begin{definition}
  Assume that all calculations are performed modulo $n$. An
  $(n,k,\lambda)$ CDS is a set $Q=\{d_{0},d_{1},\ldots,d_{J-1}\}$ of
  $J$ integers with the property that any non-zero integer $i \leq
  n$,
  is a difference of two elements in $Q$ and that there exist exactly
  $\lambda$ ways to choose these two elements of $Q$.
\end{definition}

For a CDS, the sets $Q_{r}=\{d_{0}+r,d_{1}+r,\ldots,d_{J-1}+r\}$ for
all $r=0,\ldots,n-1$, form a cyclic $2-(n,k,\lambda)$ block
design~\cite{hall98}.

According to~\cite{linetal04}, a CDS code of length $n$ is
characterized by a polynomial $z(x)=\sum_{i=0}^{k-1}x^{d_{i}}$
that gives rise to a parity-check polynomial
$h(x)={\textrm{gcd}}(z(x),x^{n}+1)$.

For this particular case of cyclic codes, one can show that if
$h(x)=z(x)$, then
$$
|OO_{\kappa}| = \lambda, \; \; \;
|OZ_{\kappa}|=|ZO_{\kappa}|=k-\lambda, \; \; \;
|ZZ_{\kappa}|=n-2k+\lambda, \; \forall\;\, \kappa>0,
$$
and that
$$
S_{\sigma,2}=\binom{m}{2} \left(\lambda\,
\binom{n-2k+\lambda}{\sigma-1} +
(k-\lambda)^{2}\binom{n-2k+\lambda}{\sigma-2}\right).
$$
As a result,
$$
\mu_{\ell} \geq  \max_{\sigma < \ell } \left\lceil
\frac{\binom{n}{\sigma} - M}{k\cdot \binom{n-k}{\sigma-1} - M}
\right\rceil,
$$
where
$$M=\lambda \cdot \binom{n-2k+\lambda}{\sigma-1}
+ (k-\lambda)^{2}\cdot \binom{n-2k+\lambda}{\sigma-2},$$
and where
$\mu_{\ell}$ is defined as in the previous example.

Consider the class of Singer difference sets~\cite{hall98}, with
parameters
$$
(n,k,\lambda)=\left(\frac{q^{a+1}-1}{q-1},\frac{q^{a}-1}{q-1},\frac{q^{a-1}-1}{q-1}\right),
$$
where we specialize $a=2$, and $q=2^{s}$. The resulting difference
set has parameters $(n,k,\lambda)=(2^{s}(2^{s}+1)+1,2^{s}+1,1)$,
and leads to a $[2^{s}(2^{s}+1)+1,2^{s}(2^{s}+1)-3^{2},2^{2}+2]$
CDS code, for which, if $h(x)=z(x)$ as defined above,
$$
\mu_{\ell} \geq   \max_{\sigma < \ell } \left\lceil
\frac{\binom{2^{2s}+2^{s}+1}{\sigma} -
\binom{2^{2s}-2^{s}}{\sigma-1} -
2^{2s}\binom{2^{2s}-2^{s}}{\sigma-2}} {(2^{s}+1) \cdot
\binom{2^{2s}}{\sigma-1} - \binom{2^{2s}-2^{s}}{\sigma-1} -
2^{2s}\cdot \binom{2^{2s}-2^{s}}{\sigma-2}} \right\rceil.
$$
\end{example}
\vspace{0.2in} The results presented in this section characterized
the smallest number of rows in a cyclic parity-check matrix needed
for resolving all (or a certain number of) stopping sets of a
given size. To obtain upper bounds on these numbers, we resort to
constructive methods that rely on identifying good choices for a
cog defining the parity-check matrix. Our cog selection procedure,
along with a comparison of all constructive redundancy
parity-check design methods discussed, is presented in the next
section.

\subsection{Upper Bounds on the Stopping Distance of Cyclic
Parity-Check Matrices} \label{sec:motivation}

In what follows, we focus on deriving upper bounds on the stopping
distance of cyclic parity-check matrices of the Golay code and BCH
codes. We review several known approaches for constructing
redundant parity-check matrices for Hamming codes with a
prescribed number of stopping sets of a given size, and then
compare them to those based on optimizing cyclic parity-check
matrices. For Hamming codes, the number of stopping sets of a
full-rank, non-redundant parity-check matrix was derived in
analytical form in \cite{ghaffaretal07}. We observed that this
number does not depend on the form of the parity-check matrix.
However, if redundant parity-check matrices are used, the number
of stopping sets clearly depends on the chosen form of the matrix.

As will be shown, for a given stopping set distribution, cyclic
parity-check matrices for the Golay and BCH codes offer
row-redundancies comparable to those achievable by the best known
methods described in~\cite{schwartzetal06}
and~\cite{hollmannetal07}, respectively. For the general class of
BCH codes, we first provide a new method for constructing
redundant parity-check matrices with prescribed stopping distance,
by extending the work in~\cite{hollmannetal07}. We then compare
these findings with the stopping redundancy parameters obtained
from parity-check matrices of cyclic form.

\subsubsection{The Binary Golay Code}

To evaluate the performance of cyclic parity-check matrices of the
$[23,12,7]$ Golay code $\mathcal{G}$, the $506$ codewords of
minimum weight eight in the dual $[23,11,8]$ code
$\mathcal{G}^{\perp}$ are grouped into cog orbits. The codewords
of weight eight in $\mathcal{G}^{\perp}$ also belong to the code
$\mathcal{G}$, and can be arranged into $22$ cog orbits with $23$
codewords each.

Different choices for a cyclic parity-check matrix-defining cog
result in different stopping distance bounds. We concentrate on
two cogs that offer the best and the worst stopping distance
properties among all investigated cogs, and refer to them by the
subscripts $\A$ and $\D$.

The cog indexed by $\A$ has the following octal representation $
{\mathrm{cog}}_{[23,12],{\mathrm{A}}}=[2\, 1\, 2\, 1\, 3\, 5\, 0\,
0]$, with the most significant bit on the left side. Its
polynomial representation is
$1+x^4+x^6+x^{10}+x^{12}+x^{13}+x^{14}+x^{16}$. The cog indexed by
$\D$ has the representation
${\mathrm{cog}}_{[23,12],{\mathrm{D}}}=[ 3\, 4\, 6\, 0\, 3\, 2\,
0\, 0]$.

Upper bounds $U(\rho_{j}({\mathcal{C}}))$ on the stopping redundancy
hierarchy obtained from cogs ${\mathrm{cog}}_{[23,12],\A}$,
${\mathrm{cog}}_{[23,12],\B}$, and ${\mathrm{cog}}_{[23,12],\D}$ of
the parity-check matrix of the $[23,12,7]$ Golay code and the general
upper bounds obtained from Theorem~\ref{highimp} with
$\epsilon=10^{-3}$ and from Equation~\eqref{eq:vardy_bound} are listed
in Table~\ref{table:stop_red_golay}.

\begin{table}[h!]
\caption{Upper bounds on the stopping redundancy hierarchy of the
$[23,12,7]$ Golay code.} \label{table:stop_red_golay}
\begin{center}
\begin{tabular}{|c|c|c|c|c|}
\hline
&$\hspace{-1mm}U(\rho_4({\mathcal{C}}))\hspace{-1mm}$&$\hspace{-1mm}U(\rho_5({\mathcal{C}}))\hspace{-1mm}$
& $\hspace{-1mm}U(\rho_6({\mathcal{C}}))\hspace{-1mm}$ &
$\hspace{-1mm}U(\rho_7({\mathcal{C}}))\hspace{-1mm}$ \\ \hline
 $\cog_{23,12,{\A}}$ & $n-k$ & $16$ & $18$ & $23$  \\ 
$\cog_{23,12,{\B}}$ &  $13$ & $15$ & $19$ & $23$  \\ 
 $\cog_{23,12,{\D}}$ & $n-k$ & $16$ & $21$ & $>23$ \\\hline 
Minimum & $n-k$ & $15$ & $18$ & $23$  \\\hline 
Upper bound acc.\ to Theorem \ref{highimp}, $\epsilon=10^{-3}$&$39$&--&--&--\\\hline
Upper bound acc.\ to Eq.~\eqref{eq:vardy_bound} & $1023$ &$1023$&$1023$& $1023$\\
\hline
\end{tabular}
\end{center}
\end{table}

As can be seen from Table~\ref{table:stop_red_golay}, the stopping
redundancy hierarchy of the Golay code is bounded as
\begin{equation}
\rho_4(\mathcal{C}) \leq n-k, \; \rho_5(\mathcal{C}) \leq 15,
\rho_6(\mathcal{C}) \leq 18, \rho_7(\mathcal{C}) \leq 23. \notag
\end{equation}
In addition, from the same table it can be seen that the bound in
Equation~\eqref{eq:vardy_bound} is extremely loose - significant reductions are possible when using the constructive
cyclic matrix design approach.

\subsubsection{The Hamming Codes}
\label{subsec:hamming_codes}

Several approaches for generating parity-check matrices of Hamming
codes that resolve correctable erasure patterns up to a given size
were recently described by Weber and Abdel-Ghaffar~\cite{weberetal05}
and Hollmann and Tolhuizen \cite{hollmannetal07}.
In~\cite{weberetal05}, the authors introduced the notion of
\emph{three-erasure correcting parity-check collections}, capable of
resolving the largest possible number of erasure patterns of weight
three. These results were extended in~\cite{hollmannetal07}, where
{\emph{generic $(\bar{m},\bar{\sigma})$ erasure correcting sets}}
$\ve{A}_{\bar{m},\bar{\sigma}}$ were defined as

$$\{\ve{a}=(a_1,a_2,\dots,a_{\bar{m}})\mid a_1=1,
{\mathrm{wt}}(\ve{a})\leq \bar{\sigma}\}.$$

In this context, if $\ve{H}$ represents a parity-check matrix of
dimension $\bar{m}\times\bar{\sigma}$, the collection of
parity-checks $\{\ve{a}\ve{H}\mid
\ve{a}\in\ve{A}_{\bar{m},\bar{\sigma}}\}$ resolves all correctable
erasure patterns up to size $\bar{\sigma}$.

For the class of $[2^s-1,2^s-s-1,d=3]$ Hamming codes, according
to~\cite{hollmannetal07}, we first construct parity-check matrices
that are generic $(\bar{m},\bar{\sigma})$ erasure correcting sets,
and then compare the stopping set distribution of these matrices
with that of optimized cyclic parity-check matrices. As for
Hamming codes $\rho_{2}(\mathcal{C})=\rho_{3}(\mathcal{C})=n-k$
and $d=3$, our comparison is performed with respect to the number
of unresolved stopping sets of size $\sigma=3$, for several fixed
values of $m$.

Assume that the parity-check matrix of a Hamming code is given by
a row vector over the finite field  $\mathbb{F}_{n+1}$,
$n=2^{n'}-1, n'>0$, namely
\begin{equation}
\ve{H}_{\mathrm{h}}=\left(\begin{array}{cccc}\alpha^{0\,\cdot
b}&\alpha^{1\,\cdot
b}&\dots&\alpha^{(n-1)\,b}\end{array}\right),\notag
\end{equation}
where $\alpha$ denotes a primitive element of the underlying
field. Henceforth, we set $b=1$. In order to form a binary
parity-check matrix for the Hamming code, the elements of
$\ve{H}_{\mathrm{h}}$ are represented as binary vectors over the
vector space $\mathbb{F}_2^{\log_2(n+1)}$, i.e.\ each element
$\alpha^{f},\, f\in\{0,\dots,n-1\}$, is described by a binary
column vector. Let us refer to this parity-check matrix as the
\emph{standard parity-check matrix} for Hamming codes and denote
it by $\ve{H}_{\Hammingstandard}$.

Start by constructing a generic erasure correcting set
$\ve{A}_{\bar{m},\bar{\sigma}}$ with $\bar{m}=\log_2(n+1)$ and
$\bar{\sigma}=3$. By applying the generic erasure set construction
to $\ve{H}_{\Hammingstandard}$, we arrive at a redundant
parity-check matrix of the code denoted by $\ve{H}_{\Hamminggeneralized}$. This
matrix  has
\begin{equation}
m^{\star}=\sum\limits_{i=0}^{\bar{\sigma}-1}{\bar{m}-1\choose
i}=\log_2(n+1)+{\log_2(n+1)-1\choose 2}
\notag
\end{equation}
rows. Note that the only unresolved stopping sets of size three in
such a parity-check matrix are the actual codewords of the Hamming
code. We then proceed by constructing redundant parity-check
matrices of cyclic form, obtained from $m^{\star}$ consecutive
cyclic shifts of a minimum weight codeword of the dual code.
Table~\ref{tab:Tolhuizen_vs_cyclic_63_57} and
Table~\ref{tab:Tolhuizen_vs_cyclic_127_120} show the number of
unresolved stopping sets of size three in the redundant
parity-check matrices described above, for $n=63$ and $n=127$. The
cog vectors of the two cyclic parity-check matrices are
\begin{equation}
\text{cog}_{[63,57]}=
[4\,1\,4\,2\,4\,7\,5\,0\,7\,1\,1\,3\,3\,5\,4\,6\,5\,3\,7\,4\,0],
\notag
\end{equation}
and
\begin{equation}
\text{cog}_{[127,120]}=[1\,0\,4\,6\,1\,3\,5\,3\,3\,0\,1\,4\,6\,5\,1\,6\,3\,6\,6\,4\,1\,2\,
5\,7\,5\,1\,2\,1\,5\,6\,1\,7\,7\,0\,3\,5\,7\,1\,3\,1\,1\,0\,0],
\notag
\end{equation}
respectively. The vectors are listed in octal form, with the most
significant bit on the left hand side.

\begin{table}[h!]
\caption{Number of unresolved stopping sets ($\sigma=3$) in the
$[63,57,3]$ Hamming code.} \label{tab:Tolhuizen_vs_cyclic_63_57}
\begin{center}
\begin{tabular}{|c|c|c|}\hline
&\multicolumn{2}{c|}{Number of unresolved stopping sets,
$\sigma=3$}\\\cline{2-3} $m$& Parity-check matrix
in~\cite{hollmannetal07}&Cyclic parity-check matrix\\\hline
$6=n-k$&$2261$&$2261$\\
$16$ &$651$&$655$\\
$17$&&$653$\\
$18$&&$651$\\\hline
\end{tabular}
\end{center}
\end{table}

\begin{table}[h!]
\caption{Number of unresolved stopping sets ($\sigma=3$) of the
$[127,120,3]$ Hamming code.}
\label{tab:Tolhuizen_vs_cyclic_127_120}
\begin{center}
\begin{tabular}{|c|c|c|}
\hline &\multicolumn{2}{c|}{Number of unresolved stopping sets,
$\sigma=3$}\\\cline{2-3} $m$& Parity-check matrix in
~\cite{hollmannetal07}&Cyclic parity-check matrix\\\hline
$7=n-k$&$11970$&$11970$\\
$22$ & $2667$&$2672$\\
$26$&&$2667$\\\hline
\end{tabular}
\end{center}

\end{table}

As one can observe from Tables \ref{tab:Tolhuizen_vs_cyclic_63_57}
and \ref{tab:Tolhuizen_vs_cyclic_127_120}, cyclic parity-check
matrices can achieve almost identical performance to the one
offered by redundant parity-check matrices specialized for the
Hamming codes. For example, there are $11970$ stopping sets of
size $\sigma=3$ in a full-rank parity-check matrix of the
[127,120,3] Hamming code. This number can be reduced to $2667$
stopping sets only, if one uses $22$ rows of the construction
described in \cite{hollmannetal07}, or $26$ rows, when using
parity-check matrices of cyclic form. These findings motivate a
further study of redundant cyclic parity-check matrices, and we
describe some illustrative examples pertaining to the family of
BCH codes in the next section.

\subsubsection{Cyclic Parity-Check Matrices of Double- and Triple-Error Correcting BCH Codes}

We present constructive upper bounds on the stopping redundancy
hierarchy of the $[127,113,5]$ double-error correcting BCH code,
and the $[31,16,7]$ triple-error correcting BCH code. Similarly as
for the Golay code, we only consider minimum-weight codewords of
the dual code for the purpose of generating redundant parity-check
matrices in cyclic form \cite{santhietal04}. In particular, we
focus on four selected cogs, indexed by $\A$ through $\D$, and
analyze their underlying parity-check matrices in detail.

The dual of the $[127,113,5]$ BCH code has $4572$ codewords of
minimum weight $56$, which can be separated into $36$ cog orbits.
The octal representations of four different cogs used in our
constructions are listed below:
\begin{eqnarray}
{\mathrm{cog}}_{[127,113],{\mathrm{A}}}=[1\,7\,6\,4\,0\,3\,0\,6\,5\,4\,4\,5\,4\,
0\,7\,5\,0\,4\,5\,4\,7\,6\,5\,1\,6\,1\,6\,0\,2\,0\,4\,2\,6\,5\,2\,4\,2\,4\,4\,0\,0\,5\,6],
\notag
\end{eqnarray}
\begin{eqnarray}
{\mathrm{cog}}_{[127,113],{\mathrm{B}}}=[1\,7\,2\,4\,2\,5\,0\,2\,6\,1\,2\,1\,5\,4\,1\,1\,1\,1\,5\,2\,6\,1\,0
\,7\,2\,1\,2\,5\,5\,1\,6\,1\,4\,0\,4\,6\,5\,4\,1\,4\,2\,7\,4],
\notag
\end{eqnarray}
\begin{eqnarray}
{\mathrm{cog}}_{[127,113],{\mathrm{C}}}=[1\,7\,5\,2\,6\,5\,5\,3\,3\,6\,4\,6
\,1\,3\,1\,2\,6\,4\,2\,1\,0\,7\,1\,1\,7\,0\,4\,0\,2\,4\,0\,2\,5\,4\,0\,3\,0\,4\,5\,2\,2\,4\,2],
\notag
\end{eqnarray}
\begin{eqnarray}
{\mathrm{cog}}_{[127,113],{\mathrm{D}}}=[1\,7\,5\,1\,7\,0\,3\,1\,2\,5\,2\,6\,7\,
3\,4\,6\,5\,0\,2\,1\,0\,2\,0\,7\,0\,3\,6\,5\,4\,0\,6\,1\,2\,2\,1\,0\,1\,4\,3\,0\,6\,4\,4].
\notag
\end{eqnarray}

The dual of the $[31,16,7]$ BCH code has $465$ codewords of
minimum weight eight, which can be partitioned into $15$ cog
orbits. Four representative cogs, indexed by $A$ through $D$, are
listed below.
\begin{eqnarray}
{\mathrm{cog}}_{[31,16],{\mathrm{A}}}=[1\,4\,1\,4\,0\,5\,0\,0\,0\,2\,2],
\notag
\end{eqnarray}
\begin{eqnarray}
{\mathrm{cog}}_{[31,16],{\mathrm{B}}}=[1\,4\,0\,6\,1\,0\,4\,1\,0\,2\,0],
\notag
\end{eqnarray}
\begin{eqnarray}
{\mathrm{cog}}_{[31,16],{\mathrm{C}}}=[1\,5\,0\,0\,0\,5\,0\,0\,4\,1\,4],
\notag
\end{eqnarray}
\begin{eqnarray}
{\mathrm{cog}}_{[31,16],{\mathrm{D}}}=[1\,5\,0\,4\,0\,2\,0\,0\,1\,3\,0].
\notag
\end{eqnarray}

We now compare the stopping distance properties of parity-check
matrices of cyclic form with that of a novel construction of
redundant parity-check matrices generalizing the method
in~\cite{hollmannetal07}. In a nutshell, the construction exploits
the fact that BCH codes are nested, and that they are sub-codes of
Hamming codes. Redundant parity-check equations are chosen in two
different steps: in the first step, they are selected to eliminate
stopping sets corresponding to certain codewords of the Hamming
code, while in the second step they are chosen to eliminate
stopping sets that do not correspond to codewords in the Hamming
code. A detailed description of this scheme is given in
Appendix~\ref{app:BCH_construction_scheme}. For simplicity, we
refer to it as the \emph{generalized HT} (Hollmann-Tolhuizen)
construction.

We next compare the stopping distances of various redundant
parity-check matrices of the aforementioned BCH codes. The
smallest values found by this comparison represent constructive
upper bounds on $\rho_{\ell}(\mathcal{C})$, $\ell\leq d$, of BCH
codes.

Table~\ref{tab:Tolhuizen_vs_cyclic_63_51_and_127_113} lists the
number of rows in a redundant parity-check matrix needed to
achieve a given stopping distance, for matrices constructed by the
generalized HT method, the upper bound from \cite{hollmannetal07},
and a bound obtained from cyclic parity-check matrices. In
addition, the upper bound from~\cite{schwartzetal06}, given in
Equation~\eqref{eq:vardy_bound}, is also shown. All results
pertain to the $[127,113,5]$ BCH code.

\begin{table}[h!]
\caption{Upper bounds on $\rho_{\ell}(\mathcal{C})$, $\ell\leq
5$.} \label{tab:Tolhuizen_vs_cyclic_63_51_and_127_113}
\begin{center}
\begin{tabular}{|c|c|c|c|c|}
\hline
&\multicolumn{4}{c|}{$U(\rho_4({\mathcal{C}}))$}\\\cline{2-5}
Code&Cyclic&Gen. HT&\cite{hollmannetal07}&General
\cite{schwartzetal06}\\\hline
$[127,113,5]$& $20$& $96$ &$92$ &$469$\\\hline\hline
&\multicolumn{4}{c|}{$U(\rho_5({\mathcal{C}}))$}\\\cline{2-5}
Code&Cyclic&Gen. HT &\cite{hollmannetal07}&General
\cite{schwartzetal06}\\\hline
$[127,113,5]$& $34$& $229$ &$378$ &$469$ \\\hline
\end{tabular}
\end{center}
\end{table}

The values presented for cyclic codes are obtained by searching
for the best cog orbit, with cogs corresponding to minimum weight
codewords. The cog indexed by $A$ has the ``best'' stopping set
properties, when both the numbers of stopping sets of size three
and four are considered. Stopping distances of all four described
cogs are given in Table~\ref{table:stopp_dist_hierarchy_127_113}.

As a consequence, for the $[127,113,5]$ BCH code $\mathcal{B}_{[127,113,5]}$ one has
\begin{equation}
\rho_4(\mathcal{B}_{[127,113,5]}) \leq 20, \; \rho_5(\mathcal{B}_{[127,113,5]}) \leq 34.
\notag
\end{equation}

Table~\ref{tab:Tolhuizen_vs_cyclic_63_51_and_127_113} shows that
the general bounds derived in~\cite{schwartzetal06} are very
loose. Although the approach from~\cite{hollmannetal07} provides
tighter bounds for this class of codes, it is still significantly
outperformed by both the generalized HT method, as well as the
cyclic (redundant) parity-check matrix construction.

\begin{table}[h!]
\caption{Upper bounds on the stopping redundancy hierarchy of the
$[127,113,5]$ BCH code.}
\label{table:stopp_dist_hierarchy_127_113}
\begin{center}
\begin{tabular}{|c|c|c|}
\hline
&$U(\rho_4({\mathcal{C}}))$&$U(\rho_5({\mathcal{C}}))$\\\hline
$\cog_{127,113,\A}$&$20$&$34$\\
$\cog_{127,113,\B}$&$22$&$34$\\
$\cog_{127,113,\C}$&$23$&$46$\\
$\cog_{127,113,\D}$&$22$&$45$\\\hline Minimum &$20$&$34$\\\hline
Eq.~\eqref{eq:vardy_bound}&$469$&$469$\\\hline
\end{tabular}

\end{center}
\end{table}

The results of a similar analysis performed on the $[31,16,7]$ BCH
code $\mathcal{B}_{[31,16,7]}$ are shown in
Table~\ref{table:stopp_dist_hierarchy_31_16}.
\begin{table}[h!]
\caption{Upper bounds on the stopping redundancy hierarchy of the
$[31,16,7]$ BCH code.} \label{table:stopp_dist_hierarchy_31_16}
\begin{center}
\begin{tabular}{|c|c|c|c|c|}
\hline &$U(\rho_4({\mathcal{C}}))$&$U(\rho_5({\mathcal{C}}))$&
$U(\rho_6({\mathcal{C}}))$&$U(\rho_7({\mathcal{C}}))$\\\hline
$\cog_{31,16,\A}$&$n-k=15$&$18$&$19$&$21$\\
$\cog_{31,16,\B}$&$n-k=15$&$16$&$20$&$22$\\
$\cog_{31,16,\C}$&$n-k=15$&$n-k=15$&$20$&$28$\\
$\cog_{31,16,\D}$&$n-k=15$&$16$&$21$&$26$\\\hline Minimum &
$n-k=15$&$n-k=15$&$19$&$21$\\\hline
Upper bound acc.\ to Theorem \ref{highimp}, $\epsilon=10^{-3}$&$45$&--&--&--\\\hline
Upper bound acc.\ to Eq.~\eqref{eq:vardy_bound},~\eqref{eq:vardy_bound_subcodes}&$91$&$175$&$4943$&$4943$\\\hline
\end{tabular}
\end{center}
\end{table}
As can be seen from the table, the stopping redundancy hierarchy
of the $[31,16,7]$ BCH codes is upper bounded as
\begin{equation}
\rho_4(\mathcal{B}_{[31,16,7]})=15, \; \rho_5(\mathcal{B}_{[31,16,7]})=15, \;
\rho_6(\mathcal{B}_{[31,16,7]}) \leq 19, \; \rho_7(\mathcal{B}_{[31,16,7]})\leq 21. \notag
\end{equation}

\section{The Automorphism Redundancy} \label{sec:analysis}

One important observation motivates our subsequent study of
stopping redundancy properties of cyclic (redundant) parity-check
matrices. It is the simple fact that using a collection of
cyclically shifted rows in $\ve{H}$ for resolving a stopping set
has the same effect as using only \emph{one of these rows} and
then \emph{cyclically shifting the received codeword}.

In general, the same observation is true for parity-check matrices
that consist of rows that represent different images of one given
row-vector under a set of coordinate permutations.
Such coordinate permutations must necessarily preserve the
codeword structure, i.e.\ they must correspond to
\emph{automorphisms} of the code. Furthermore, if properly chosen,
such automorphisms may be used to relocate the positions of the
erasures in the received codeword to coordinates that do not
correspond to stopping sets.

We propose to use this observation in order to construct
parity-check matrices of codes that, jointly with a new decoding
technique, allow the edge-removal algorithm to avoid detrimental
effects of stopping sets on its performance. For this purpose, we
first recall the definition of the {\emph{automorphism group}} of
a code.
\begin{definition}(\cite[Ch.\ 8]{macwilliamsetal77})
The set of permutations which send a code $\mathcal{C}$ into
itself, i.e.\ permutations that map codewords into (possibly
different) codewords, are called the automorphism group of the
code $\mathcal{C}$, and are denoted by $\Aut(\mathcal{C})$.
\end{definition}

It is straightforward to see that
$\Aut(\mathcal{C})=\Aut(\mathcal{C}^{\perp})$, a fact that we
exploit in our subsequent derivations.

Decoding procedures that use information about the automorphism
group of a code have a long history~\cite[Ch.\
16]{macwilliamsetal77}. Such procedures are known as
\emph{permutation decoding algorithms} and they are traditionally
restricted to decoding of messages transmitted over the binary
symmetric channel (BSC)\footnote{Recently, permutation decoders
were also used for decoding of messages transmitted over the AWGN
channel~\cite{halfordetal06}, although not for the purpose of
finding error-free information sets nor for the purpose of
eliminating pseudocodewords such as stopping sets.}. Within the
framework of permutation decoding, a codeword
$\ve{c}\in\mathcal{C}$, corrupted by an error vector
$\ve{e}=[e_0,\;e_1,\;\dots,\;e_{n-1}]$ of weight less than or
equal to $t$, where $2t+1\leq d$, is subjected to the following
process. First, a parity-check matrix of the code of the form
$\ve{H}=[\ve{A}|\ve{I}]$, along with the corresponding generator
matrix $\ve{G}=[\ve{I}|\ve{A}^{\mathrm{T}}]$, is chosen. Here,
$\ve{I}$ is used to denote the identity matrices of order $n-k$
and $k$, respectively. Then, the syndrome of the received vector
$\ve{y}=\ve{c}+\ve{e}$, $\ve{z}=\ve{H}\ve{y}^{\transposed}$, is
computed. If the weight of $\ve{z}$ is greater than $t$, the
vector $\ve{y}$ is permuted according to a randomly chosen
automorphism. This process is repeated until either all
automorphisms are tested or until the syndrome has weight less
than or equal to $t$. In the former case, the decoder declares an
error. In the latter case, all decoding errors are provably
confined to parity-check positions, so that decoding terminates by
recovering the uncorrupted information symbols.

For the purpose of permutation decoding, one would like to
identify the smallest set of automorphisms that moves any set of
not more than $t$ positions in $\{{0,\ldots,n-1\}}$ into the
parity-check positions $\{{k,k+1,\ldots,n-1\}}$ of
$[\ve{A}|\ve{I}]$.
\begin{definition} (\cite{key06})
If $\mathcal{C}$ is a $t$-error correcting code with an
information bit index set $\mathcal{I}$ and a parity-check index
set $\mathcal{P}$, then a \textbf{PD}($\mathcal{C}$)-set
(permutation decoding set of $\mathcal{C}$) is a set $S$ of
automorphisms of $\mathcal{C}$ such that every $t$-set of
coordinate positions is moved by at least one member of $S$ into
the check-positions $\mathcal{P}$. For $s \leq t$, an
\textbf{s-PD}($\mathcal{C}$)-set is a set of automorphisms of
$\mathcal{C}$ such that every $b$-set of coordinate positions, for
all $ b \leq s$, is moved by at least one member of $S$ into
$\mathcal{P}$.
\end{definition}

Throughout the remainder of this paper, we will be concerned with
\textbf{PD} and \textbf{s-PD} sets of \emph{smallest possible
size}, and we simply refer to them as \textbf{PD} sets. Clearly,
\textbf{PD} and \textbf{s}-\textbf{PD} sets may not exist for a
given code, and complete or partial knowledge about \textbf{PD}
sets is available for very few codes~\cite{key06}. Nevertheless,
even this partial information can be used to derive useful results
regarding the analogues of \textbf{PD} sets for iterative decoders
operating on stopping sets.


We introduce next the notion of a \textbf{S}topping \textbf{A}utomorphism Group \textbf{D}ecoding (\textbf{SAD})
set, a generalization of the notion of a \textbf{PD} set for the
edge-removal iterative decoder. We then proceed to relate
\textbf{SAD} sets to both \textbf{PD} sets and to the stopping
redundancy hierarchy of a code.
\begin{definition}
  Let $\ve{H}$ be a parity-check matrix of an error-correcting code
  $\mathcal{C}$ with minimum distance $d$. A \textbf{SAD}($\ve{H}$)
  set of $\ve{H}$ is the smallest set $S$ of automorphisms of
  $\mathcal{C}$ such that every $b$-set of coordinate positions, $1
  \leq b \leq d-1$, is moved by at least one member of $S$ into a set
  of positions that do not correspond to a stopping set of $\ve{H}$.
  Similarly, if $s \leq d-1$, an \textbf{s-SAD}($\ve{H}$)-set is the
  smallest set of automorphisms of $\mathcal{C}$ such that every
  $b$-set of coordinate positions, for all $b \leq s$, is moved by at least
  one member of $S$ into positions that do not correspond to a
  stopping set in $\ve{H}$. Without loss of generality, we assume that
  an \textbf{s-SAD}($\ve{H}$)-set contains the identity permutation.
  For a given code $\mathcal{C}$, we also define
\begin{equation}
\begin{split}
&\textbf{S}_s^{\star}(\mathcal{C})=\min_{\ve{H}(\mathcal{C})}
\;|\textbf{s-SAD}(\ve{H}(\mathcal{C}))|, \notag \\
&\textbf{S}^{\star}(\mathcal{C})=\min_{\ve{H}(\mathcal{C})}
\;|\textbf{SAD}(\ve{H}(\mathcal{C}))|,
\end{split}
\notag 
\end{equation}
and refer to $\textbf{S}_s^{\star}(\mathcal{C})$ and
$\textbf{S}^{\star}(\mathcal{C})$ as to the \emph{s-automorphism
redundancy} and \emph{automorphism redundancy} of $\mathcal{C}$.
\end{definition}

For a given code and parameter \textbf{s}, an
\textbf{s}-\textbf{SAD} set may not exist. This is a consequence
of the fact that there may be no automorphisms that move all
arbitrary collections of not more than $s$ coordinates into
positions that do not correspond to a stopping set in one given
parity-check matrix. But whenever such sets exist, they can be
related to the stopping redundancy hierarchy and \textbf{PD} sets
of the code. First, it is straightforward to show that for all $1
\leq s \leq d-1$, one has $\textbf{S}_s^{\star}(\mathcal{C}) \leq
|\textbf{s-PD}(\mathcal{C})|$, whenever such sets exist. This
follows from considering parity-check matrices in systematic form,
and from the Singleton bound, which asserts that for any linear
code $d-1 \leq n-k$. Furthermore, it is straightforward to see
that for a restricted set of parity-check matrices, automorphism
group decoders trade redundant rows with automorphisms. This is
formally described by the following lemma, the proof of which is
straightforward and hence omitted.
\begin{lemma} Let $\mathcal{C}$ be an $[n,k,d]$ code. Then
\begin{equation}\label{eq:S_s_star_upper_bound}
\rho_{s+1}(\mathcal{C}) \leq (n-k)\,\times \,
\textbf{S}_{s}^{\star}(\mathcal{C}),
\end{equation}
for all $1 \leq s \leq d-1 $, provided that an
\textbf{s-SAD}($\mathcal{C}$) set exists.
\end{lemma}

One class of codes for which it is straightforward to prove the
existence of certain \textbf{SAD} sets is the class of codes with
\emph{transitive} automorphism groups, described below.
\begin{definition}
A group $\Gamma$ of permutations of the symbols
$\{{0,1,\ldots,n-1\}}$ is transitive if for any two symbols $i,j$
there exists a permutation $\pi \in \Gamma$ such that $i \pi=j$. A
group is said to be $t$-fold transitive if for any two collections
of $t$ distinct numbers $i_1,\ldots,i_t \in \{{0,\ldots,n-1\}}$
and $j_1,\ldots,j_t \in \{{0,\ldots,n-1\}}$, there exists a $\pi \in
\Gamma$ such that $i_1\,\pi=j_1,\ldots,i_t\,\pi=j_t$.
\end{definition}

\begin{lemma} Let $\mathcal{C}$ be a code with an
$s$-transitive automorphism group. Then there exist \textbf{b-SAD}
sets of $\mathcal{C}$ for all $b \leq s$. \end{lemma}

\begin{proof} Let $\ve{H}$ be of the form $[\ve{A}|\ve{I}]$. Clearly,
the positions of $\ve{H}$ indexed by $k$ through $n-1$ are free of
stopping sets of size $s \leq n-k$. Since the automorphism group
of $\mathcal{C}$ is $s$-transitive, any collection of not more
than $s$ coordinates in $\{{0,\ldots,n-1\}}$ is moved by some
element of $\Aut(\mathcal{C})$ into the positions indexed by $k$
to $n-1$. Consequently, the automorphism group itself represents a
(possibly non-minimal) \textbf{s-SAD} set.
\end{proof}

Finding \textbf{SAD} sets of codes is a very complicated task, so
that we focus our attention on deriving bounds on the size of such
sets for specific examples of codes.

\begin{example}
Consider the $[24,12,8]$ extended Golay code $\mathcal{G}_{24}$.
Since the automorphism group of the extended Golay code is $5$-fold
transitive, a \textbf{5-SAD} set exists, and $\textbf{S}_5^{\star}
(\mathcal{G}_{24})\leq |\mathcal{M}_{24}|$, where
$|\mathcal{M}_{24}|$ denotes the order of the Mathieu group
$\mathcal{M}_{24}$. It is well known that
$|\mathcal{M}_{24}|=244823040$.

One can actually show a much stronger result, described below.

\begin{theorem} \label{thm:S5_Golay24 } The automorphism redundancy
of the $\mathcal{G}_{24}$ code is upper bounded by $14$, i.e.\
$\textbf{S}_7^{\star}(\mathcal{G}_{24}) \leq 14$.
\end{theorem}
\begin{proof}
The proof of the result is constructive. Below, we list the
particular form of $\ve{H}$ used to meet the claimed result, as
well as the corresponding set of {\textbf{SAD}} automorphisms. The matrix in
question is $\ve{H}(\mathcal{G}_{24})=[\ve{I}_{12}|\ve{M}]$, where
\begin{equation} \label{eq:gol-wolf}
\begin{split}
\ve{M}=\left[\begin{array}{cccc}
  I_3 & A   & A^2 & A^4 \\
  A   & I_3 & A^4 & A^2 \\
  A^2 & A^4 & I_3 & A \\
  A^4 & A^2 & A   & I_3 \\
\end{array} \right],\;\;
A=\left[ \begin{array}{ccc}
  1 & 1 & 1 \\
  1 & 0 & 0 \\
  1 & 0 & 1 \\
\end{array} \right].
\end{split}
\end{equation}
The automorphisms are of the form $\theta^{i} \times \psi^{j}$,
$i=0,1$, $j=0,1,\ldots,5,6$, with
\begin{equation} 
\begin{split}
&\theta=(0,12)(1,13)(2,14)(3,15)\ldots(10,22)(11,23);\\
&\psi=(3,6,15,9,21,18,12)(4,7,16,10,22,19,13)(5,8,17,11,23,20,14);
\end{split} \notag
\end{equation}
Both permutations are listed in standard cycle form. Note that
these automorphisms include the identity permutation for $i=j=0$.
Incidentally, this is the same matrix that was studied
in~\cite{wolfmann83}, where it was used to show that
\textbf{PD}($\mathcal{G}_{24}$)=14.
\end{proof}

Henceforth, the matrix obtained by combining the
$\ve{H}(\mathcal{G}_{24})$-images of all the listed automorphisms
is denoted by $\ve{H}_{\W}$. The subscript $W$ refers to the name
of the first author to have studied this matrix -
\textbf{W}olfmann~\cite{wolfmann83}.

Equation~\eqref{eq:S_s_star_upper_bound} allows one to bound
$\rho_6(\mathcal{G}_{24}) \leq 12 \times 14=168$, which is
significantly larger than the constructive bound provided by the
matrix in Equation~\eqref{H_golay_24_12_cyclic_group_19_21_shifts}
below, which consists of $21$ rows only. The matrix was
constructed by identifying the cog of the $[23,12,7]$ code that
leads to optimal stopping distance properties, generating a $m=21$
cyclic parity-check matrix, and then adding a single (fixed)
parity-check position to each of the rows.

{

\ifTEXTDOUBLESPACING

\baselineskip=0.55555555\baselineskip

\fi

\begin{equation}
\ve{H}_{[24,12]}=\left(\begin{split}
1 \,\, 0 \,\, 0 \,\, 0 \,\, 0 \,\, 1 \,\, 1 \,\, 0 \,\, 0 \,\, 1 \,\, 0 \,\, 1 \,\, 1 \,\, 0 \,\, 0 \,\, 0 \,\, 1 \,\, 0 \,\, 1 \,\, 0 \,\, 0 \,\, 0 \,\, 0 \,\, 0  \\[-1.7mm]
0 \,\, 1 \,\, 0 \,\, 0 \,\, 0 \,\, 0 \,\, 1 \,\, 1 \,\, 0 \,\, 0 \,\, 1 \,\, 0 \,\, 1 \,\, 1 \,\, 0 \,\, 0 \,\, 0 \,\, 1 \,\, 0 \,\, 1 \,\, 0 \,\, 0 \,\, 0 \,\, 0  \\[-1.7mm]
0 \,\, 0 \,\, 1 \,\, 0 \,\, 0 \,\, 0 \,\, 0 \,\, 1 \,\, 1 \,\, 0 \,\, 0 \,\, 1 \,\, 0 \,\, 1 \,\, 1 \,\, 0 \,\, 0 \,\, 0 \,\, 1 \,\, 0 \,\, 1 \,\, 0 \,\, 0 \,\, 0  \\[-1.7mm]
0 \,\, 0 \,\, 0 \,\, 1 \,\, 0 \,\, 0 \,\, 0 \,\, 0 \,\, 1 \,\, 1 \,\, 0 \,\, 0 \,\, 1 \,\, 0 \,\, 1 \,\, 1 \,\, 0 \,\, 0 \,\, 0 \,\, 1 \,\, 0 \,\, 1 \,\, 0 \,\, 0  \\[-1.7mm]
0 \,\, 0 \,\, 0 \,\, 0 \,\, 1 \,\, 0 \,\, 0 \,\, 0 \,\, 0 \,\, 1 \,\, 1 \,\, 0 \,\, 0 \,\, 1 \,\, 0 \,\, 1 \,\, 1 \,\, 0 \,\, 0 \,\, 0 \,\, 1 \,\, 0 \,\, 1 \,\, 0  \\[-1.7mm]
1 \,\, 0 \,\, 0 \,\, 0 \,\, 0 \,\, 1 \,\, 0 \,\, 0 \,\, 0 \,\, 0 \,\, 1 \,\, 1 \,\, 0 \,\, 0 \,\, 1 \,\, 0 \,\, 1 \,\, 1 \,\, 0 \,\, 0 \,\, 0 \,\, 1 \,\, 0 \,\, 0  \\[-1.7mm]
0 \,\, 1 \,\, 0 \,\, 0 \,\, 0 \,\, 0 \,\, 1 \,\, 0 \,\, 0 \,\, 0 \,\, 0 \,\, 1 \,\, 1 \,\, 0 \,\, 0 \,\, 1 \,\, 0 \,\, 1 \,\, 1 \,\, 0 \,\, 0 \,\, 0 \,\, 1 \,\, 0  \\[-1.7mm]
1 \,\, 0 \,\, 1 \,\, 0 \,\, 0 \,\, 0 \,\, 0 \,\, 1 \,\, 0 \,\, 0 \,\, 0 \,\, 0 \,\, 1 \,\, 1 \,\, 0 \,\, 0 \,\, 1 \,\, 0 \,\, 1 \,\, 1 \,\, 0 \,\, 0 \,\, 0 \,\, 0  \\[-1.7mm]
0 \,\, 1 \,\, 0 \,\, 1 \,\, 0 \,\, 0 \,\, 0 \,\, 0 \,\, 1 \,\, 0 \,\, 0 \,\, 0 \,\, 0 \,\, 1 \,\, 1 \,\, 0 \,\, 0 \,\, 1 \,\, 0 \,\, 1 \,\, 1 \,\, 0 \,\, 0 \,\, 0  \\[-1.7mm]
0 \,\, 0 \,\, 1 \,\, 0 \,\, 1 \,\, 0 \,\, 0 \,\, 0 \,\, 0 \,\, 1 \,\, 0 \,\, 0 \,\, 0 \,\, 0 \,\, 1 \,\, 1 \,\, 0 \,\, 0 \,\, 1 \,\, 0 \,\, 1 \,\, 1 \,\, 0 \,\, 0  \\[-1.7mm]
0 \,\, 0 \,\, 0 \,\, 1 \,\, 0 \,\, 1 \,\, 0 \,\, 0 \,\, 0 \,\, 0 \,\, 1 \,\, 0 \,\, 0 \,\, 0 \,\, 0 \,\, 1 \,\, 1 \,\, 0 \,\, 0 \,\, 1 \,\, 0 \,\, 1 \,\, 1 \,\, 0  \\[-1.7mm]
1 \,\, 0 \,\, 0 \,\, 0 \,\, 1 \,\, 0 \,\, 1 \,\, 0 \,\, 0 \,\, 0 \,\, 0 \,\, 1 \,\, 0 \,\, 0 \,\, 0 \,\, 0 \,\, 1 \,\, 1 \,\, 0 \,\, 0 \,\, 1 \,\, 0 \,\, 1 \,\, 0  \\[-1.7mm]
1 \,\, 1 \,\, 0 \,\, 0 \,\, 0 \,\, 1 \,\, 0 \,\, 1 \,\, 0 \,\, 0 \,\, 0 \,\, 0 \,\, 1 \,\, 0 \,\, 0 \,\, 0 \,\, 0 \,\, 1 \,\, 1 \,\, 0 \,\, 0 \,\, 1 \,\, 0 \,\, 0  \\[-1.7mm]
0 \,\, 1 \,\, 1 \,\, 0 \,\, 0 \,\, 0 \,\, 1 \,\, 0 \,\, 1 \,\, 0 \,\, 0 \,\, 0 \,\, 0 \,\, 1 \,\, 0 \,\, 0 \,\, 0 \,\, 0 \,\, 1 \,\, 1 \,\, 0 \,\, 0 \,\, 1 \,\, 0  \\[-1.7mm]
1 \,\, 0 \,\, 1 \,\, 1 \,\, 0 \,\, 0 \,\, 0 \,\, 1 \,\, 0 \,\, 1 \,\, 0 \,\, 0 \,\, 0 \,\, 0 \,\, 1 \,\, 0 \,\, 0 \,\, 0 \,\, 0 \,\, 1 \,\, 1 \,\, 0 \,\, 0 \,\, 0  \\[-1.7mm]
0 \,\, 1 \,\, 0 \,\, 1 \,\, 1 \,\, 0 \,\, 0 \,\, 0 \,\, 1 \,\, 0 \,\, 1 \,\, 0 \,\, 0 \,\, 0 \,\, 0 \,\, 1 \,\, 0 \,\, 0 \,\, 0 \,\, 0 \,\, 1 \,\, 1 \,\, 0 \,\, 0  \\[-1.7mm]
0 \,\, 0 \,\, 1 \,\, 0 \,\, 1 \,\, 1 \,\, 0 \,\, 0 \,\, 0 \,\, 1 \,\, 0 \,\, 1 \,\, 0 \,\, 0 \,\, 0 \,\, 0 \,\, 1 \,\, 0 \,\, 0 \,\, 0 \,\, 0 \,\, 1 \,\, 1 \,\, 0  \\[-1.7mm]
1 \,\, 0 \,\, 0 \,\, 1 \,\, 0 \,\, 1 \,\, 1 \,\, 0 \,\, 0 \,\, 0 \,\, 1 \,\, 0 \,\, 1 \,\, 0 \,\, 0 \,\, 0 \,\, 0 \,\, 1 \,\, 0 \,\, 0 \,\, 0 \,\, 0 \,\, 1 \,\, 0  \\[-1.7mm]
1 \,\, 1 \,\, 0 \,\, 0 \,\, 1 \,\, 0 \,\, 1 \,\, 1 \,\, 0 \,\, 0 \,\, 0 \,\, 1 \,\, 0 \,\, 1 \,\, 0 \,\, 0 \,\, 0 \,\, 0 \,\, 1 \,\, 0 \,\, 0 \,\, 0 \,\, 0 \,\, 0  \\[-1.7mm]
0 \,\, 1 \,\, 1 \,\, 0 \,\, 0 \,\, 1 \,\, 0 \,\, 1 \,\, 1 \,\, 0 \,\, 0 \,\, 0 \,\, 1 \,\, 0 \,\, 1 \,\, 0 \,\, 0 \,\, 0 \,\, 0 \,\, 1 \,\, 0 \,\, 0 \,\, 0 \,\, 0  \\[-1.7mm]
1 \,\, 0 \,\, 1 \,\, 0 \,\, 1 \,\, 1 \,\, 1 \,\, 0 \,\, 0 \,\, 0 \,\, 1 \,\, 1 \,\, 0 \,\, 0 \,\, 0 \,\, 0 \,\, 0 \,\, 0 \,\, 0 \,\, 0 \,\, 0 \,\, 0 \,\, 0 \,\, 1  \\[-1.7mm]
\end{split} \right)
\label{H_golay_24_12_cyclic_group_19_21_shifts}
\end{equation}
}

For the extended Golay code, one can show an even stronger result
which allows for achieving decoding performance close to that of
maximum likelihood (ML) decoders. This result will be discussed in
detail in the next section.

In this setting, the parity-check matrix of the code
$\ve{H}_{[24,12],\star}$ is of the form shown below, and the {\textbf{SAD}} set
consists of a set of $23$ automorphisms
$\epsilon,\tau,\tau^2,\ldots, \tau^{22}$, where $\epsilon$ denotes
the identity element of $S_{24}$ (the symmetric group of order
$24$), and $\tau=(0\;1\;2\,\ldots\,21\;22)(23)$. The matrix
$\ve{H}_{[24,12],\star}$ consists of a subset of distinct cogs of
the extended Golay code, which allows for covering all parity-check
equations in the orbits of the chosen cogs.

It can be shown that for this particular combination of
parity-check matrix and {\textbf{SAD}} set, all uncorrectable erasure
patterns of weight up to $11$ correspond to codewords of the code
(see Section VII for more details).

\vspace{-0.3cm}

{

\ifTEXTDOUBLESPACING

\baselineskip=0.55555555\baselineskip

\fi

\begin{equation}
\ve{H}_{[24,12],\star}=\left(\begin{split} 1 \,\, 1 \,\, 1 \,\, 0
\,\, 0 \,\, 0 \,\, 0 \,\, 0 \,\, 1 \,\, 0 \,\, 0 \,\,
1 \,\, 1 \,\, 0 \,\, 0 \,\, 0 \,\, 0 \,\, 0 \,\, 1 \,\, 0 \,\, 0 \,\, 0 \,\, 0 \,\, 1  \\[-1.7mm]
1 \,\, 1 \,\, 0 \,\, 0 \,\, 0 \,\, 0 \,\, 1 \,\, 0 \,\, 0 \,\, 0
\,\, 0 \,\,
1 \,\, 0 \,\, 0 \,\, 1 \,\, 1 \,\, 1 \,\, 0 \,\, 0 \,\, 0 \,\, 0 \,\, 0 \,\, 0 \,\, 1  \\[-1.7mm]
1 \,\, 1 \,\, 0 \,\, 1 \,\, 0 \,\, 0 \,\, 1 \,\, 0 \,\, 1 \,\, 0
\,\, 1 \,\,
0 \,\, 0 \,\, 1 \,\, 0 \,\, 0 \,\, 0 \,\, 0 \,\, 0 \,\, 0 \,\, 0 \,\, 0 \,\, 0 \,\, 1  \\[-1.7mm]
1 \,\, 1 \,\, 1 \,\, 0 \,\, 0 \,\, 0 \,\, 1 \,\, 1 \,\, 0 \,\, 0
\,\, 0 \,\,
0 \,\, 0 \,\, 0 \,\, 0 \,\, 0 \,\, 0 \,\, 0 \,\, 0 \,\, 1 \,\, 0 \,\, 1 \,\, 0 \,\, 1  \\[-1.7mm]
1 \,\, 1 \,\, 0 \,\, 0 \,\, 0 \,\, 1 \,\, 0 \,\, 0 \,\, 0 \,\, 1
\,\, 0 \,\,
1 \,\, 0 \,\, 1 \,\, 0 \,\, 0 \,\, 0 \,\, 0 \,\, 0 \,\, 0 \,\, 0 \,\, 1 \,\, 0 \,\, 1  \\[-1.7mm]
1 \,\, 1 \,\, 0 \,\, 1 \,\, 0 \,\, 0 \,\, 0 \,\, 1 \,\, 0 \,\, 1
\,\, 0 \,\,
0 \,\, 0 \,\, 0 \,\, 0 \,\, 0 \,\, 1 \,\, 0 \,\, 1 \,\, 0 \,\, 0 \,\, 0 \,\, 0 \,\, 1  \\[-1.7mm]
0 \,\, 1 \,\, 1 \,\, 0 \,\, 0 \,\, 0 \,\, 1 \,\, 0 \,\, 0 \,\, 0
\,\, 0 \,\,
1 \,\, 0 \,\, 1 \,\, 0 \,\, 0 \,\, 0 \,\, 1 \,\, 0 \,\, 0 \,\, 1 \,\, 0 \,\, 0 \,\, 1  \\[-1.7mm]
1 \,\, 1 \,\, 0 \,\, 1 \,\, 1 \,\, 0 \,\, 0 \,\, 0 \,\, 0 \,\, 0
\,\, 0 \,\,
1 \,\, 0 \,\, 0 \,\, 0 \,\, 0 \,\, 0 \,\, 0 \,\, 0 \,\, 1 \,\, 1 \,\, 0 \,\, 0 \,\, 1  \\[-1.7mm]
1 \,\, 1 \,\, 1 \,\, 1 \,\, 0 \,\, 1 \,\, 0 \,\, 0 \,\, 0 \,\, 0
\,\, 0 \,\,
0 \,\, 0 \,\, 0 \,\, 1 \,\, 0 \,\, 0 \,\, 1 \,\, 0 \,\, 0 \,\, 0 \,\, 0 \,\, 0 \,\, 1  \\[-1.7mm]
1 \,\, 1 \,\, 0 \,\, 0 \,\, 1 \,\, 0 \,\, 0 \,\, 0 \,\, 0 \,\, 0
\,\, 1 \,\,
0 \,\, 0 \,\, 0 \,\, 1 \,\, 0 \,\, 0 \,\, 0 \,\, 1 \,\, 0 \,\, 0 \,\, 1 \,\, 0 \,\, 1  \\[-1.7mm]
0 \,\, 0 \,\, 1 \,\, 1 \,\, 0 \,\, 0 \,\, 1 \,\, 0 \,\, 1 \,\, 0
\,\, 0 \,\,
1 \,\, 0 \,\, 0 \,\, 1 \,\, 0 \,\, 0 \,\, 0 \,\, 0 \,\, 1 \,\, 0 \,\, 0 \,\, 0 \,\, 1  \\[-1.7mm]
0 \,\, 0 \,\, 1 \,\, 1 \,\, 0 \,\, 1 \,\, 0 \,\, 1 \,\, 0 \,\, 0
\,\, 1 \,\, 1 \,\, 0 \,\, 0 \,\, 1 \,\, 0 \,\, 0 \,\, 0 \,\, 0
\,\, 0 \,\, 0 \,\, 0 \,\, 1 \,\, 0   \notag
\end{split}\right)
\end{equation}
\vspace{-0.2cm}
}
\end{example}

We conclude this brief overview by pointing out that there exists
a strong connection between the problem of set
coverings~\cite{schoenheim64} and the problem of finding
\textbf{PD} and \textbf{SAD} sets. This relationship can be used
to develop simple heuristic search strategies for parity-check
matrices that have good \textbf{SAD}-set properties.

For this purpose, assume that the parity-check matrix is in
systematic form, with parity positions confined to the coordinates
$k,\ldots,n-1$. The number of automorphisms $N_A$ needed to move
any erasure patterns of cardinality $\leq d-1$ into the last $n-k$
positions (free of stopping sets of any size) has to satisfy
\begin{equation} \label{schoen}
N_A \geq \left\lfloor \frac{n}{k} \left\lfloor \frac{n-1}{k-1} \ldots
\left\lfloor \frac{n-d+2}{k-d+2} \right\rfloor \ldots \right\rfloor \right\rfloor,
\end{equation}
which follows from a result by Schoenheim~\cite{schoenheim64}, who
derived it in the context of set coverings.

The crux of the heuristic {\textbf{SAD}}-set search approach lies in
identifying subsets of columns of $\ve{H}$ that have cardinality
larger than $n-k$ and that are free of stopping sets of size up to
and including $d-1$, and in combining these results with ideas
borrowed from set covering theory.

For example, it is straightforward to show that the first $15$
columns of the parity-check matrix in Equation~\eqref{eq:gol-wolf}
are free of stopping sets of size less than eight, and that any
collection of more than $15$ columns must contain a stopping set
of size smaller than eight. Formula~\eqref{schoen} shows that in
this case, the smallest number of automorphisms required to map
any set of not more than seven coordinates of a codeword into the
first $15$ positions is at least $60$ (the same bound, for $12$
positions only, equals $498$). Clearly, there is no guarantee that
there exist sufficiently many automorphisms that map arbitrary
collections of coordinates to these positions. Nevertheless,
extensive computer simulations reveal that a good strategy for
identifying small {\textbf{SAD}} sets is to use a parity-check matrix in
systematic form, to find a large collection of columns $K$ free of
stopping sets of size less than $d$, and then perform a search for
automorphisms that map subsets of positions in $I-K$ to positions
in $K$. This method usually produces good results when the
underlying codes have a large automorphism group.

\section{Automorphism Group Decoders for the BEC}
\label{sec:algos}

In order to distinguish between iterative decoders that use
automorphisms to reduce errors due to stopping sets and standard
permutation decoders, we refer to the former as \emph{automorphism
group decoders}. Automorphism group decoders offer one significant
advantage over iterative decoders operating on redundant
parity-check matrices: they have low hardware complexity (since
only $n-k$ rows of the parity-check matrix are stored, along with
very few permutations) and at the same time excellent decoding
performance. This is, to a certain degree, offset by the slightly
increased computational complexity of the decoders. The results
presented in Section~\ref{sec:results} show that these additional
expenses are negligible for error rates of interest.

We describe next automorphism group decoders (AGD) for cyclic
codes, and also provide an example pertaining to extended cyclic
codes. We restrict our attention to these classes of codes since
large subgroups of the automorphism group of such codes are known
and since the implementation complexity of AGD decoders in this
case is very small. Nevertheless, the described decoding
techniques can be applied to other classes of codes for which some
information about the automorphism group is available. For the
case of extended cyclic codes, it is tacitly assumed that the
overall parity-check bit is confined to the last position of the
codewords and that its index is $n-1$.

Throughout the section, we make use of the following result.
\begin{theorem}(\cite[Ch.\ 8]{macwilliamsetal77})
Let $\ve{c}=(c_0,c_1,\ldots,c_{n-1})$ be a codeword of an
$[n,k,d]$ cyclic code. The automorphism group of the code contains
the following two sets of permutations, denoted by $C_1$ and
$C_2$:
\begin{itemize}
\item[$C_1$:]{The set of cyclic permutations
$\gamma^0,\gamma^1,\ldots,\gamma^{n-1}$, where $\gamma: i \to
i+1\, \mymod n$}; \item[$C_2$:]{The set of permutations
$\zeta^0,\zeta^1,\ldots,\zeta^{c-1}$, where $\zeta: i \to 2\cdot
i\, \mymod n$, and where $c$ denotes the cardinality of the
cyclotomic coset (of the $n$-th roots of unity) that contains the
element one.}
\end{itemize}
\label{theorem:permutations}
\end{theorem}

For extended cyclic codes, we use the same notation $C_1$ and
$C_2$ to describe permutations that fix $c_{n-1}$ and act on the
remaining coordinates as described in the theorem above. All
automorphisms can be decomposed into products of disjoint cycles.
Permutations in $C_1$ have one single cycle (or two cycles, for
the case of extended cyclic codes), while permutations in $C_2$
have a number of cycles that equals the number of cyclotomic
cosets $r$ of the $n$-roots of unity (or $r+1$ cycles, for the
case of extended cyclic codes). The number of cycles in the
automorphisms used for decoding influences the hardware complexity
of the scheme and should be kept small.

\textbf{$\AGDA$ Decoders}: These decoders use permutations drawn
from the set $C_1$, which reduces the permuter architecture to one
single shift register. One way to perform automorphism group
decoding is to set the permuter to $\gamma^0$, until the presence
of a stopping set is detected. In that case, the $\AGDA$ decoder
applies a randomly chosen cyclic shift $\gamma^i$ to the current
word, $i \neq 0$. If the iterative decoder encounters another
stopping set, the whole process is repeated with a (yet another)
randomly chosen cyclic permutation. The decoding process
terminates if either all permutations in $C_1$ are tested or if
the decoder successfully recovers the codeword.

Assume that the number of cogs of a cyclic code is at least $n-k$.
In this case, the parity-check matrix used for decoding consists
of $n-k$ different cogs, provided that such a matrix has full
row-rank. The redundant parity-check matrix consisting of the
collection of all vectors in the $C_1$-orbits of the cogs is
henceforth denoted by $\ve{H}_{\AGDA}$.

\textbf{$\AGDB$ Decoders}: These decoders use permutations drawn
from both $C_1$ and $C_2$, so that the resulting permuter
architecture is slightly more complex than that of $\AGDA$
decoders. If a stopping set is encountered, the decoder first
tries to resolve this set by applying a randomly chosen
permutation from $C_1$. Only after the whole set $C_1$ is
exhausted, a randomly chosen permutation from $C_2$ is applied to
the current decoder output. The parity-check matrix used for this
decoder is generated in terms of a greedy procedure. First, an
arbitrary cog is chosen. Then, another cog that is not in the
orbit of the first cog under any permutation in $C_2$ is chosen,
if such a cog exists. The procedure is repeated until either the
set of cyclic orbit generators is exhausted or until the matrix
contains $n-k$ rows. In the former case, additional rows are
chosen from the set of second-smallest weight codewords of the
dual code, provided that they give rise to a matrix of row-rank
$n-k$.

\begin{figure}[h!]
\begin{center}
\subfigure[$C_1$]{\label{AGD_A_extended_cyclic}
\psfrag{dots}[c][c]{$\vdots$}
\psfrag{ldots}[l][l]{$\vdots$}
\psfrag{0}[r][r][0.8]{$0$}
\psfrag{1}[r][r][0.8]{$1$}
\psfrag{n-3}[r][r][0.8]{$n-3$}
\psfrag{n-2}[r][r][0.8]{$n-2$}
\psfrag{n-1}[r][r][0.8]{$n-1$}
\includegraphics[scale=0.35]{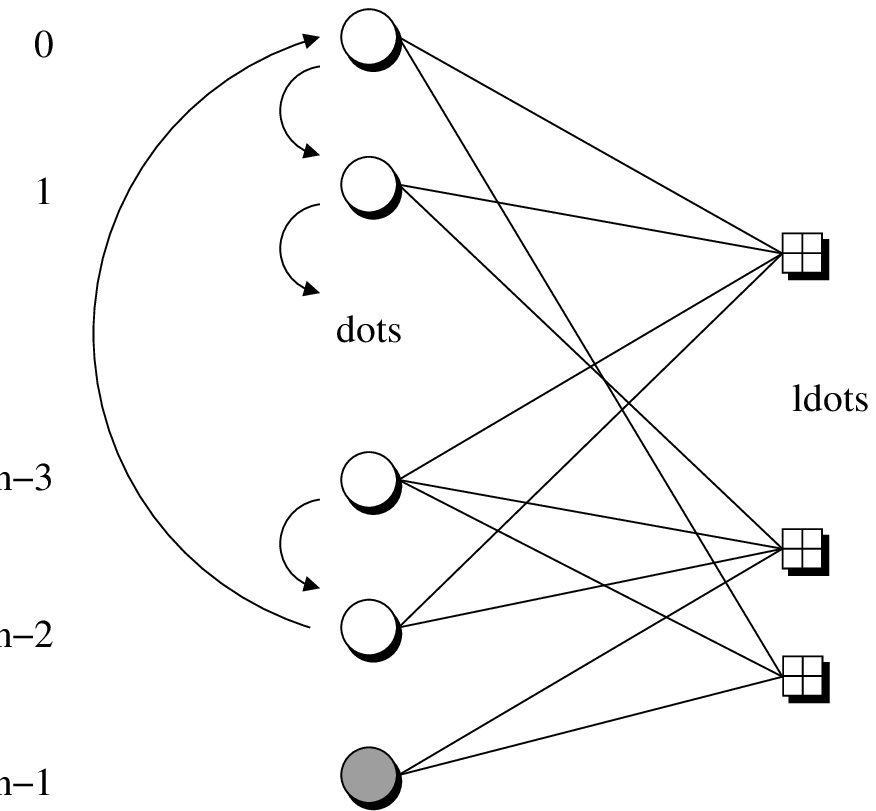}}
\hspace{1.5cm}\subfigure[$C_2$]{\label{AGD_B_extended_cyclic}
\psfrag{dots}[c][c]{$\vdots$}
\psfrag{ldots}[l][l]{$\vdots$}
\psfrag{0}[r][r][0.8]{$0$}
\psfrag{1}[r][r][0.8]{$1$}
\psfrag{n-3}[r][r][0.8]{$n-3$}
\psfrag{n-2}[r][r][0.8]{$n-2$}
\psfrag{n-1}[r][r][0.8]{$n-1$}
\includegraphics[scale=0.35]{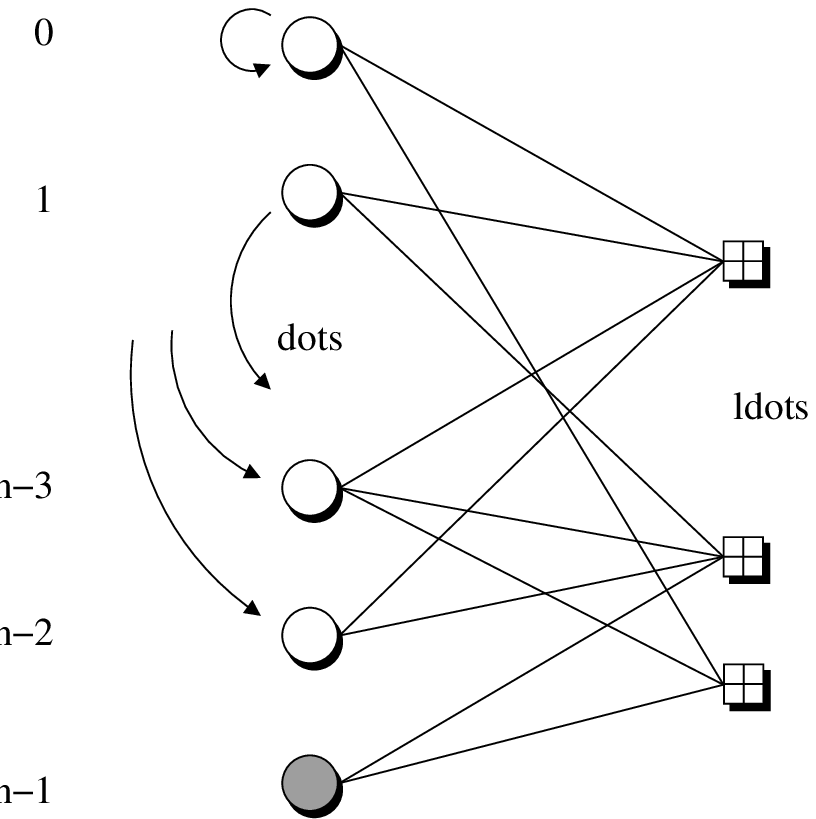}}
\caption{Permutations of $C_1$ and $C_{2}$ for extended cyclic codes.}\label{AGD_extended_cyclic}
\end{center}
\end{figure}

Figure~\ref{AGD_A_extended_cyclic} and
Figure~\ref{AGD_B_extended_cyclic} illustrate the permutation
operations used in the described decoding architectures. Here, the
check nodes and edges in the graph of an extended cyclic code are
fixed, while the variable nodes are permuted according to the
permutations $C_1$ and $C_2$ defined in Theorem
\ref{theorem:permutations}.

The set of minimum weight cogs can be partitioned into cog
families ($CF$), where all members of one given family have
identical intersection numbers. More precisely, a $CF$ is a set of
cogs with identical intersection cardinalities $|\Sigma_{\sigma,i}
\cap \Sigma_{\sigma,j}|$, and consequently, identical values of
$S_{\sigma,2}$, as given in Equation~\eqref{eq:bound2_cyclic_1}.
This result is a consequence of the fact that cogs within the same
family have the same parameters $|XY|_{\kappa}$, introduced in
Lemma~\ref{lemma:num_stopsets_cyclic_pair_of_rows}.

Cogs from the family with optimal intersection numbers are good
candidates for constructing parity-check matrices amenable for
automorphism group decoding. In this case, it is important to
select subsets of cogs that are not images of each other under the
automorphisms used by the decoder. For example, for the
$[23,12,7]$ Golay code, the $22$ cogs of minimum weight eight can
be grouped into two families $CF_{1}$ and $CF_{2}$, each
consisting of $11$ cogs. Both families are closed under the
permutations in $C_{2}$.


We partition the set of shift indices $\kappa$, $\kappa \in \{ 1, 2, \ldots, n-1\},$ into subsets $\Psi_i$ which have equal parameters $|OO_{\kappa}|$, $|OZ_{\kappa}|$, $|ZO_{\kappa}|$, and $|ZZ_{\kappa}|$.
Table~\ref{table:cog_group_intersections_23_12} lists a selected
subset of intersection numbers for cogs in the $CF_{1}$ and
$CF_{2}$ families. It can be seen from the table that every cog in $CF_{1}$ has parameters $|OO_{\kappa}|=0$, $|OZ_{\kappa}|=|ZO_{\kappa}|=8$, and $|ZZ_{\kappa}| = 7$  for the two shift indices forming $\Psi_1$.

\begin{table}
\caption{Intersection numbers of the $CF_{1}$ and $CF_{2}$
families of the $[23,12,7]$ Golay code.}
\label{table:cog_group_intersections_23_12}
\begin{center}
\begin{tabular}{|c|cccc|}
\hline
$CF_{1}$  &\multicolumn{4}{c|}{Intersection numbers}\\
$\kappa \in $& $|OO_{\kappa}|$ & $|OZ_{\kappa}|$ & $|ZO_{\kappa}|$
& $|ZZ_{\kappa}|$ \\ \hline
$\Psi_1$, $|\Psi_1|=2$ & 0 & 8 & 8 & 7 \\
$\Psi_2$, $|\Psi_2|=8$  & 4 & 4 & 4 & 11\\
$\Psi_3$, $|\Psi_3|=12$  & 2 & 6 & 6 & 9\\ \hline
\end{tabular}
\end{center}
\begin{center}
\begin{tabular}{|c|cccc|}
\hline
$CF_{2}$  &\multicolumn{4}{c|}{Intersection numbers}\\
$\kappa \in $ & $|OO_{\kappa}|$ & $|OZ_{\kappa}|$ &
$|ZO_{\kappa}|$ & $|ZZ_{\kappa}|$ \\ \hline
$\Psi_4$, $|\Psi_4|=6$  & 4 & 4 & 4 & 11\\
$\Psi_4$, $|\Psi_4|=16$ & 2 & 6 & 6 & 9 \\ \hline
\end{tabular}
\end{center}
\end{table}%

\begin{table}
\caption{Intersection numbers of the unique $CF$ of the
$[31,16,7]$ BCH code.} \label{table:cog_group_intersections_31_16}
\begin{center}
\begin{tabular}{|c|cccc|}
\hline
$CF$  &\multicolumn{4}{c|}{Intersection numbers}\\
$\kappa \in $ & $|OO_{\kappa}|$ & $|OZ_{\kappa}|$ &
$|ZO_{\kappa}|$ & $|ZZ_{\kappa}|$ \\ \hline
$\Psi^1$, $|\Psi^1|=2$ & 4 & 4 & 4 & 19\\
$\Psi^2$, $|\Psi^2|=4$  & 0 & 8 & 8 & 15\\
$\Psi^3$, $|\Psi^3|=24$ & 2 & 6 & 6 & 17 \\ \hline
\end{tabular}
\end{center}
\end{table}%

For the $[31,16,7]$ BCH code, the set of $15$ cogs of weight eight
forms one single family $CF$. A selected set of intersection
numbers of this family is shown in
Table~\ref{table:cog_group_intersections_31_16}.

\section{Performance Results}
\label{sec:results}
\label{sec:simulation}

We present next a selected set of simulation results for the bit
error rate (BER) and frame error rate (FER) performance of both
redundant parity-check matrices and various automorphism group
decoders. The BER equals half of the fraction of the residual
erasures, as at the end of the decoding procedure one can guess
all uncorrectable bits. A frame error is declared if at least one
symbol in the frame estimated by the decoder does not match the
corresponding symbol in the transmitted frame. Our results
indicate that the FER performance of codes operating in the above
described manner can be improved by adding an additional
``guessing'' feature, which is described in more details in the
concluding part of the section.

\subsection{The Golay and Extended Golay Code}

\begin{figure}[h!]
\begin{minipage}{9cm}
\begin{center}
\subfigure[]{\label{fig:golay_23_wt8_performance}
\psfrag{BER}[cb][cb]{$\BER\,\rightarrow$}
\psfrag{EP}[ct][ct]{$\leftarrow\,\EP$}
\psfrag{wt8gr10row11_____________}[l][l][1]{\scriptsize{$\cog_{23,12,{\A}}$,
$m=11$, BP}}
\psfrag{wt8gr10row16}[l][l][1]{\scriptsize{$\cog_{23,12,{\A}}$,
$m=16$, BP}}
\psfrag{wt8gr10row18}[l][l][1]{\scriptsize{$\cog_{23,12,{\A}}$,
$m=18$, BP}}
\psfrag{wt8gr10row23}[l][l][1]{\scriptsize{$\cog_{23,12,{\A}}$,
$m=23$, BP}} \psfrag{ML estimation}[l][l][1]{\scriptsize{ML
estimation}}
\includegraphics[scale=0.48]{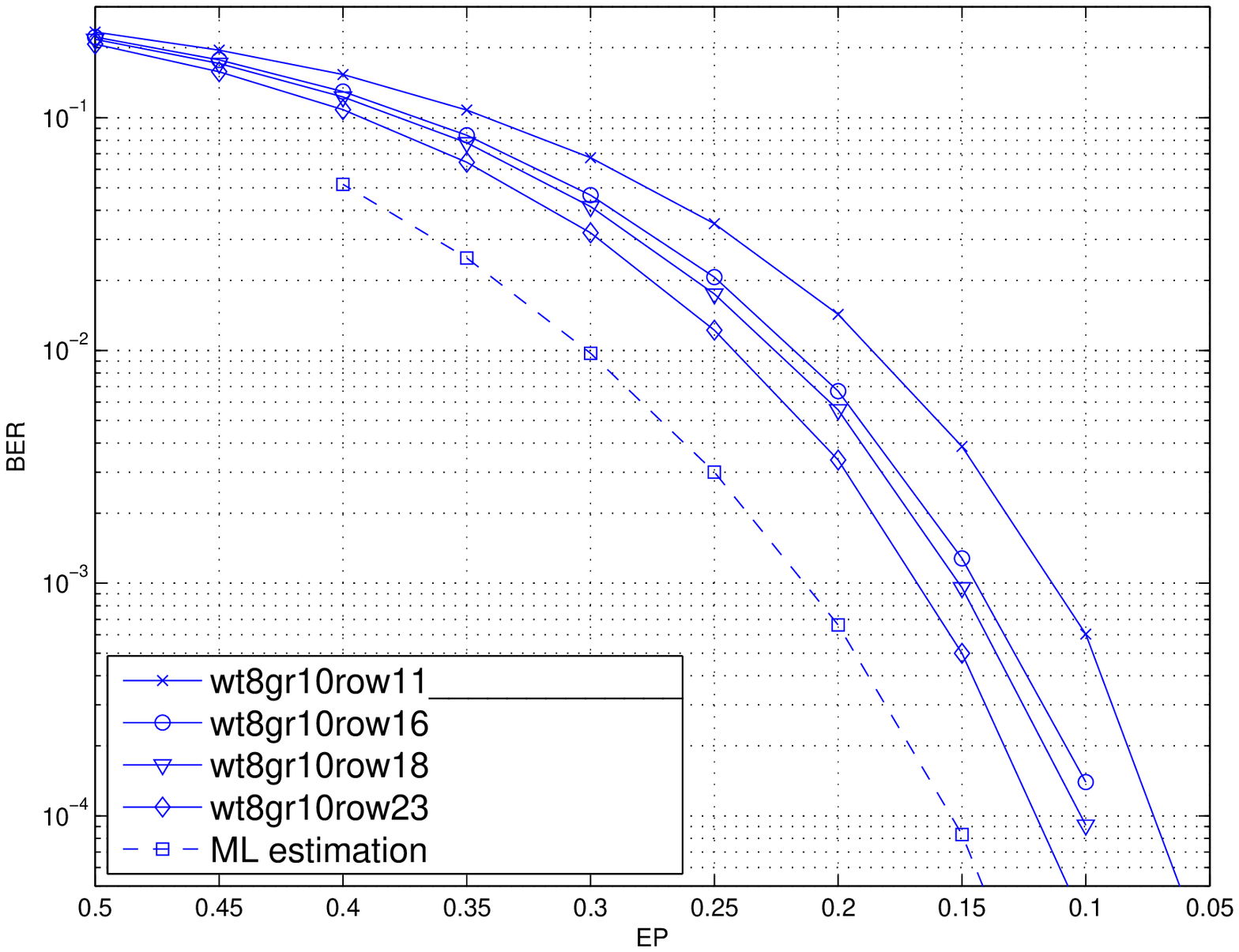}
}
\subfigure[]{\label{ber_golay_24_12_different_cog_standard_and_extended_cyclic_compare_to_papers}
\psfrag{EP}[ct][ct][1]{$\leftarrow\,\EP$}
\psfrag{BER}[cb][cb][1]{$\BER\,\rightarrow$} \psfrag{ber 24 12
diff. cog standard}[l][l][1]{\scriptsize{$\ve{H}_{[24,12],\star}$, $m=12$,
BP}} \psfrag{ber 24 12 wolfmann
decoding}[l][l][1]{\scriptsize{$\ve{H}_{\W}$, $m=168$, BP}}
\psfrag{ber 24 12 siegel 34
rows}[l][l][1]{\scriptsize{$\ve{H}_{\HS}$, $m=34$, BP}}
\psfrag{ber 24 12 diff. cog ext.
cyclic___}[l][l][1]{\scriptsize{$\ve{H}_{[24,12],\star}$, $m=12$, $\AGDA$}}
\psfrag{ber 24 12 diff. cog ext.
auto.}[l][l][1]{\scriptsize{$\ve{H}_{[24,12],\star}$, $m=12$, $\AGDB$}}
\psfrag{ber 24 12 ML estimation}[l][l][1]{\scriptsize{ML
estimation}}
\includegraphics[scale=0.48]{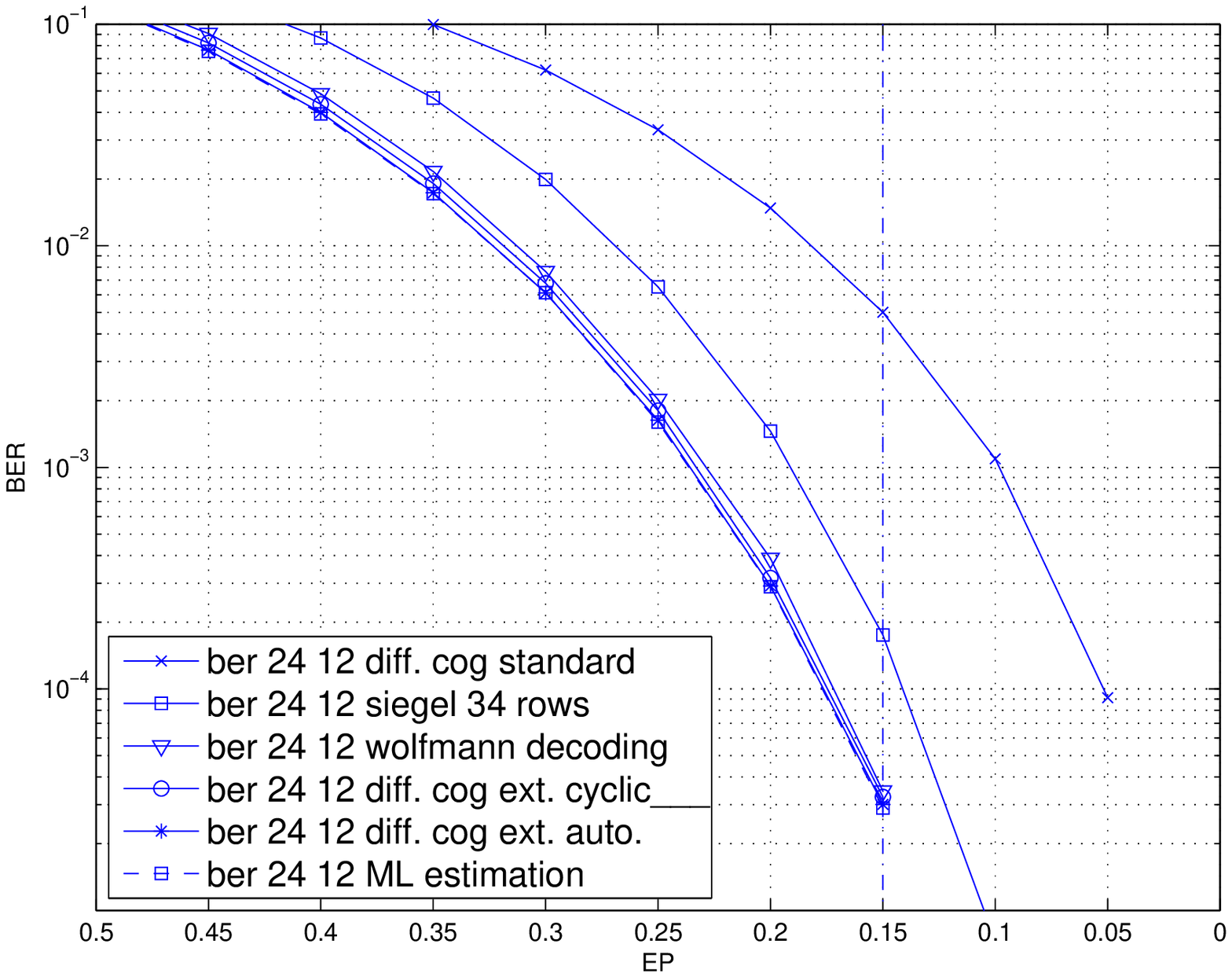}
}
\end{center}
\end{minipage}
\begin{minipage}{9cm}
\begin{center}
\subfigure[]{\label{fer_golay_24_12_different_cog_standard_and_extended_cyclic_compare_to_papers}
\psfrag{EP}[ct][ct][1]{$\leftarrow\,\EP$}
\psfrag{FER}[cb][cb][1]{$\FER\,\rightarrow$} \psfrag{fer 24 12
diff. cog standard}[l][l][1]{\scriptsize{$\ve{H}_{[24,12],\star}$, $m=12$,
BP}} \psfrag{fer 24 12 wolfmann
decoding}[l][l][1]{\scriptsize{$\ve{H}_{\W}$, $m=168$, BP}}
\psfrag{fer 24 12 siegel 34
rows}[l][l][1]{\scriptsize{$\ve{H}_{\HS}$, $m=34$, BP}}
\psfrag{fer 24 12 diff. cog ext.
cyclic___}[l][l][1]{\scriptsize{$\ve{H}_{[24,12],\star}$, $m=12$, $\AGDA$}}
\psfrag{fer 24 12 diff. cog ext.
auto.}[l][l][1]{\scriptsize{$\ve{H}_{[24,12],\star}$, $m=12$, $\AGDB$}}
\psfrag{fer 24 12 ML}[l][l][1]{\scriptsize{ML estimation}}
\includegraphics[scale=0.48]{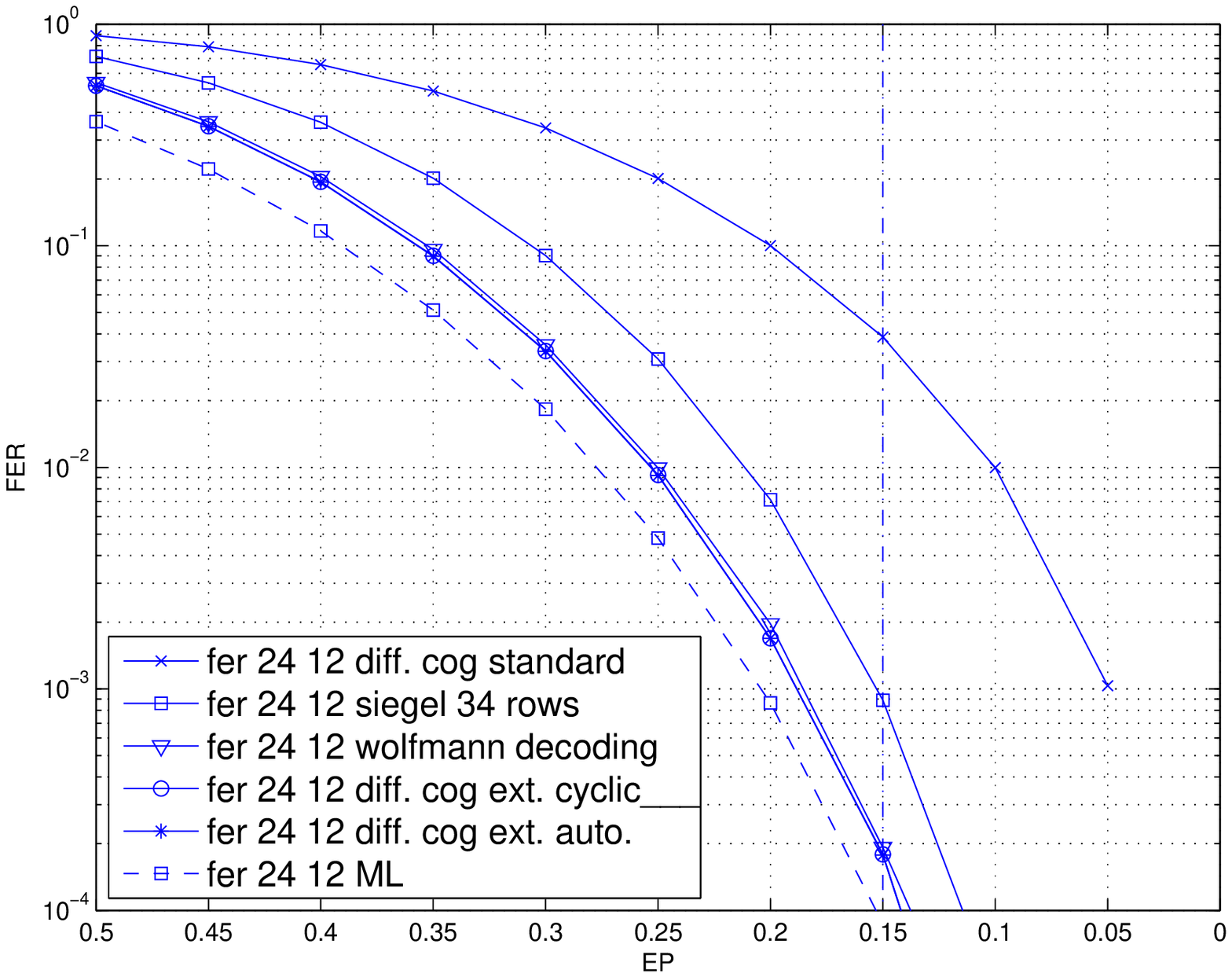}
}
\subfigure[]{\label{iter_golay_24_12_different_cog_standard_and_extended_cyclic_compare_to_papers}
  \psfrag{EP}[ct][ct][1]{$\leftarrow\,\EP$}
  \psfrag{Iterations}[cb][cb][1]{Number of iterations$\,\rightarrow$}
  \psfrag{iter 24 12 diff. cog
    standard}[l][l][1]{\scriptsize{$\ve{H}_{[24,12],\star}$, $m=12$, BP}}
  \psfrag{iter 24 12 wolfmann
    decoding}[l][l][1]{\scriptsize{$\ve{H}_{\W}$, $m=168$, BP}}
  \psfrag{iter 24 12 siegel 34
    rows}[l][l][1]{\scriptsize{$\ve{H}_{\HS}$, $m=34$, BP}}
  \psfrag{iter 24 12 diff. cog ext.
    cyclic___}[l][l][1]{\scriptsize{$\ve{H}_{[24,12],\star}$, $m=12$, $\AGDA$}}
  \psfrag{iter 24 12 diff. cog ext.
    auto.}[l][l][1]{\scriptsize{$\ve{H}_{[24,12],\star}$, $m=12$, $\AGDB$}}
\includegraphics[scale=0.48]{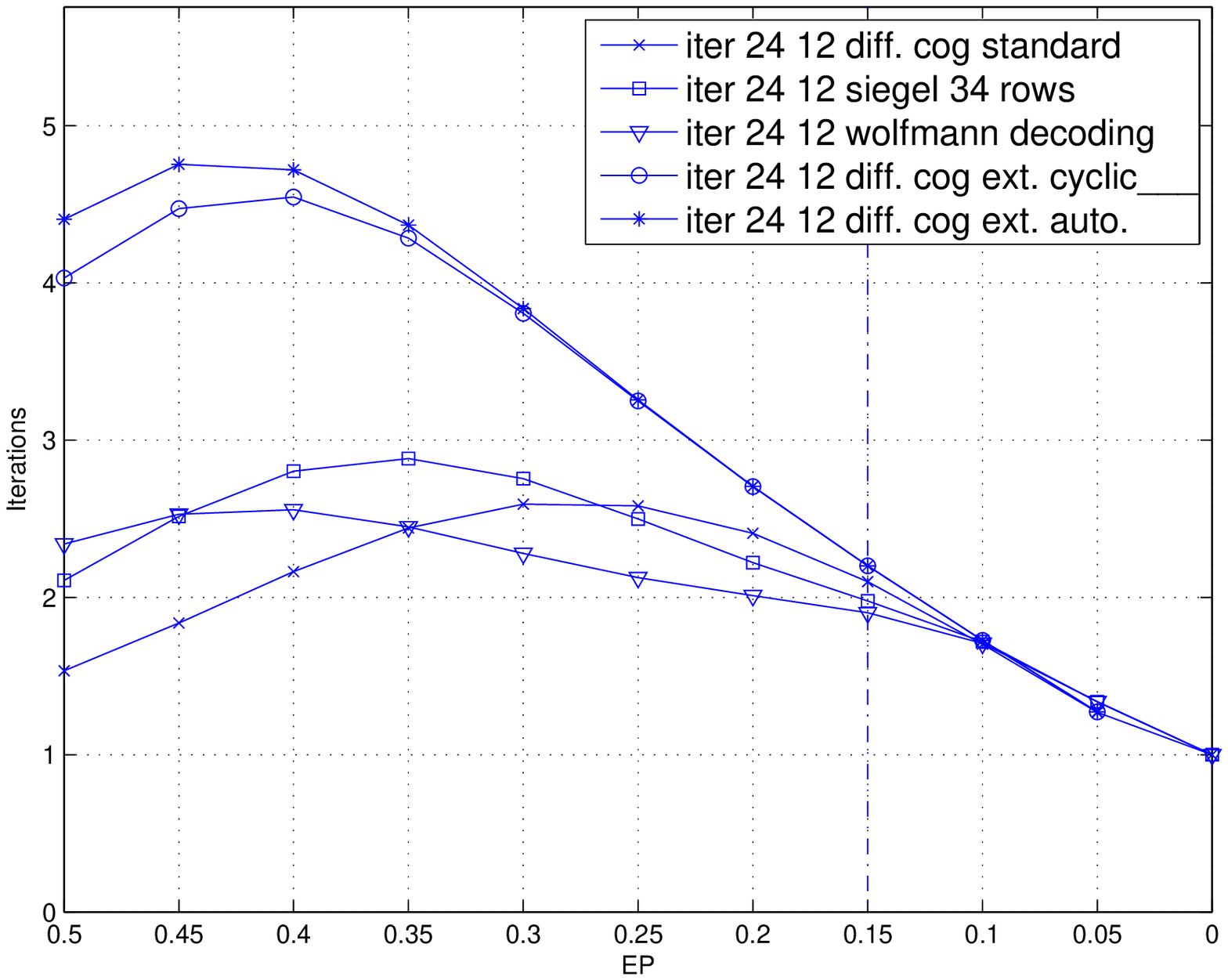}
}
\end{center}
\end{minipage}
\caption{\label{golay_24_12_different_cog_standard_and_extended_cyclic_compare_to_papers}Performance
of the $[23,11,7]$ and $[24,12,8]$ (extended) Golay codes. }
\end{figure}

The residual bit error rate performance of edge-removal decoding
on the cyclic parity-check matrix representation based on
$\cog_{23,12,{\A}}$ is shown in
Figure~\ref{fig:golay_23_wt8_performance}. In this figure, the
cyclic $23 \times 23$ matrix is truncated to $11$, $16$, $18$, and
$23$ rows, corresponding to the upper bounds on the stopping
redundancy hierarchy obtained from this cyclic representation. For
comparison, the performance of the $\ML$ decoder is also shown in
the same figure.

In
Table~\ref{table:golay_24_12_compare_uncorrectable_erasure_patterns}
we list the number of uncorrectable erasure patterns of size up to
$\sigma=12$ in several parity-check matrices of the
$[24,12,8]$ extended Golay code; $\ve{H}_{[24,12],\star}$ and $\ve{H}_{\W}$ were
defined in Section~\ref{sec:analysis}, while the matrix
$\ve{H}_{\HS}$ corresponds to the matrix of dimension $34 \times
24$ described in~\cite{hanetal07}. The index $\ML$ refers to the
erasure patterns that cannot be decoded by an ML decoder.
Figures~\ref{ber_golay_24_12_different_cog_standard_and_extended_cyclic_compare_to_papers}, \ref{fer_golay_24_12_different_cog_standard_and_extended_cyclic_compare_to_papers}, and \ref{iter_golay_24_12_different_cog_standard_and_extended_cyclic_compare_to_papers}
show the performance of iterative decoders operating on a
selection of the parity-check matrices in
Table~\ref{table:golay_24_12_compare_uncorrectable_erasure_patterns}
using standard BP decoding on $\ve{H}_{[24,12],\star}$ and using
$\AGDA$ and $\AGDB$ decoders.

As can be seen, there is a significant performance gain of $\AGDA$
or $\AGDB$ decoders when compared to that of standard edge-removal
decoders operating on the redundant parity-check matrix described
in~\cite{hanetal07} or on any other non-redundant Tanner graph. In
fact, the BER performance of $\AGDA$ and $\AGDB$ decoders
approaches the performance of ML decoders. For $\EP\leq 0.15$, all
matrix representations require an almost identical average number
of iterations, indicated by the vertical bar in
Figures~\ref{ber_golay_24_12_different_cog_standard_and_extended_cyclic_compare_to_papers},
\ref{fer_golay_24_12_different_cog_standard_and_extended_cyclic_compare_to_papers},
and
\ref{iter_golay_24_12_different_cog_standard_and_extended_cyclic_compare_to_papers}.

\begin{center}
\begin{table}[htb]
\caption{Uncorrectable erasure patterns of the $[24,12,8]$
extended Golay code.
\label{table:golay_24_12_compare_uncorrectable_erasure_patterns}}
\begin{center}
{\footnotesize{
\begin{tabular}{|c|c|c|c|c|c|}\hline
&\multicolumn{5}{c|}{Number of uncorrectable erasure
patterns}\\\cline{2-6} $\sigma$&$\ve{H}_{[24,12],\star}$ on BP&
$\ve{H}_{[24,12],\star}$ on $\AGDA$& $\ve{H}_{\W}$ on BP& $\ve{H}_{\HS}$ \cite{hanetal07} on BP&$\ML$\\\hline
$3$&$7$&$0$&$0$&$0$&$0$\\
$4$&$190$&$0$&$0$&$0$&$0$\\
$5$&$2231$&$0$&$0$&$0$&$0$\\
$6$&$15881$&$0$&$0$&$0$&$0$\\
$7$&$79381$&$0$&$0$&$0$&$0$\\
$8$&$293703$&$759$&$759$&$3284$&$759$\\
$9$&$805556$&$12144$ &$12158$&$78218$&$12144$\\
$10$& $1613613$ & $91080$ & $93477$ & $580166$&$91080$\\
$11$& $2378038$ & $425040$ & $481764$ & $1734967$  &$425040$\\
$12$& $2690112$ & $1322178$ & $1547590$ & $2569618$ &$1313116$\\
$\geq 13$& $24\choose\sigma$ &$24\choose\sigma$ &$24\choose\sigma$ & $24\choose\sigma$ &$24\choose\sigma$\\[0.05cm]\hline
\end{tabular}
}}
\end{center}
\end{table}
\end{center}

\subsection{BCH Codes}

\begin{figure}[h!]
\begin{minipage}{9cm}
\begin{center}
\subfigure{\label{ber_bch_127_113_different_cog_standard_and_cyclic_and_ML_bound}
\psfrag{EP}[ct][ct][1]{$\leftarrow\,\EP$}
\psfrag{BER}[cb][cb][1]{$\BER\,\rightarrow$} \psfrag{ber 127 113
cog 14 14 rows
st.}[l][l][1]{\scriptsize{${\mathrm{cog}}_{127,113,{\mathrm{A}}}$,
$m=14$, BP}} \psfrag{ber 127 113 different cog 14 rows
st.}[l][l][1]{\scriptsize{Various cogs, $m=14$, BP}} \psfrag{ber
127 113 cog 14 34 rows
st.}[l][l][1]{\scriptsize{${\mathrm{cog}}_{127,113,{\mathrm{A}}}$,
$m=34$, BP}} \psfrag{ber 127 113 different cog 14 rows
cyclic}[l][l][1]{\scriptsize{Var. cogs, $m=14$, $\AGDA$}}
\psfrag{ber 127 113 MLSE Union bound}[l][l][1]{\scriptsize{ML
Union Bound}}
\includegraphics[scale=0.48]{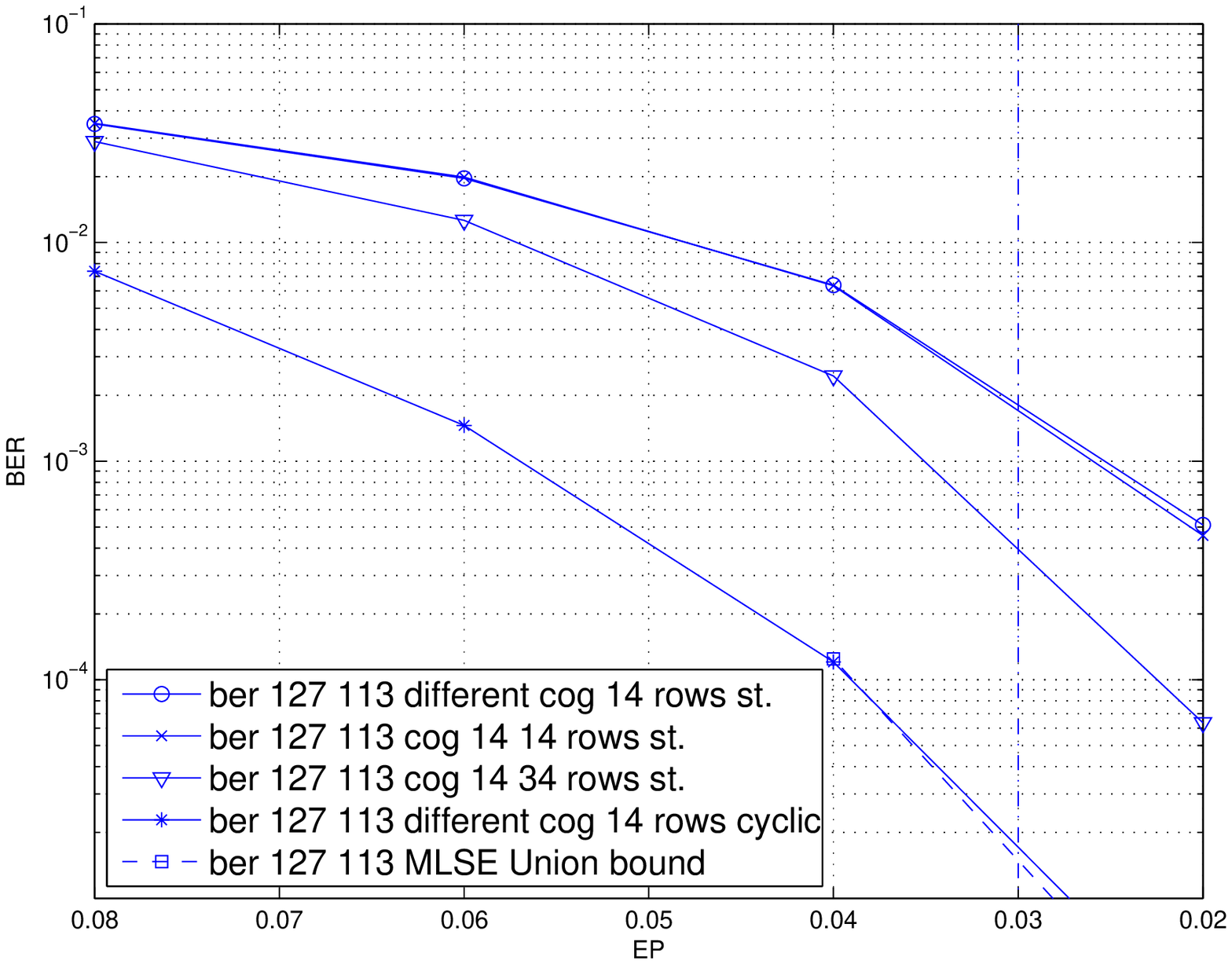}
}
\end{center}
\end{minipage}\hspace{0.1cm}
\begin{minipage}{9cm}
\begin{center}
\subfigure{\label{fer_bch_127_113_different_cog_standard_and_cyclic_and_ML_bound}
\psfrag{EP}[ct][ct][1]{$\leftarrow\,\EP$}
\psfrag{FER}[cb][cb][1]{$\FER\,\rightarrow$} \psfrag{fer 127 113
cog 14 14 rows
st.}[l][l][1]{\scriptsize{${\mathrm{cog}}_{127,113,{\mathrm{A}}}$,
$m=14$, BP}} \psfrag{fer 127 113 different cog 14 rows
st.}[l][l][1]{\scriptsize{Various cogs, $m=14$, BP}} \psfrag{fer
127 113 cog 14 34 rows
st.}[l][l][1]{\scriptsize{${\mathrm{cog}}_{127,113,{\mathrm{A}}}$,
$m=34$, BP}} \psfrag{fer 127 113 different cog 14 rows
cyclic}[l][l][1]{\scriptsize{Var. cogs, $m=14$, $\AGDA$}}
\psfrag{fer 127 113 MLSE Union bound}[l][l][1]{\scriptsize{ML
Union Bound}}
\includegraphics[scale=0.48]{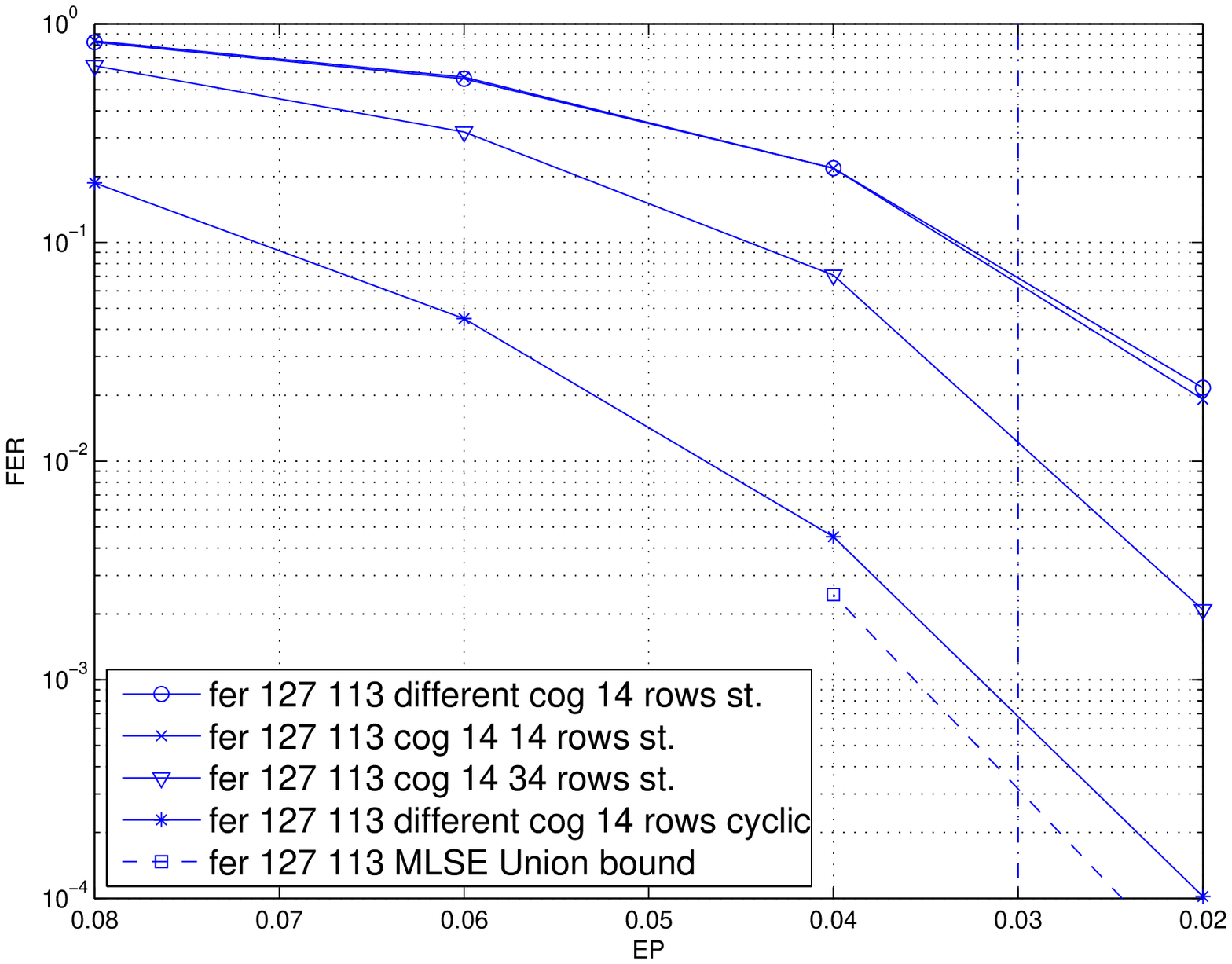}
}
\end{center}
\end{minipage}
\caption{\label{bch_127_113_different_cog_standard_and_cyclic_and_ML_bound}Performance
of the $[127,113,5]$-BCH code.}
\end{figure}

Figure~\ref{bch_127_113_different_cog_standard_and_cyclic_and_ML_bound}
plots the performance of iterative edge-removal and automorphism
group decoders with and without permutation features, operating on
several (redundant) parity-check matrices of the $[127,113,5]$ BCH
code.

The standard edge-removal decoder uses parity-check matrices of
cyclic form, generated by the cog
${\mathrm{cog}}_{127,113,{\mathrm{A}}}$. The matrices have
$m=n-k=14$ and $m=34$ rows, respectively. The cyclic parity-check
matrix with $36$ rows offers significant performance improvements,
when compared to its counterpart with $14$ rows. Excellent BERs
can also be achieved with the $\AGDA$ decoder that operates on a
parity-check matrix containing $14$ rows of distinct
${\mathrm{cog}}$s.

Figure
\ref{bch_127_113_different_cog_standard_and_cyclic_and_ML_bound}
shows that the AGD$_A$ decoder performs very close to the $\ML$
union bound, which is known to be fairly tight for the $\BER$s
considered. Again, a vertical bar (placed at $\EP= 0.03$) is
used to indicate the BER region for which the average number of
iterations needed for successful decoding is approximately the
same for all tested decoders.

\begin{figure}
\begin{minipage}{9cm}
\begin{center}
\subfigure[]{\label{ber_bch_31_16_all_cog_standard_and_cyclic_and_ML}
\psfrag{EP}[ct][ct][1]{$\leftarrow\,\EP$}
\psfrag{BER}[cb][cb][1]{$\BER\,\rightarrow$} \psfrag{ber 31 16 cog
3 15 rows
st.}[l][l][1]{\scriptsize{${\mathrm{cog}}_{31,16,{\mathrm{A}}}$,
$m=15$, BP}} \psfrag{ber 31 16 all cog 15 rows
st.}[l][l][1]{\scriptsize{Var. cogs, $m=15$, BP}}
\psfrag{ber 31 16 cog 3 21 rows
st.}[l][l][1]{\scriptsize{${\mathrm{cog}}_{31,16,{\mathrm{A}}}$,
$m=21$, BP}} \psfrag{ber 31 16 cog 3 31 rows
st.}[l][l][1]{\scriptsize{${\mathrm{cog}}_{31,16,{\mathrm{A}}}$,
$m=31$, BP}} \psfrag{ber 31 16 all cog 15 rows
cyclic}[l][l][1]{\scriptsize{Var. cogs, $m=15$, $\AGDA$}}
\psfrag{ber 31 16 ML estimation}[l][l][1]{\scriptsize{ML
estimation}}
\includegraphics[scale=0.48]{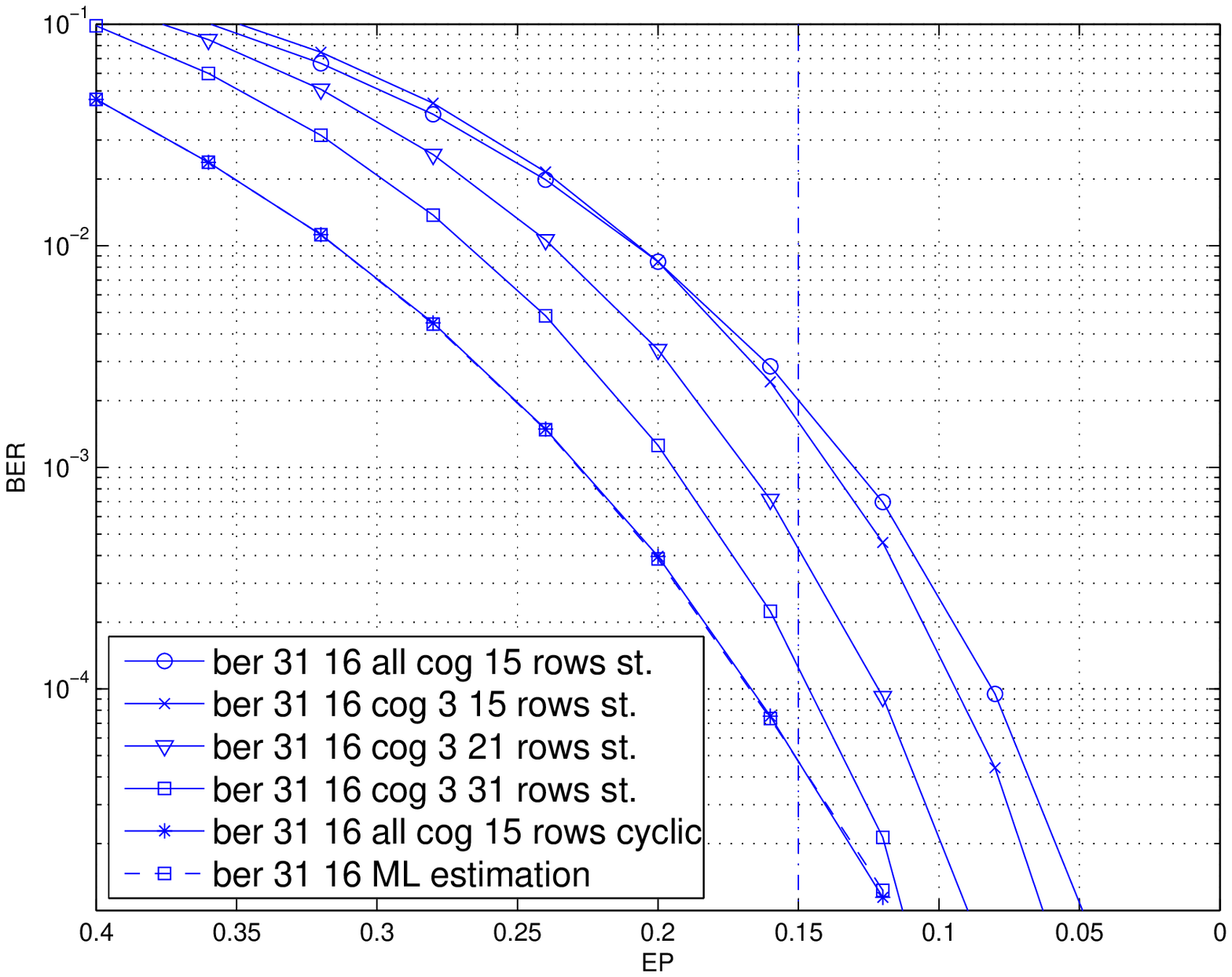}
}
\subfigure[]{\label{fer_bch_31_16_all_cog_standard_and_cyclic_and_ML}
\psfrag{EP}[ct][ct][1]{$\leftarrow\,\EP$}
\psfrag{FER}[cb][cb][1]{$\FER\,\rightarrow$} \psfrag{fer 31 16 cog
3 15 rows
st.}[l][l][1]{\scriptsize{${\mathrm{cog}}_{31,16,{\mathrm{A}}}$,
$m=15$, BP}} \psfrag{fer 31 16 all cog 15 rows
st.}[l][l][1]{\scriptsize{Various cogs, $m=15$, BP}} \psfrag{fer
31 16 cog 3 21 rows
st.}[l][l][1]{\scriptsize{${\mathrm{cog}}_{31,16,{\mathrm{A}}}$,
$m=21$, BP}} \psfrag{fer 31 16 cog 3 31 rows
st.}[l][l][1]{\scriptsize{${\mathrm{cog}}_{31,16,{\mathrm{A}}}$,
$m=31$, BP}} \psfrag{fer 31 16 all cog 15 rows
cyclic}[l][l][1]{\scriptsize{Var. cogs, $m=15$, $\AGDA$}}
\psfrag{fer 31 16 ML}[l][l][1]{\scriptsize{ML estimation}}
\includegraphics[scale=0.48]{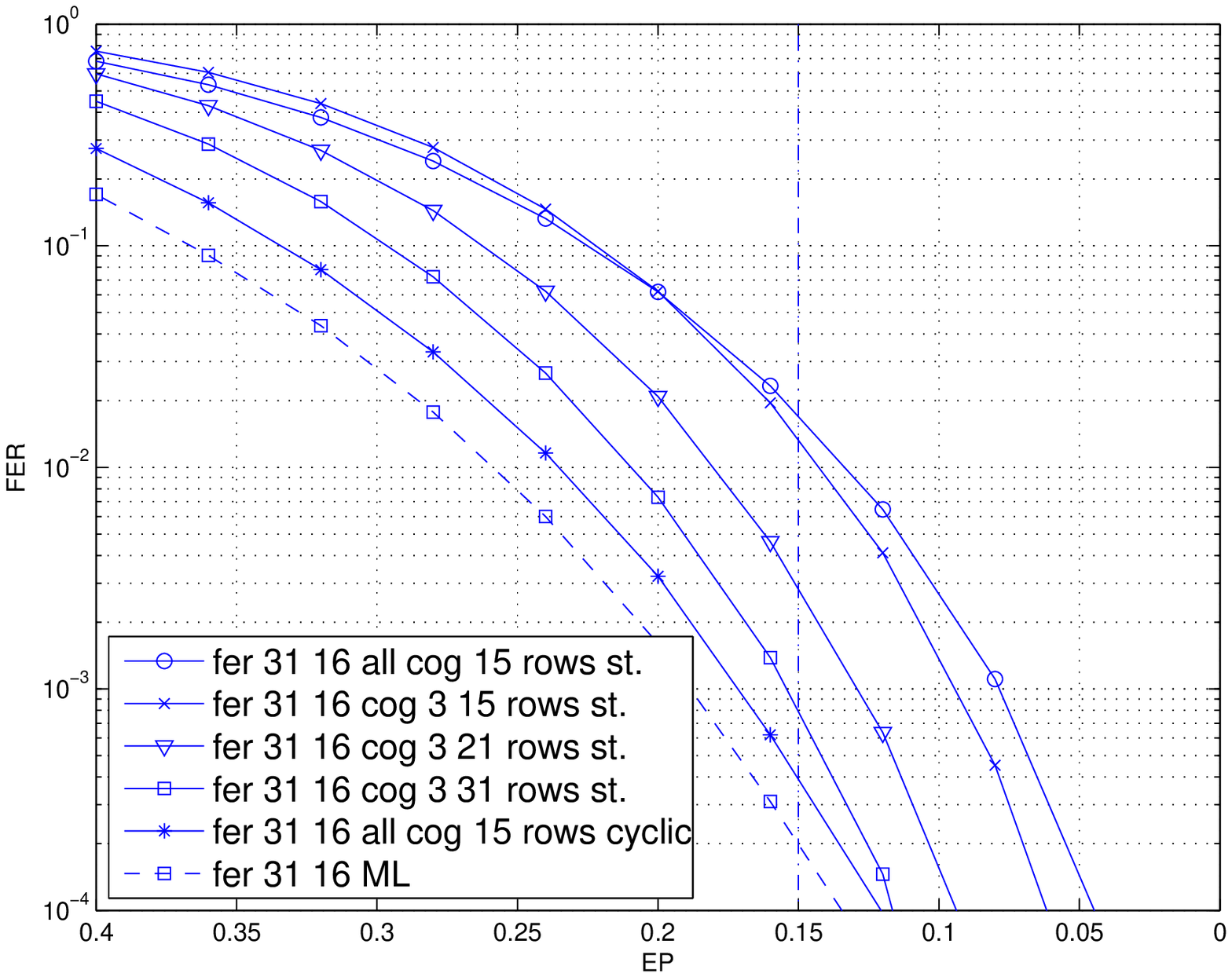}
}
\end{center}
\end{minipage}
\begin{minipage}{9cm}
\begin{center}
\subfigure[]{\label{iter_bch_31_16_all_cog_standard_and_cyclic_and_ML}
\psfrag{EP}[ct][ct][1]{$\leftarrow\,\EP$}
\psfrag{Iterations}[cb][cb][1]{Number of
iterations$\,\rightarrow$} \psfrag{iter 31 16 cog 3 15 rows
st.}[l][l][1]{\scriptsize{${\mathrm{cog}}_{31,16,{\mathrm{A}}}$,
$m=15$, BP}} \psfrag{iter 31 16 all cog 15 rows
st.}[l][l][1]{\scriptsize{Various cogs, $m=15$, BP}} \psfrag{iter
31 16 cog 3 21 rows
st.}[l][l][1]{\scriptsize{${\mathrm{cog}}_{31,16,{\mathrm{A}}}$,
$m=21$, BP}} \psfrag{iter 31 16 cog 3 31 rows
st.}[l][l][1]{\scriptsize{${\mathrm{cog}}_{31,16,{\mathrm{A}}}$,
$m=31$, BP}} \psfrag{iter 31 16 all cog 15 rows
cyclic}[l][l][1]{\scriptsize{Var. cogs, $m=15$, $\AGDA$}}
\psfrag{iter 31 16 ML}[l][l][1]{\scriptsize{ML estimation}}
\includegraphics[scale=0.48]{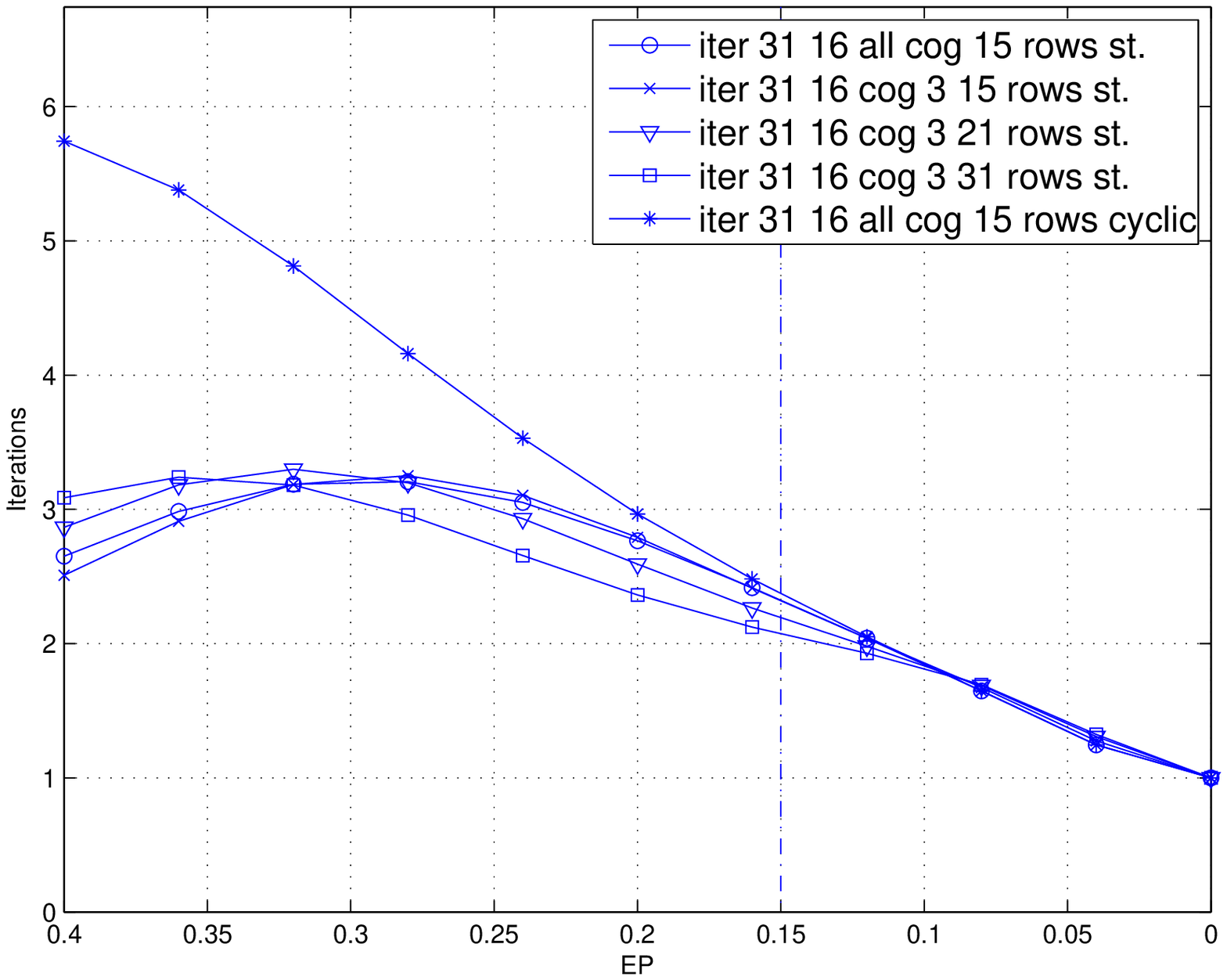}
}
\subfigure[]{\label{fig:ber_bch_31_16_cog_3_and_cog_6}
\psfrag{BER}[cb][cb]{$\BER\,\rightarrow$}
\psfrag{EP}[ct][ct]{$\leftarrow\,\EP$} \psfrag{group 3 15
rows__________}[l][l][1]{\scriptsize{${\mathrm{cog}}_{31,16,{\mathrm{A}}}$,
$m=15$, BP}} \psfrag{group 3 18
rows}[l][l][1]{\scriptsize{${\mathrm{cog}}_{31,16,{\mathrm{A}}}$,
$m=18$, BP}} \psfrag{group 3 19
rows}[l][l][1]{\scriptsize{${\mathrm{cog}}_{31,16,{\mathrm{A}}}$,
$m=19$, BP}} \psfrag{group 3 21
rows}[l][l][1]{\scriptsize{${\mathrm{cog}}_{31,16,{\mathrm{A}}}$,
$m=21$, BP}} \psfrag{group 6 15
rows}[l][l][1]{\scriptsize{${\mathrm{cog}}_{31,16,{\mathrm{C}}}$,
$m=15$, BP}} \psfrag{group 6 21
rows}[l][l][1]{\scriptsize{${\mathrm{cog}}_{31,16,{\mathrm{C}}}$,
$m=21$, BP}} \psfrag{ML estimation}[l][l][1]{\scriptsize{ML
estimation}}
\includegraphics[scale=0.48]{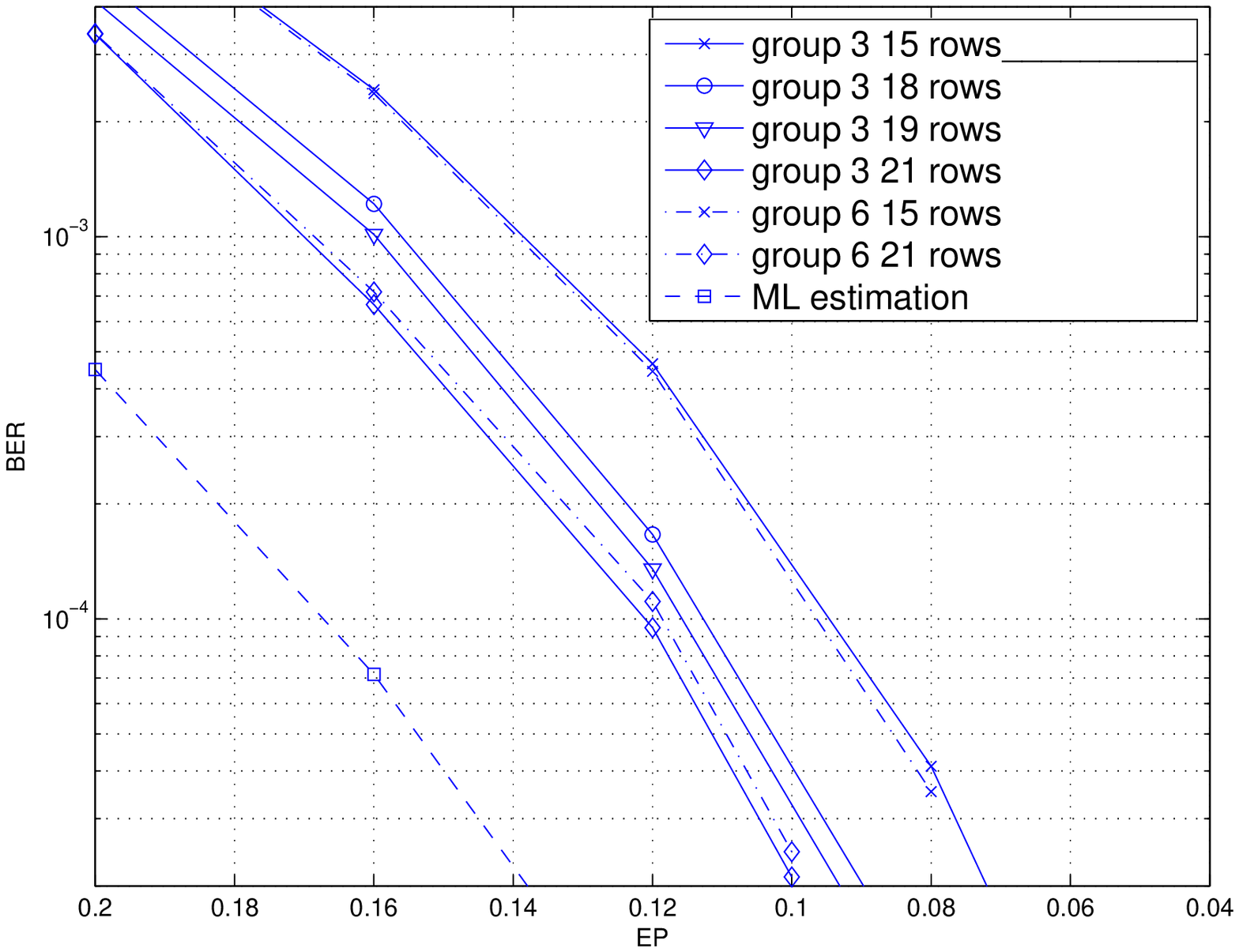}
}
\end{center}
\end{minipage}
\caption{\label{bch_31_16_all_cog_standard_and_cyclic_and_ML}
Performance of the $[31,16,7]$-BCH code.}
\end{figure}

Figures~\ref{ber_bch_31_16_all_cog_standard_and_cyclic_and_ML},
~\ref{fer_bch_31_16_all_cog_standard_and_cyclic_and_ML},
and~\ref{iter_bch_31_16_all_cog_standard_and_cyclic_and_ML} show
the performance of several classes of iterative decoders operating
on the $[31,16,7]$ BCH code. The $\AGDA$ decoder uses a
parity-check matrix that consists of all $15$ cogs of the code,
and for comparison, standard iterative decoding is performed on
the same matrix. In addition, the performance of standard
iterative decoders on three different cyclic parity-check matrices
with $m=n-k=15$, $m=21$, and $m=31$ rows is plotted as well. The
generating cog of these matrices is
${\mathrm{cog}}_{31,16,{\mathrm{A}}}$.

Performance results for the BCH code under consideration are presented
in Figure~\ref{fig:ber_bch_31_16_cog_3_and_cog_6}, pertaining to the
cyclic parity-check matrices based on $\cog_{31,16,\A}$ and
$\cog_{31,16,\C}$ in Table~\ref{table:stopp_dist_hierarchy_31_16}. In
all simulations, the number of rows in the parity-check matrix equals
the smallest listed value in the table. By considering only the
representation indexed by $\cog_{31,16,\A}$, it can be observed that the most
significant BER improvement is achieved when the matrix has $m=18$,
rather than $m=15$ rows. The results also show that for short block
codes, like the ones considered in this example, almost all cogs give
rise to similar residual bit error rates. Nevertheless, for codes of
longer length and larger co-dimension, the particular choice of the
cog may have a significant bearing on the stopping set
characteristics.

\begin{figure}
\begin{center}
\psfrag{EP}[ct][ct][1]{$\leftarrow\,\EP$}
\psfrag{Iterations}[cb][cb][1]{Number of iterations\,$\rightarrow$} \psfrag{iter 127 113 cog 14 14 rows
st.}[l][l][1]{\scriptsize{${\mathrm{cog}}_{127,113,{\mathrm{A}}}$,
$m=14$, BP}} \psfrag{iter 127 113 different cog 14 rows
st.}[l][l][1]{\scriptsize{Various cogs, $m=14$, BP}} \psfrag{iter
127 113 cog 14 34 rows
st.}[l][l][1]{\scriptsize{${\mathrm{cog}}_{127,113,{\mathrm{A}}}$,
$m=34$, BP}} \psfrag{iter 127 113 different cog 14 rows
cyclic}[l][l][1]{\scriptsize{Var. cogs, $m=14$, $\AGDA$}}
\includegraphics[scale=0.48]{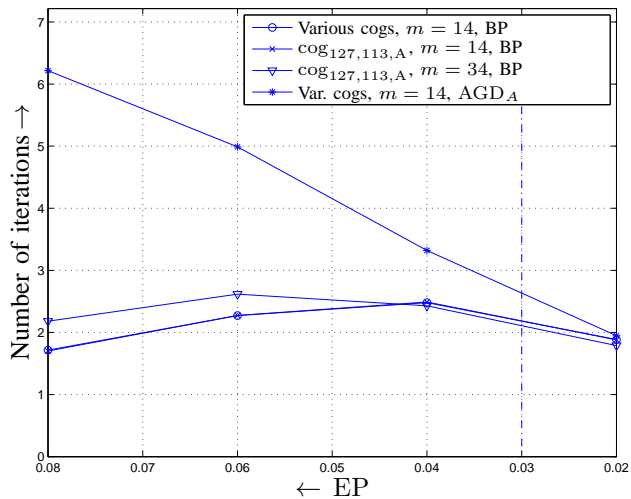}
\caption{\label{iter_bch_127_113_different_cog_standard_and_ML_bound} Number of iterations of the
automorphism group decoder for the $[127,113,5]$ BCH code.}
\end{center}
\end{figure}

In all the examples investigated, the improved performance of
automorphism group decoders comes at the cost of increased
decoding complexity, when compared to edge-removal techniques.
Clearly, to correct more erasures than standard iterative
decoders, automorphism group decoders have to go through
additional decoding iterations, and in addition, perform a certain
number of permutation operations. A measure of computational
complexity of automorphism group decoders is plotted in
Figure~\ref{iter_bch_127_113_different_cog_standard_and_ML_bound}.
Here, the number of excess iterations (when compared to
edge-removal decoding operating on the same channel output) is
plotted versus the $\EP$ of the BEC. As one can see, at most four
additional iterations suffice to resolve the decoding ambiguities
of most error patterns that are uncorrectable by belief-propagation techniques.

The results in
Figures~\ref{ber_bch_31_16_all_cog_standard_and_cyclic_and_ML}
and~\ref{fer_bch_31_16_all_cog_standard_and_cyclic_and_ML} show
that there exists a significant gap between the performance of
automorphism group decoders and ML decoders when considering FERs,
rather than BERs.

\begin{figure}
\begin{center}
\psfrag{EP}[ct][ct][1]{$\leftarrow\,\EP$}
\psfrag{FER}[cb][cb][1]{$\FER\,\rightarrow$} \psfrag{fer 31 16 all
cog 15 rows st.}[l][l][1]{\scriptsize{Various cogs, $m=15$, BP}}
\psfrag{fer 31 16 all cog 15 rows cyclic no
sg}[l][l][1]{\scriptsize{Var. cogs, $m=15$, $\AGDA$}} \psfrag{fer
31 16 all cog 15 rows cyclic sg}[l][l][1]{\scriptsize{Var. cogs,
$m=15$, $\AGDA$, sg}} \psfrag{fer 31 16 ML}[l][l]{\scriptsize{ML
estimation}}
\includegraphics[scale=0.48]{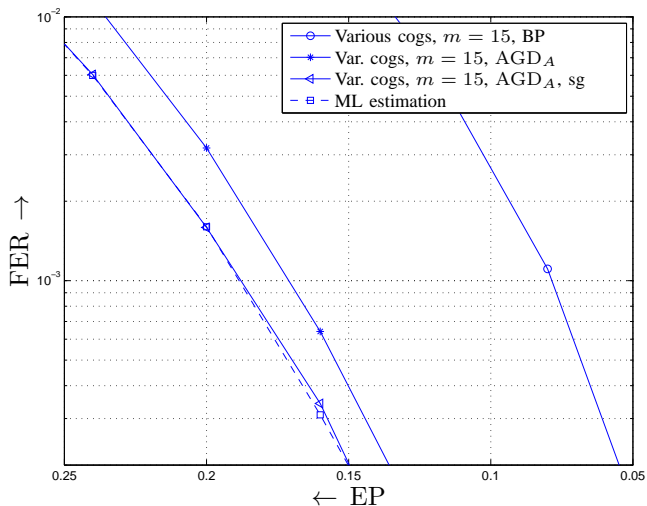}
\caption{\label{comparison_no_sg_and_sg} Performance
improvements achievable by the guessing method.}
\end{center}
\end{figure}

To mitigate this problem, we propose to include an additional
feature into the edge-removal and automorphism group decoding
algorithms. The feature in question is \emph{guessing}: when the
decoder fails to correct a given erasure, it continues decoding by
guessing the value of one of the erased bits. The idea of
incorporating \emph{guessing} methods into belief propagation
algorithms was first described in~\cite{niketal04} and, in the
context of stopping set analysis in unpublished work by Han and
Siegel~\cite{milenkovic06a}. Figure \ref{comparison_no_sg_and_sg}
shows the FER performance improvement for the $[31,16,7]$ BCH code
achieved through combined automorphism group decoding and bit
guessing. As it can be observed, the $\FER$ performance of $\AGDA$
decoders with the guessing feature approaches the performance of
$\ML$ decoding.

\section{Conclusions}
\label{sec:conclusions}

We introduced the notion of the \emph{stopping redundancy
hierarchy} of linear block codes, and derived lower and upper
bounds on the elements of this ordered list. We also investigated
the stopping redundancy of cyclic-parity check matrices, and
proposed new redundant parity-check matrix constructions and
automorphism group decoding techniques. In this setting, we
introduced the notion of a stopping set automorphism group
decoding set, and related this new code invariant to the stopping
redundancy hierarchy.

\begin{appendices}

\renewcommand{\theequation}{\thesection.\arabic{equation}}
\renewcommand{\thetheorem}{\thesection.\arabic{theorem}}

\section{Proof of Theorem~\ref{improve}}\label{app:proof_improve}

We start the proof by introducing the notion of an orthogonal
array~\cite{milenkovicetal06}.

\begin{definition} (\cite[p.5]{hedayatetal99}) \label{co-def}
An orthogonal array of strength $t$ is a matrix of $m \times n$
elements with the property that every $m \times t$ subarray
contains \emph{each} possible $t$-tuple \emph{the same number of
times}. The codewords of an $[n,k,d]$ code $\mathcal{C}$ form an
orthogonal array of dimension $2^{k}\times n$ and strength
$d^{\perp}-1$, where, as before, $d^{\perp}$ denotes the dual
distance of $\mathcal{C}$.
\end{definition}
It can be shown that the set of all codewords of a code forms an
orthogonal array of strength $d^{\perp}-1$~\cite{hedayatetal99}.
\begin{definition} Let $\mathcal{C}$ be an $[n,k,d]$ code
and $\mathcal{C}^\bot$ its dual code. The ensemble
$\mathcal{M}_{\mathcal{C}}(m)$ is defined as the set of all
$m\times n$ matrices with rows chosen uniformly, independently,
and with replacement, from the set of $2^{n-k}$ codewords of
$\mathcal{C}^\bot$.
\end{definition}
Let $\mathcal{A}$ be the set of all binary $t$-tuples of weight
one. Let $E_i$ be the event that the $i$-th collection of $t$
columns in an array randomly drawn from the
$\mathcal{M}_{\mathcal{C}}(m)$ ensemble contains no element from
$\mathcal{A}$, i.e.\ that the $i$-th collection of columns
corresponds to a stopping set of size $t$. If $\{{S(t)=0\}}$
denotes the event that a given array does not contain stopping
sets of size $t$, then
$P\{{\bigcap\,\overline{E}_i\}}=P\{{S(t)=0\}}$. Observe next that
\begin{equation}
\begin{split} \label{eq:E_i}
P\{{E_i\}}&=P\,\{\;\bigcap_{k=1}^m\,\{{k^{\text{th}}\;\text{row of
the}\;i^{\text{th}}\; \text{restriction}\; \notin
\mathcal{A}\}}\;\} \\
&=\left(\,P \,\{\,1^{\text{st}}\;\text{row of
the}\;i^{\text{th}}\;
\text{restriction}\; \notin \mathcal{A}\,\} \right)^m\\
&=\left(1-t\cdot 2^{-t}\right)^m,
\end{split}
\end{equation}
where the last step follows from the fact that the occurrence of
every $t$-tuple in the orthogonal array is equally likely, with
probability $2^{-t}$. The dependence number $\tau$ of the events
$E_i$, defined in Equation~\eqref{lovasz_local_lemma}, equals
\begin{equation}
(\tau+1) = \binom{n}{t} - \binom{n-t}{t}.
\label{eq:dep-no}
\end{equation}
The above result is a consequence of the fact that two collections
of $t$ columns are co-dependent if and only if they share at least one
column. If one collection of $t$ columns is fixed, another
collection is independent of it if its columns are chosen from the
remaining $n-t$ columns.

By inserting the expressions~\eqref{eq:E_i} and~\eqref{eq:dep-no}
into the defining inequality~\eqref{eq:lll_tau_ineq} of LLL gives the
following sufficient condition for the existence of a parity-check
matrix with $m$ rows, free of stopping sets of size $t$:
\begin{equation}
{\mathrm{e}}\; \left(1-t\cdot 2^{-t}\right)^m\;\left(\binom{n}{t} -
\binom{n-t}{t}\right)\, \leq 1.
\notag 
\end{equation}
This implies that, if $m$ satisfies the condition above, there
exists at least one parity-check matrix of dimension $m$ free of
stopping sets of size $t$. Consequently, this bound on $m$
represents an upper bound on $\rho_{\ell}(\mathcal{C})$.

A sufficient condition for the existence of a matrix with $m$ rows
that does not contain stopping sets of size $j \in
\{{1,2,\ldots,t\}}$, is
\begin{equation}
\sum\limits_{j=1}^{t} {\mathrm{e}}\; \left(1-j\cdot
2^{-j}\right)^m \;\left(\binom{n}{j} - \binom{n-j}{j}\right)\,
\leq 1. \notag
\end{equation}
By observing that
\begin{equation}
\max\limits_{1 \leq j \leq t}   \left(1-j\cdot 2^{-j}\right)^m =
\left(1-t\cdot 2^{-t}\right)^m,
\notag 
\end{equation}
we arrive at
\begin{equation}\label{eq:ext-condition}
{\mathrm{e}}\; \left(1-t\cdot 2^{-t}\right)^m \;\sum\limits_{j=1}^{t}
\left(\binom{n}{j} - \binom{n-j}{j}\right)\, \leq 1.
\end{equation}
Note that Equation~\eqref{eq:ext-condition} can be rewritten as
\begin{equation}
m \geq
\frac{1+\log\sum\limits_{j=1}^{t} \left(\binom{n}{j} - \binom{n-j}{j}\right)}
{-\log\left(1-\frac{t}{2^{t}}\right)}. \notag
\end{equation}
In order to ensure that the chosen matrix has full row-rank, it
may be necessary to add $n-k-\ell+1$ additional rows.

Observe that any collection of at most $d-1$ columns of a
parity-check matrix is linearly independent. This follows from the
fact that the minimum distance of the code is $d$. As a
consequence, in order for two collections of $t$ columns to be
independent, one must have $2\cdot t < d$. This restricts the
validity of the above bound to stopping distances $\ell \leq
\lfloor \frac{d+1}{2}\rfloor$ only. Setting $t=\ell-1$ completes
the proof.

\section{Asymptotic Behavior of the Bound in Theorem~\ref{improve}}\label{app:asymptotic}

We investigate the asymptotic behavior of the bound in Theorem~\ref{improve}. Let $m'$ be the first summand of
Eq.~\eqref{eq:improve}, i.e.\
\begin{equation}
m' = \frac{1+\log\sum\limits_{j=1}^{\ell-1} \left(\binom{n}{j} -
\binom{n-j}{j}\right)}
{-\log\left(1-\frac{\ell-1}{2^{\ell-1}}\right)}.
\end{equation}
Let us consider the case $\ell-1=\lambda\,n,$ for $0<\lambda<1$.
In this case, the logarithm in the numerator grows at most
linearly with $n$, since
\begin{equation}
\sum_{j=1}^{\ell-1} \binom{n}{j} < 2^n,
\end{equation}
and the second sum under the logarithm never exceeds the first.

As a result, the asymptotic behavior of the given upper bound is
dominated by the expression in the denominator, which, for large
$n$, takes the form
\begin{equation}
- \log\left(1-\frac{\ell-1}{2^{\ell-1}}\right) \simeq
\frac{\lambda\,n}{2^{\lambda\,n}},\notag
\end{equation}
and therefore
\begin{equation}
m'\simeq \frac{(1+n)\,2^{\lambda\,n}}{\lambda\,n}\simeq
\frac{2^{\lambda\,n}}{\lambda}. \notag
\end{equation}
Consequently, for codes with minimum distance $d=\mathrm{const.}\cdot
n$, the upper bound on $m$ is exponential in $n$.

\section{Proof of Theorem~\ref{highimp}}\label{app:proof_highimp}
Theorem~\ref{highimp} can be proved using arguments similar to
those described in the proof of Theorem~\ref{improve}. First,
observe that the number of distinct collections of columns, and
consequently, the number of events $E_i$ used in LLL equals
\begin{equation}
N=\sum\limits_{j=1}^{t}\binom{n}{j}.
\notag
\end{equation}
Since the dependency number for the high probability variation of
Lov{\'a}sz Local Lemma is the same as the one given in
Theorem~\ref{improve}, it follows that
\begin{equation}
\left(1-t\cdot 2^{-t}\right)^m \leq
\frac{\epsilon}{\sum_{j=1}^{t}\binom{n}{j}}\left(1-\frac{\epsilon}
{\sum_{j=1}^{t}\binom{n}{j}}\right)^{\sum\limits_{j=1}^{t}
\left[\binom{n}{j}
- \binom{n-j}{j} -1\right]}.
\notag
\end{equation}
After some simple algebraic manipulation, one can show that the
above expression gives rise to the following bound:
\begin{equation}
m \geq \frac{\log\frac{\epsilon}{\sum_{j=1}^{t}\binom{n}{j}} +
\left({\sum\limits_{j=1}^{t} \left[\binom{n}{j} - \binom{n-j}{j}
-1\right]}\right) \cdot \log
\left(1-\frac{\epsilon}{\sum_{j=1}^{t}\binom{n}{j}}\right)}
{\log{\left(1-\frac{t}{2^{t}}\right)}}.
\notag
\end{equation}
As a result, for a given set of code parameters, with probability
greater than $1-\epsilon$ \emph{every} parity-check matrix with
$m$ rows has stopping distance at least $t+1$. Substituting
$t=\ell-1$ completes the proof.

\section{Cyclic Parity-Check Matrices of QR Codes}\label{sec:app_qr_codes}

Let $\mathcal{Q}$ be a QR code of prime length $n=3\,(\mymod 4)$,
and let $N$ and $Q$ denote the set of quadratic non-residues and
residues of the underlying finite field, respectively. The
idempotent of the code is the polynomial $\sum_{i \leq n} f_{i}
x^{i}$, where $f_i=1$ if $i \in N \cup \{ 0\}$, and is zero
otherwise. The idempotent can be used as the first row of a
(redundant) cyclic parity-check matrix of $\mathcal{Q}$.

In what follows, we assume for two fixed indices $i$ and $j$ that
$\kappa = i - j\, (\mymod n)$. For a prime $n=4t-1$, $|Q|=2t-1$,
and for each integer $\kappa \in Q$, $\kappa \neq 0 \,(\mymod n)$,
there exist exactly $t-1$ ordered pairs $(i,j)$, $i \neq j,$ such
that $i,j\, \in Q$. Similarly, for fixed $\kappa$ and $i \in Q$,
there exist $t-1$ distinct values $j \in N$ that result in the
given value of $\kappa$.

The lower bound of Example~\ref{example:qr_codes} can be derived
as follows. First, observe that the number of overlapping zeros
between the first row and its $\kappa$-th cyclic shift,
$|ZZ_{\kappa}|$, equals
\begin{equation}
|ZZ_{\kappa}| = \sum_{i=0}^{n-1} (1-f_{i})(1-f_{i-\kappa}). \notag
\end{equation}

Let us now determine the number of  pairs $(i,j)$, $i, j \in Q$,
for which $\kappa \in N$. As already pointed out, for each $\kappa
\in Q$, there are $t-1$ choices for pairs $i, j \in Q$, and $t-1$
choices for $i \in Q$, $j \in N$. The $(2t-1)^{2}$ pairs $i,j \in
Q$ arise from $(2t-1)(t-1)$ different values of $\kappa$, and
there are $(2t-1)$ choices for $\kappa=0$ that lead to $i \in Q$.
In addition, $(2t-1)(t-1)$ values of $\kappa$ result in $i,j \in
Q$, $\kappa \in N$, while for $t-1$ values of $\kappa \in N$ one
has $i,j \in Q$. Therefore, the number of pairs $(i,j)$ such that
$i,j \in Q$ equals $t-1$ for $\kappa \in Q$, as well as for
$\kappa \in N$. As there are $t-1$ quadratic residues, for $\kappa
\in Q$ as well as for $\kappa \in N$, it follows that
$|ZZ_{\kappa}|=t-1$ (excluding $\kappa=0$).

The row-weight of the cyclic parity-check matrix equals
$\omega=|N|+1=2t$, where $\omega=|OO_{\kappa}| +
|OZ_{\kappa}|=|OO_{\kappa}| + |ZO_{\kappa}|$. Therefore,
$|OZ_{\kappa}|=|ZO_{\kappa}|$. Furthermore, one has $|OO_{\kappa}|
+ |OZ_{\kappa}|+|ZO_{\kappa}| + |ZZ_{\kappa}|=n$, and the same
result can be deduced from the fact that
$$
|OZ_{\kappa}|=\sum_{i=0}^{n-1} f_{i} (1-f_{i-\kappa}) =
\sum_{i=0}^{n-1} f_{i} - \sum_{i=0}^{p-1} f_{i}f_{i-\kappa} = 2t -
|OO_{\kappa}|.
$$

One can now derive an expression for $|OO_{\kappa}|$ as follows:
\begin{eqnarray}
|OO_{\kappa}| &=& n- |ZZ_{\kappa}| - |OZ_{\kappa}| - |ZO_{\kappa}|
=
n- (t-1) - 2\cdot |OZ_{\kappa}| = n - t+1-2\cdot (2t-|OO_{\kappa}|) \notag\\
       &=& n-t+1-4t+2\cdot |OO_{\kappa}|, \; \text{i.e.} \notag \\
|OO_{\kappa}| &=& 5t-1-n = 5t-1-(4t-1)=t,
\notag
\end{eqnarray}
so that
$$
|OZ_{\kappa}|=|ZO_{\kappa}|=2t-|OO_{\kappa}|=2t-(t-2)=t.
$$
Since the intersection numbers do not depend on the value of
$\kappa$ and on this number being a quadratic residue or
non-residue, we henceforth omit the subscript $\kappa$. Also, note
that we do not make use of the intersection numbers for which
$\kappa=0$, since we are interested in counting the number of
stopping sets resolved by distinct rows of a (redundant)
parity-check matrix.

Inserting the expressions above into the formula
of~Lemma~\ref{lemma:num_stopsets_cyclic_pair_of_rows} shows
$$|\Sigma_{\sigma, \mu} \cap \Sigma_{\sigma, \mu+\kappa}|
= |OO| \cdot \binom{|ZZ|}{\sigma-1} + |OZ|\cdot |ZO| \cdot
\binom{|ZZ|}{\sigma-2} = \frac{n+1}{4}\cdot
\binom{\frac{n-3}{4}}{\sigma-1} + \left(\frac{n+1}{4}\right)^{2}
\binom{\frac{n-3}{4}}{\sigma-2},$$ so that
$$S_{\sigma,2}=\sum_{\mu=1}^{m-1} \sum_{\kappa=1}^{m-\mu}|\Sigma_{\sigma, \mu}\cap
\Sigma_{\sigma, \mu+\kappa}| = \binom{m}{2} \cdot
\frac{n+1}{4}\cdot \left( \binom{\frac{n-3}{4}}{\sigma-1} +
\frac{n+1}{4} \binom{\frac{n-3}{4}}{\sigma-2}\right).$$

It is also straightforward to show that
\begin{equation}
\begin{split}
m \; |\Sigma_{\sigma,1}| &- \frac{2}{m}\cdot
\sum\limits_{\kappa=1}^{m-1} (m-\kappa) | \Sigma_{\sigma,1} \cap
\Sigma_{\sigma, ({(1+\kappa)}\modstar m)} | = \\
 m \cdot
\frac{n+1}{2}\cdot \binom{\frac{n-1}{2}}{\sigma-1} - &(m-1) \cdot
\frac{n+1}{4}\cdot \left( \binom{\frac{n-3}{4}}{\sigma-1} +
\frac{(n+1)}{4} \binom{\frac{n-3}{4}}{\sigma-2} \right). \notag
\end{split}
\notag
\end{equation}

From the condition $|\bigcup_{i=1}^m \Sigma_{\sigma, i}| \geq
\binom{n}{\sigma}$ and $\sigma=\ell-1$, we obtain a lower bound
for the number of rows $\mu_{\ell}$ that a parity-check matrix
consisting of cyclic shifts of the idempotent must have in order
to have stopping distance at least $\ell$:
\begin{equation}\label{eq:m_ell_lower_bound}
\mu_{\ell} \geq \max_{\sigma < \ell } \,\left \lceil
\frac{\binom{n}{\sigma} - M}
{\frac{n+1}{2}\binom{\frac{n-1}{2}}{\sigma-1} - M} \right \rceil.
\end{equation}
In the above equation, we used
$$
M=\frac{n+1}{4}\cdot  \left( \binom{\frac{n-3}{4}}{\sigma-1} +
\frac{(n+1)}{4} \binom{\frac{n-3}{4}}{\sigma-2} \right). \notag
$$

Note that the set of quadratic residues also forms a cyclic
difference set with parameters
$(n,k,\lambda)=(n,\frac{n-1}{2},\frac{n-3}{4})$ ~\cite{tarakanov},
so that the results of Section IV can be used in this
context as well.\\

It is also of interest to find an upper bound on the stopping
distance $\ell$ of a redundant $n\times n$ parity-check matrix
that consists of \emph{all} cyclic shifts of the idempotent of a
QR code. For this purpose, we use
Equation~\eqref{eq:m_ell_lower_bound} with $m=n$, and ask for the
largest value of $\sigma$ for which
\begin{equation}
\binom{n}{l} \leq  n \cdot \frac{n+1}{2}\cdot
\binom{\frac{n-1}{2}}{l-1} - (n-1) \cdot \frac{n+1}{4}\cdot \left(
\binom{\frac{n-3}{4}}{l-1} + \frac{(n+1)}{4}
\binom{\frac{n-3}{4}}{l-2} \right),\; 1 \leq l \leq \sigma.
\notag 
\end{equation}
We point out that although the value $\sigma+1$ serves as an upper
bound on the stopping distance of the parity-check matrix under
consideration, there is no guarantee that the matrix itself has a
stopping distance that meets this bound.

To this end, we only seek a solution for $l=\sigma$, since it can
be shown that this value of the parameter imposes the tightest
restriction on the inequality.


Also, for both $n$ and $\sigma$
sufficiently large (where $\sigma=o(n)$), we can approximate the above
inequality by

\begin{equation} \label{eq:cond-simple}
\binom{n}{\sigma} \lesssim n\,\sigma\,
\binom{n/2}{\sigma}-n\,\sigma^2 \binom{n/4}{\sigma},
\end{equation}

since
\begin{equation}
\frac{n^2}{2} \binom{n/2}{\sigma-1} \simeq n\, \sigma\, \binom{n/2}{\sigma},
\notag
\end{equation}

and

\begin{equation}
\frac{n^2}{4} \left( \binom{n/4}{\sigma-1} +\frac{n}{4}\,
\binom{n/4}{\sigma-2}\right)
\simeq n\, \sigma^2\, \binom{n/4}{\sigma}. \notag
\end{equation}

As the family of QR codes contains an infinite number of codes, we
provide next an asymptotic analysis that allows us to upper bound
the achievable stopping distance of redundant parity-check
matrices based on the idempotent of the code.

To find an asymptotic lower bound for the stopping redundancy of
cyclic parity-check matrices, we use the bound~\cite[p. 309, Eq.
16]{macwilliamsetal77}
\begin{equation}
f_{1}(\beta,n)\cdot \text{e}^{n\cdot H(\beta)} \leq
\binom{n}{\beta n} \leq f_{2}(\beta,n) \text{e}^{n\cdot H(\beta)},
\notag 
\end{equation}
where $\beta\,n$ is an integer, $H(\beta)=-\beta \ln \beta -
(1-\beta)\ln (1-\beta)$ stands for Shannon's entropy in natural
units, $f_{1}(\beta,n)=\frac{1}{\sqrt{8 n\beta (1-\beta)}}$, and
$f_{2}(\beta,n)=\frac{1}{\sqrt{2\pi n\beta (1-\beta)}}$.

For $\beta \to 0$, the following asymptotic formula holds for
$H(\beta)$:
\begin{eqnarray*}
H(\beta)&=&-\beta \ln \beta - (1-\beta)\ln (1-\beta)\\
&=& -\beta \ln \beta -\beta + \beta^2/2+ O(\beta^3).
\end{eqnarray*}

In the same asymptotic domain, $f_{1}(\beta,n)$ and
$f_{2}(\beta,n)$ take the form
$$
f_{1}(\beta,n)=\frac{1}{\sqrt{8 n\beta (1-\beta)}} =
\frac{1}{\sqrt{8\,n\,\beta}}\;\left(1+O(\beta)\right),
$$
and
$$
f_{2}(\beta,n)=\frac{1}{\sqrt{2\pi n\beta (1-\beta)}} =
\frac{1}{\sqrt{2\,\pi\, n\,\beta}}\;\left(1+O(\beta)\right) .
$$
Denote the lower bound for the left hand side of
Equation~\eqref{eq:cond-simple} by $L(n,\sigma)$. Similarly, let
the upper bound for the two terms on the right hand side of the
same equation be denoted by $R_1(n,\sigma)$ and $R_2(n,\sigma)$,
respectively.

Invoking the asymptotic expressions described above shows that
\begin{equation}\label{eq:l-bound}
\binom{n}{\sigma} \gtrsim
\frac{n^{\sigma}}{\sqrt{8\,\sigma^{2\,\sigma+1}}}\;\;
\text{e}^{-\sigma+\sigma^2/(2\,n)+O(\sigma^3/n^2)}\;
\left(1+O(\sigma/n)\right)=L(n,\sigma), \notag
\end{equation}
and that
\begin{equation}
n\, \sigma\,\binom{n/2}{\sigma} \lesssim
n^{2\,\sigma+1}\, \sigma \;\frac{1}
{\sqrt{4\,\pi\,2^{4\,\sigma}\,\sigma^{4\,\sigma+1}}}\;\;
\text{e}^{-2\,\sigma+2\,\sigma^2/n+O(\sigma^3/n^2)}\;
\left(1+O(\sigma/n)\right)=R_1(n,\sigma), \notag
\end{equation}
\begin{equation}
-n\, \sigma^2\,\binom{n/4}{\sigma} \lesssim
-n^{4\,\sigma+1}\, \sigma^2\,
\frac{1}{\sqrt{32\;4^{8\,\sigma}\,\sigma^{8\,\sigma+1}}}\;\;
\text{e}^{-4\,\sigma+8\,\sigma^2/n+O(\sigma^3/n^2)}\;\left(1+O(\sigma/n)\right)=R_2(n,\sigma).
\notag
\end{equation}

Tedious, but straightforward algebraic manipulation reveals that
the above inequalities imply
\begin{equation}
1 \lesssim
\frac{\sqrt{2}\,\sigma\,n^{\sigma+1}}{\sqrt{\pi\,2^{4\sigma}\,\sigma^{2\,\sigma}}}\,
\text{e}^{(-\sigma+3\sigma^2/2n)} -
\frac{n^{3\sigma+1}\,
\sigma^2}{2\,\sqrt{4^{8\sigma}\,\sigma^{6\,\sigma}}}\,\text{e}^{(-3\sigma+15\sigma^2/2n)}.
\notag
\end{equation}

The \emph{square-root bound} for the minimum distance of QR codes
asserts that their minimum distance is lower bounded by
$\sqrt{n}$. Although the minimum distance of many quadratic
residue codes exceeds this bound, little is known about the
asymptotic behavior of the exact minimum distance. We henceforth
define the \emph{designed minimum} distance, which equals
$\sqrt{n}$, and proceed to evaluate the right hand side of the
expression in the above equation for $\sigma =\sqrt{n}$. In this
case, the resulting formula equals
\begin{equation}
c_1\,
n^{\sqrt{n}+3/2}\;\text{e}^{-\sqrt{n}-2\sqrt{n}\,\ln\,2}-c_2\,
n^{3\sqrt{n}+2}\;\text{e}^{-3\sqrt{n}-4\sqrt{n}\,\ln\,4},
\notag
\end{equation}
for some positive constants $c_1$ and $c_2$. For $n \to \infty$,
this expression diverges to $-\infty$. As a consequence, the
stopping distance of redundant cyclic parity-check matrices
generated by the idempotent of a QR code cannot meet the
square-root bound. Furthermore, no positive fraction of this bound
is attainable by a matrix of this form.

This finding implies that, rather than using idempotents as
generators, cogs should be used instead, since they have
significantly lower density.

\section{Construction of Redundant Parity-Check Matrices: Double-Error Correcting BCH Codes}
\label{app:BCH_construction_scheme}

We describe the generalized Hollmann-Tolhuizen (HT) method for
generating redundant parity-check matrices of double-error correcting
BCH codes. The procedure consists of four steps. In the first step,
the algorithm outlined in~\cite{hollmannetal07} is used to create a
redundant parity-check matrix for the Hamming supercode. This matrix
is used as the first sub-block in the parity-check matrix of the BCH
code.  In the second step, a second sub-block of rows is added,
consisting of codewords of the dual of the BCH code that represent a
basis of the code. The third step consist of adding redundant rows
that resolve stopping sets corresponding to codewords of weight three
and a subclass of codewords of weight four in the Hamming code. The
fourth step introduces into the parity-check matrix a set of redundant
rows obtained by a greedy search strategy, designed to completely
resolve stopping sets of size four.

The generalized HT algorithm is summarized below.

\begin{description}
\item[{\bfseries{Step 1}})]{\hspace{0.1in} A redundant
parity-check matrix for a Hamming code that resolves all stopping
sets that correspond to correctable erasure patterns of size up to
(but not including) $\sigma=4$ is created first. To this end, a
generic $(\bar{m},\bar{\sigma})$ erasure correcting set with
$\bar{m}=\log_2(n+1)$ and $\bar{\sigma}=4$ is multiplied with the
standard parity-check matrix of the Hamming code. The resulting
matrix contains only unresolved stopping sets of size $\sigma=3$
and $\sigma=4$, which either correspond to a codeword of the
Hamming code or contain the support of one such codeword in their
support. We refer to this matrix as a generic erasure correcting
matrix, and denote it by $\ve{H}_{\Hamminggeneralized}$.
Figure~\ref{fig:vis_structured_approach_1} visualizes the first
step of the construction procedure.}

\item[{\bfseries{Step 2}})]{\hspace{0.1in} The matrix obtained in
Step 1 is augmented to form a parity-check matrix of a BCH-code
with minimum distance $d=5$. This is achieved by adding the binary
expansion of the row vector
\begin{equation}
\ve{h}_{\mathrm{BCH}}=\left(\begin{array}{cccc}\alpha^{0\cdot
(1+c)}&\alpha^{1\cdot(1+c)}&
\dots&\alpha^{(n-1)(1+c)}\end{array}\right), \notag
\end{equation}
where $c$ is chosen according to the standard procedure for
generating BCH codes (one common choice being $c=2$). Here,
$\alpha$ is as defined in Section \ref{subsec:hamming_codes}. We
refer to this second component of the redundant parity-check
matrix of the BCH code as $\ve{H}_{\BCH}$.
Figure~\ref{fig:vis_structured_approach_all_steps} visualizes the
second step of the construction procedure.}

\begin{figure}[h!]
\begin{center}
\psfrag{1}[c][c]{Step $1$}
\psfrag{H1}[c][c]{\scriptsize{$\ve{H}_{\Hammingstandard}$}}
\psfrag{H2}[c][c]{\scriptsize{$\ve{H}_{\Hamminggeneralized}$}}
\includegraphics[scale=0.4]{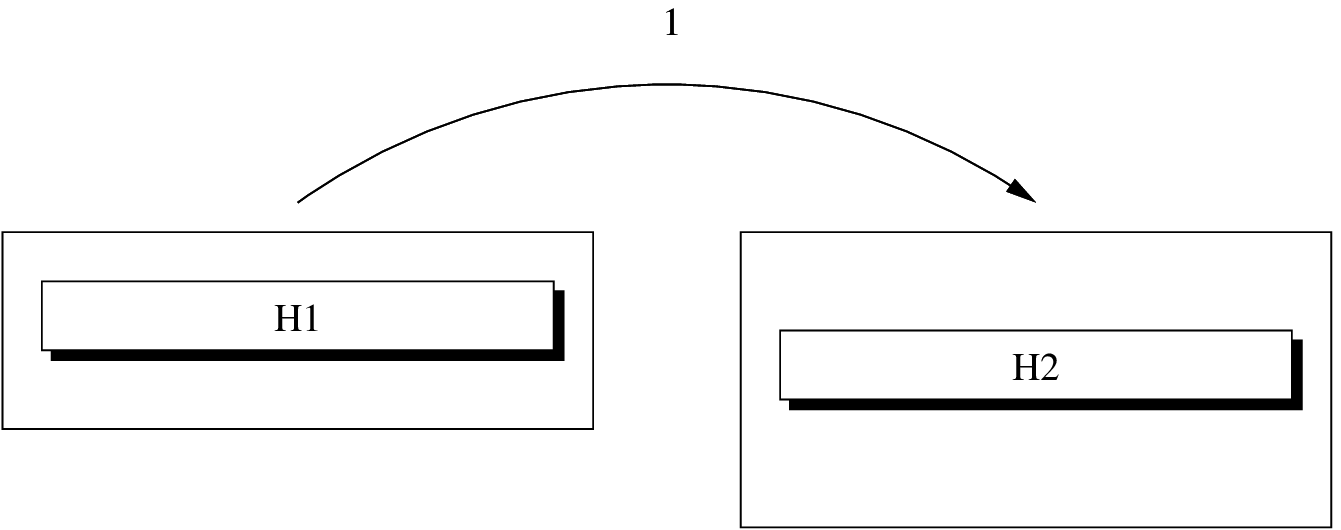}
\caption{\label{fig:vis_structured_approach_1} Step 1: Transform
$\ve{H}_{\Hammingstandard}$ into a generic
$(\bar{m},\bar{\sigma})$ erasure correcting set $\ve{H}_{\Hamminggeneralized}.$}
\end{center}
\end{figure}

\begin{figure}[h!]
\begin{center}
\psfrag{2}[l][l]{Step $2$} \psfrag{3}[l][l]{Step $3$}
\psfrag{4}[l][l]{Step $4$}
\psfrag{H2}[c][c]{\scriptsize{$\ve{H}_{\Hamminggeneralized}$}}
\psfrag{H3}[c][c]{\scriptsize{$\ve{H}_{\BCH}$}}
\includegraphics[scale=0.4]{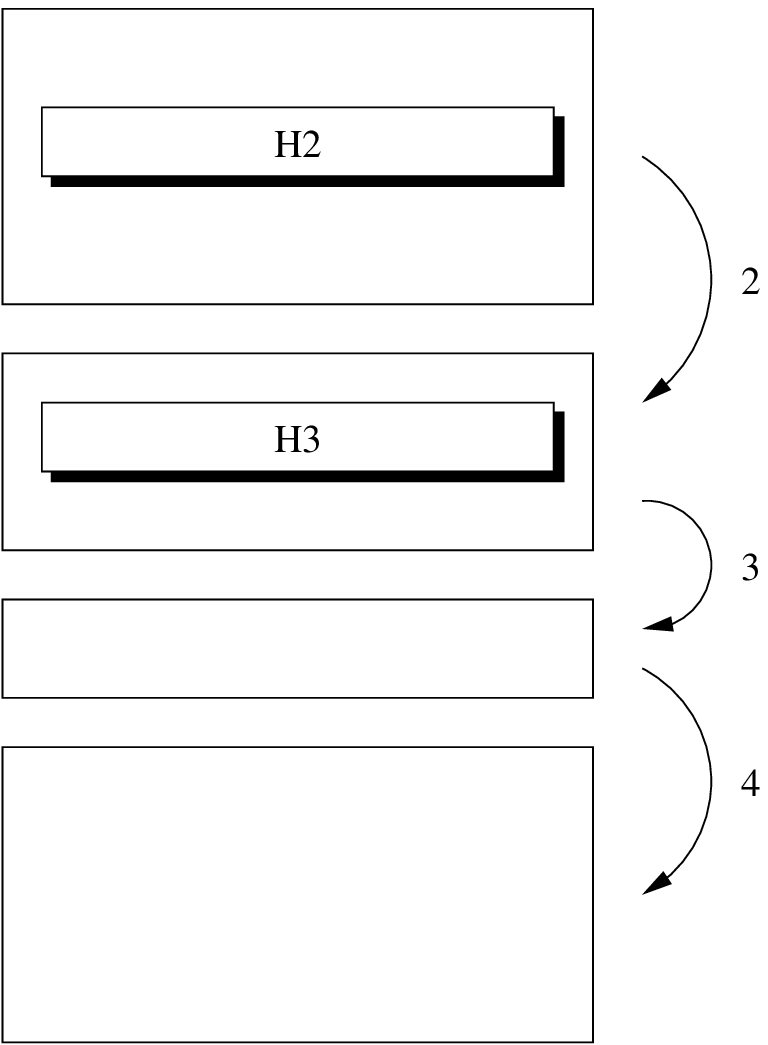}
\caption{\label{fig:vis_structured_approach_all_steps} Steps 2-4:
Append $\ve{H}_{\BCH}$ (Step 2), and add code-defining
parity-check equations to $\ve{H}_{\Hamminggeneralized}$ (Steps 3
and 4).}
\end{center}
\end{figure}

\item[{\bfseries{Step 3}})] {\hspace{0.1in} In this step, an
additional set of redundant rows is added to the matrix in order
to resolve all stopping sets of size $\sigma=3$ and a subset of
those of weight $\sigma=4$. To this end, we use the fact that all
unresolved stopping sets of size three and four correspond to
codewords of the Hamming code.

Let $\ve{t}$ denote the support of an unresolved stopping set of
size $\sigma=3$. The restriction of $\ve{H}_{\Hamminggeneralized}$
to $\ve{t}$ equals
\[ \left[ \left(\alpha^a\right)\,\,
\left(\alpha^b\right)\,\,\left(\alpha^a + \alpha^b\right)\right]
\]
with $0\leq a,b <n-2$, $\alpha^a \neq \alpha^b$, whereas the restriction of
$\ve{H}_{\BCH}$ to $\ve{t}$ equals
\[
\left[\left(\alpha^{3a}\right)\,\, \left(\alpha^{3b}\right)\,\,
\left((\alpha^a + \alpha^b)^3\right)\right],
\]
where we have set $c=2$.

Consider first all stopping sets for which
$\alpha^{3a}=\alpha^{3b}$, but $\alpha^a\not=\alpha^b$.

\begin{lemma} \label{lemma:3aplus3b} In a finite field with
characteristic two and order $n+1$, for any primitive element
$\alpha$, $\alpha^{3a}=\alpha^{3b}$ implies
$(\alpha^a+\alpha^b)^3=\alpha^{3a}$ iff $\alpha^a\not=\alpha^b$.
\end{lemma}

\begin{proof}
If $\alpha^a=\alpha^b$, the lemma does not hold, as
$(\alpha^a+\alpha^b)^3=0$. We henceforth consider the case
$\alpha^a\not=\alpha^b$ only. Since the field has characteristic
two, one can write
\begin{equation}
(\alpha^a+\alpha^b)^3 = \alpha^{3a}+ 3\alpha^{2a+b} +
3\alpha^{a+2b}+ \alpha^{3b} = \alpha^{2a+b}+\alpha^{a+2b}. \notag
\end{equation}
Consequently,
\begin{equation}
\alpha^{2a+b}+\alpha^{a+2b}=\alpha^{3a}\left(\alpha^{b-a}+\alpha^{a-b}\right).
\notag
\end{equation}

As $3a-3b=j\,n$, for some integer $j \geq 0$, it follows that
$a-b=j\,n/3$ and $b-a=2j\,n/3$. In order for
$\alpha^a\not=\alpha^b$ to hold, $j$ has to be restricted to
$j\not=0\,(\mymod 3)$.
For these values of $j$, one has
$\left(\alpha^{b-a}+\alpha^{a-b}\right)=\left(\alpha^{n/3}+\alpha^{2n/3}\right)$.
To complete the proof, we first prove the following more general
claim: for each $u\in\left\{2k+1\mid k\in {\mathbb{N}}\right\}$,
it holds that $\sum_{i=1}^{u-1}\, \alpha^{in/u}=1$. The result of
the lemma then follows by setting $u=3$.

We prove the claimed result by showing that
\begin{equation}
\left(\sum_{i=1}^{u-1}\, \alpha^{in/u}\right)^2=
\sum_{i=1}^{u-1}\alpha^{2in/u}=\sum_{i=1}^{u-1}\alpha^{in/u}.
\notag
\end{equation}
We observe that
\begin{eqnarray}
\sum_{i=1}^{u-1}\alpha^{2in/u}\hspace{-0.25cm}&=&\hspace{-0.25cm}
\sum_{i=1}^{(u-1)/2}\alpha^{2in/u}+\sum_{i=(u+1)/2}^{u-1}\alpha^{2in/u}\notag\\
&=&\hspace{-0.25cm}\sum_{i=1}^{(u-1)/2}\alpha^{2in/u}+
\sum_{i'=1}^{(u-1)/2}\alpha^{2(i'+u/2-1/2)n/u}\notag\\
&=&\hspace{-0.25cm}\sum_{i=1}^{(u-1)/2}\alpha^{2in/u}+
\sum_{i'=1}^{(u-1)/2}\alpha^{(2i'-1)n/u}\notag\\
&=&\hspace{-0.25cm}\sum_{i=1}^{u-1}\alpha^{in/u},\notag
\end{eqnarray}
where we have introduced $i'=i-u/2+1/2$ and used the fact that
$\alpha^n=1$. This completes the proof.
\end{proof}

Assume that $n/3$ is an integer. Then there exist exactly $n/3$
stopping sets corresponding to Hamming codewords of weight three
for which $\alpha^{3a}=\alpha^{3b}$, as the free parameter $a$ can only be chosen in
the range $0<a<n/3$. The variable $b$ needs to be chosen accordingly within $n/3\leq b<n$. If we switch to the binary representation of
the parity-check matrix, it is easy to see that the restriction of
$\ve{H}_{\Hamminggeneralized}$ to $\ve{t}$ only contains rows of
weight zero or two, whereas the same restriction of
$\ve{H}_{\BCH}$ only contains rows of weight zero or three. As a
result, each stopping set corresponding to a weight-three codeword
of the Hamming code with $\alpha^{3a}=\alpha^{3b}$, can be
resolved by an appropriate linear combination of one row from
$\ve{H}_{\Hamminggeneralized}$ and one row of $\ve{H}_{\BCH}$. We
conduct the search for the appropriate rows in these matrices in a
greedy fashion, i.e.\ by eliminating as many stopping sets as
possible with each redundant row.
Figure~\ref{fig:vis_structured_approach_all_steps} visualizes this
step.}

\item[{\bfseries{Step 4}})] { \hspace{0.1in} Step 3 gives rise to
parity-check matrices that resolve all stopping sets of size
$\sigma=3$, for which $\alpha^{3a}=\alpha^{3b}$, but $\alpha^a\not=\alpha^b$, and $n/3$ is an
integer. As an example, this claim is true when $n=63$ and the
corresponding code is a $[63,51,5]$ BCH code. In order to resolve
stopping sets that do not satisfy these constraints, we use greedy
computer search techniques to identify a collection of redundant
parity-checks suitable for accomplishing this goal. These checks
are added to the concatenation of $\ve{H}_{\Hamminggeneralized}$,
$\ve{H}_{\BCH}$, and the parity-checks found in Step $3$.}

\end{description}

\end{appendices}

\textbf{Acknowledgment:} The authors are grateful to the anonymous
reviewers for their constructive comments that significantly
improved the exposition of the work. In addition, they would like
to thank Dr. Richardson for handling the manuscript.

\bibliography{LDPC_Group_Bibfile}
\bibliographystyle{IEEEtran}


\end{document}